\documentclass[sn-mathphys]{sn-jnl}% Math and Physical Sciences Reference Style

\pdfoutput=1
\jyear{2022}%

\newcommand{\pcoreth}{p_\mathrm{core,th}}
\newcommand{\pgrossel}{p_\mathrm{gross,el}}
\newcommand{\pgrosselbinned}{p_\mathrm{gross,el}^\mathrm{binned}}

\newcommand{\tdwell}{t_\mathrm{dw}}
\newcommand{\tpulse}{t_\mathrm{pu}}
\newcommand{\ract}{r_\mathrm{act}}
\newcommand{\rpass}{r_\mathrm{pass}}
\newcommand{\rstart}{r_\mathrm{start}}
\newcommand{\estart}{e_\mathrm{start}}
\newcommand{\thermeff}{\eta_\mathrm{th}}

\newcommand{\corevom}{C_\mathrm{vom,core}}
\newcommand{\geninv}{C_\mathrm{inv,PCS}}
\newcommand{\genvom}{C_\mathrm{vom,PCS}}
\newcommand{\genfom}{C_\mathrm{fom,PCS}}
\newcommand{\totvom}{\pi^{\mVOM, total}}
\newcommand{\netavgcap}{\overline{\textsc{cap}}_\mathrm{net}^\mathrm{el}}
\newcommand{\peakthcap}{\textsc{cap}_\mathrm{peak}^\mathrm{th}}

\newcommand{\plantcapex}{\mathcal{C}_\mathrm{plant}}
\newcommand{\fixedchargerate}{f_\mathrm{CR}}
\newcommand{\slcoe}{\textsc{lcoe}}

\newcommand{\capacityfactor}{f_\mathrm{capacity}}

\renewcommand{\corevom}{\coreVOMCost}
\renewcommand{\geninv}{\genInvCost}
\renewcommand{\genfom}{\genFOMCost}
\renewcommand{\thermeff}{\dischargeEfficiency}
\renewcommand{\genvom}{\genVOMCost}

\newcommand{\techAndZone}{}

\newcommand{\mCore}{th}

\newcommand{\mInvest}{invest}
\newcommand{\mFOM}{FOM}
\newcommand{\mVOM}{VOM}

\newcommand{\genInvCost}{\pi^{\mInvest}_{\techAndZone}} % investment cost of power conversion system
\newcommand{\genFOMCost}{\pi^{\mFOM}_{\techAndZone}} % FO&M cost of power conversion system
\newcommand{\coreVOMCost}{\pi^{\mVOM, \mCore}_{\techAndZone}} % VO&M cost of the thermal core
\newcommand{\genVOMCost}{\pi^{\mVOM}_{\techAndZone}} % VO&M cost of power conversion system
\newcommand{\dischargeEfficiency}{\eta^{discharge}_{\techAndZone}} % Efficiency of converting thermal energy to electrical energy
\newcommand{\genMinStableOutput}{\rho^{min}_{\techAndZone}} % minimum stable power output level

\usepackage[version=4]{mhchem}
\usepackage{dcolumn}
\newcolumntype{d}[1]{D{.}{.}{#1}}

\usepackage{siunitx}
\usepackage{xfrac}

\DeclareSIUnit{\year}{yr}

\begin{document}

\title[Valuing fusion for the US]{The value of fusion energy to a decarbonized United States electric grid}

%%=============================================================%%
%% Prefix	-> \pfx{Dr}
%% GivenName	-> \fnm{Joergen W.}
%% Particle	-> \spfx{van der} -> surname prefix
%% FamilyName	-> \sur{Ploeg}
%% Suffix	-> \sfx{IV}
%% NatureName	-> \tanm{Poet Laureate} -> Title after name
%% Degrees	-> \dgr{MSc, PhD}
%% \author*[1,2]{\pfx{Dr} \fnm{Joergen W.} \spfx{van der} \sur{Ploeg} \sfx{IV} \tanm{Poet Laureate}
%%                 \dgr{MSc, PhD}}\email{iauthor@gmail.com}
%%=============================================================%%

\author*[1,2]{\fnm{Jacob} \sur{Schwartz}}\email{jacobas@princeton.edu}
%\equalcont{These authors contributed equally to this work.}

\author[1,2]{\fnm{Wilson} \sur{Ricks}}\email{wricks@princeton.edu}
%\equalcont{These authors contributed equally to this work.}

\author*[1,2,3]{\fnm{Egemen} \sur{Kolemen}}\email{ekolemen@princeton.edu}
%\equalcont{These authors contributed equally to this work.}

\author[1,2]{\fnm{Jesse} \sur{Jenkins}}\email{jdj2@princeton.edu}
%\equalcont{These authors contributed equally to this work.}

\affil*[1]{\orgdiv{Department of Mechanical and Aerospace Engineering}, \orgname{Princeton University}, \orgaddress{\street{Olden St.}, \city{Princeton}, \postcode{08540}, \state{NJ}, \country{USA}}}

\affil[2]{\orgdiv{Andlinger Center for Energy and the Environment}, \orgname{Princeton University}, \orgaddress{\street{Olden St.}, \city{Princeton}, \postcode{08540}, \state{NJ}, \country{USA}}}
\affil[3]{\orgname{Princeton Plasma Physics Laboratory}, \orgaddress{\street{100 Stellarator Rd.}, \city{Princeton}, \postcode{08543}, \state{NJ}, \country{USA}}}

%%==================================%%
%% sample for unstructured abstract %%
%%==================================%%

\abstract{
Fusion could be a part of future decarbonized electricity systems, but it will need to compete with other technologies.
In particular, pulsed tokamak plants have a unique operational mode, and evaluating
which characteristics make them economically competitive can help select between design pathways.
Using a capacity expansion and operations model,
we determined cost thresholds for pulsed tokamaks to reach a range of penetration levels in a future decarbonized US Eastern Interconnection.
The required capital cost to reach a fusion capacity of \SI{100}{\giga\watt} varied from \$3000 to \$\SI{7200}{\per\kilo\watt},
and the equilibrium penetration increases rapidly with decreasing cost.
The value per unit power capacity depends on the variable operational cost and on the cost of its competition, particularly fission, much more than on the pulse cycle parameters.
These findings can therefore provide initial cost targets for fusion more generally in the United States.
}

\keywords{nuclear fusion, tokamaks, capacity expansion, technology assessment}

\maketitle

\section{Introduction}\label{sec1}

%The Introduction section, of referenced text expands on the background of the work (some overlap with the Abstract is acceptable). The introduction should not include subheadings\cite{moscato_thermal_2019, minucci_electrical_2020, barucca_pre-conceptual_2021}.

Technology for the production of electrical power via nuclear fusion is under development by governments and private companies around the world\cite{windsor_can_2019}.
In fusion reactors, light atomic nuclei undergo exothermic reactions in a hot plasma, and
the kinetic energy of the products heats a working fluid\cite{okazaki_fusion_2021} or is converted directly to electrical energy\cite{miley_review_1991}.
Fusion would be a firm energy resource\cite{sepulveda_role_2018} without operational \ce{CO2} emissions, and could contribute to the deep decarbonization of the electricity sector in the United States and elsewhere.
However, even if fusion's physics and engineering challenges are overcome, there are many other technologies for electricity production.
Fusion will need to compete economically with these other technologies in order to be part of the US energy mix.
Within fusion itself there are multiple reactor concepts, with wide plausible ranges for operational parameters.
While it is difficult to determine the cost of a particular design when much of the underlying technology has yet to be developed, it is possible to set cost targets by determining the value of a design with a particular set of operational parameters in a simulated future electricity system.
This value-driven approach\cite{de_sisternes_value_2016, mallapragada_long-run_2020, heuberger_systems_2017} can help to determine which fusion concepts or technology pathways would be most useful when deployed alongside future competing and complementary energy technologies.% that lead to plants with various operational characteristics.

There have been general studies on markets for fusion energy\cite{handley_potential_2021}, on fusion costing\cite{sheffield_generic_2016, waganer_aries_2013}, its integration into energy systems\cite{tynan_how_2020}, and on the value of fusion for future electricity systems in Europe\cite{muller_analysis_2019}.
However, this work is the first in which plants with fusion-specific operational constraints have been integrated into a temporally-resolved system-scale model.
This allows for investigation of the costs of these constraints and of how fusion interacts with other resources on an hour-by-hour basis.
This is also the first study of the equilibrium value of fusion at various levels of capacity penetration for the United States, and
the first investigation of the value of integrated thermal storage for fusion plants in an hourly model.

In this study we used GenX, an electricity system capacity expansion model, to capture the operational and economic interactions between fusion plants and a number of competing energy technologies in simulated future electricity systems \cite{genx_software}. GenX is a linear optimization model that identifies an optimal set of energy technology deployment, retirement, and operational decisions to minimize total electricity system cost over a specified planning horizon. As is done in real power systems, GenX solves an `economic dispatch'\cite{al_farsi_economic_2015} problem that dispatches grid assets so as to minimize the cost of exactly meeting electricity demand at every hour over the course of a modeled year, subject to a number of operational constraints. GenX and similar capacity expansion models also capture the declining value of energy technologies with increasing deployment, arriving at a long-run economic equilibrium in which each non-constrained technology's marginal cost equals its marginal value to the system. This reflects the assumption that an effective central planner will only make investments that reduce the total cost of serving electricity demand, or equivalently that developers in a deregulated market will continue to deploy a given technology until marginal capacity additions become unprofitable. Capturing both these short- and long-run system-level interactions is necessary in order to accurately assess market sizes and cost targets for an emerging energy technology like fusion.

There are multiple existing concepts for fusion power plants, including pulsed\cite{sorbom_arc_2015, creely_overview_2020, gryaznevich_pulsed_2022} and steady-state\cite{buttery_advanced_2021} tokamaks, stellarators\cite{najmabadi_aries-cs_2008, warmer_w7-x_2017, alonso_physics_2022}, laser-driven inertial confinement devices\cite{dunne_timely_2011, tikhonchuk_progress_2020}, magnetized target fusion systems\cite{laberge_magnetized_2019, laberge_magnetized_2021},
mirror machines\cite{fowler_new_2017}, field-reversed configurations\cite{rostoker_colliding_1997}, and Z-pinches\cite{meier_nuclear_2006, shumlak_z-pinch_2020}.
For the present study we developed an abstracted operational model for a fusion plant using a pulsed tokamak design that could be linearized and implemented in GenX (see Sections~\ref{sec:fusionmodel} and Note~S3).
The tokamak was chosen because it is the most mature fusion concept---ITER\cite{shimada_progress_2007}, DEMO\cite{federici_demo_2018-1}, and one of the largest private fusion companies employ it---and because it has the most general set of performance characteristics when examined with hourly resolution.

Tokamaks confine a hot plasma in a toroidal chamber using magnetic fields, some externally imposed, and some produced by an electric current flowing through the plasma.
Pulsed tokamaks use magnetic induction from a component called the central solenoid\cite{libeyre_detailed_2009} (CS) to drive this current.
During the `flat-top' of the plasma pulse, which is typically designed to last half an hour\cite{sorbom_recent_2020} to several hours\cite{federici_demo_2018-1}, a constant rate of change in magnetic flux in the CS drives a constant voltage (and current) in the plasma.
The flat-top is when most of the fusion occurs, which generates heat.
The CS has a maximum magnetic flux that it can hold,
and the plasma cannot be sustained without a driving voltage, so the flat-top must end.
During the following `dwell period', typically a fraction of an hour, the solenoid and other systems are reset, and no fusion occurs.
Restarting the plasma requires significant electrical power, with peak levels, for seconds or minutes, comparable to the output capacity of the plant\cite{minucci_electrical_2020}, which would likely be buffered by some storage on-site.
The net output of the plant therefore varies over the pulse cycle.
While this study focuses on pulsed tokamaks, the range of performance characteristics considered can represent a wide range of potential alternative concepts.

We used the fusion model as implemented in GenX to study the value and role of fusion power in a decarbonized US Eastern Interconnection circa the 2040s, optimizing electricity technology investments and hourly operations across 20 model zones to minimize total system cost.
In order to understand the design space of model tokamaks, we varied their behavior from pulsed to nearly steady-state, and varied the variable operations and maintenance cost to reflect uncertainty in the costs of replaceable components such as the blanket and divertor.
We explored the inclusion of integrated thermal storage with a range of capacity costs,
and we varied assumptions about the cost and availability of other resources to understand the sensitivity of fusion's value to these market uncertainties and how fusion interacts with these resources.
\sisetup{per-mode = symbol, sticky-per=true}
In the present work we optimized the electricity system with respect to an exogenously fixed total fusion capacity, using GenX outputs to calculate the cost threshold at which fusion could achieve this level of deployment. See the Methods section for a more detailed description of the procedure used to identify cost thresholds.
The GenX model as configured for this study includes the following constraints:
\begin{itemize}
\item zero direct carbon emissions from electricity system operations;
\item a `capacity reserve margin' policy to ensure sufficient generation capacity in a number of reliability regions;
\item supply curves reflecting resource constraints on solar and wind power in each zone;
\item operational constraints for thermal generators that specify a minimum output level while committed, limits on how often plants can cycle on or off, and maximum power ramp rates; 
\item limits on the expansion of transmission lines between zones and losses when power is transmitted between zones;
\item a fusion reactor core which can operate flexibly, without limits on thermal cycling;
\item no maintenance scheduling or unscheduled outages for fusion nor other resources;
\item no other specific policies, subsidies or external regulations.
\end{itemize}
Across a range of fusion plant designs and market scenarios, we find that reaching \SI{100}{\giga\watt} of fusion capacity (which is about 10\% of the peak demand, and similar to the present-day US fission fleet capacity) requires that the capital cost of the plant falls below \SI{3000}[\$]{\per\kilo\watt} to \SI{7200}[\$]{\per\kilo\watt} of net electric output capacity.
Roughly half of this range results from the space of internal fusion operational parameters, and half from uncertainty in the cost and performance of competing generation and storage technologies in future electricity markets.
The former half can be mostly attributed to the variation in the marginal cost of net generation, rather than to the variation in operational constraints such as the pulse length.
This implies that the results should generalize from pulsed tokamaks to other concepts, and simplifies prediction of the value of a design in a given scenario.
Between scenarios, the value of fusion differs in large part because of the differing costs of fission, and fusion's competition with other resources becomes significant only after fission has been displaced.
Including the option to build thermal storage increases the value of the fusion core by up to \SI{1000}[\$]{\per\kilo\watt},
and in some scenarios, adding thermal storage increases the relative value of solar and wind.

\section{A representation of pulsed tokamaks for electricity systems modeling}
\label{sec:fusionmodel}
\begin{figure}%
\centering
\includegraphics[width=0.95\textwidth]{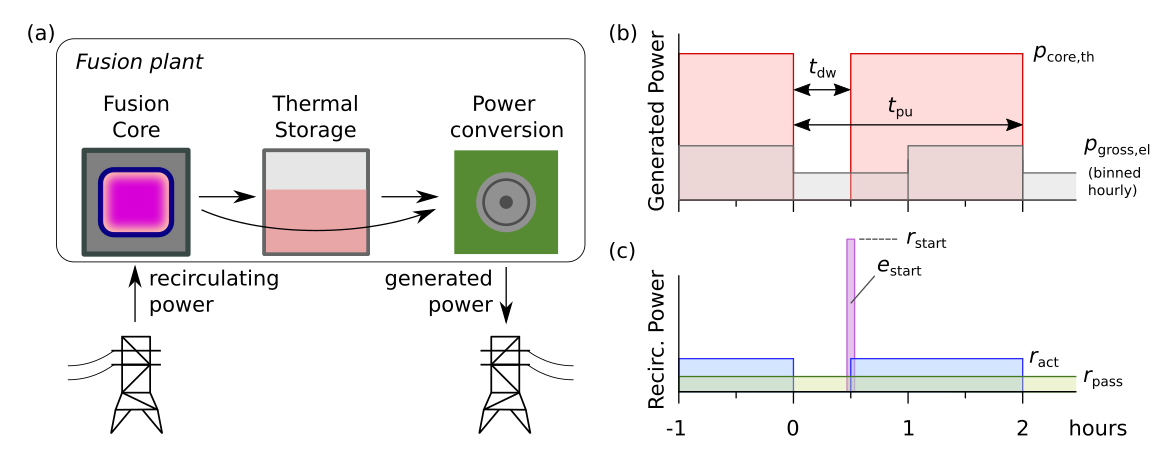}
\caption{Part (a): Fusion plant diagram. Parts (b) and (c) show qualitatively the unresolved, sub-hourly structure of the plasma pulse and recirculating power flows in the pulsed tokamak fusion core.
These plots provide one interpretation for the mathematical models implemented; in reality there would be additional structure and detail\cite{minucci_electrical_2020}.
Part (b) plots the $p_\mathrm{core,th}$, the thermal power generated by a core with a total pulse cycle time $\tpulse = \SI{2}{\hour}$ and a dwell time $\tdwell = \SI{0.5}{\hour}$.
It also shows, in gray, the gross hourly electric power generated if thermal storage is not employed, $\pgrosselbinned$.
Part (c) shows the three types of recirculating power: passive, active and the startup power. The associated startup energy fraction is also labeled, $\estart$.
}\label{fig:fusionplantlayout}
\end{figure}
For this study we developed a fusion plant model for integration into GenX.
Figure~\ref{fig:fusionplantlayout} part (a) illustrates the modeled components of the plant.
There are three parts: a fusion core which takes in parasitic `recirculating' power from the grid and makes heat,
an (optional) thermal storage system (TSS), which stores heat between hourly periods, and a power conversion system (PCS) that takes heat from the core or the TSS and generates electricity for the grid.

Parts (b) and (c) illustrate the operation of the core which is then binned into hourly increments.
The core operates on cycles of length $\tpulse$, an integer number of hours.
During the first hourly period it rests for the `dwell period'  $\tdwell$, a fraction of an hour during which time it generates no heat.
Plasma ramp-up and ramp-down periods are not explicitly accounted for since they are typically much shorter than an hour; these could be modeled by adjusting the length of the dwell period.
While the core is operating it generates a peak thermal power $\pcoreth$.
This corresponds to a peak gross electric power $\pgrossel = \thermeff \pcoreth$ where $\thermeff$ is the thermal efficiency of the PCS.
If there is no thermal storage system, the heat generated in an hourly period must be converted to electrical energy during the same period; the gross electric power $\pgrosselbinned$ therefore varies from hour to hour.

The recirculating electrical power, shown in (c), is described by four dimensionless parameters which denote fractions of the peak gross electric power: $\rpass$, $\ract$, $\estart$, and $\rstart$.
The passive recirculating power $\rpass \pgrossel$ is drawn regardless of the core's status, the active recirculating power $\ract \pgrossel$ is drawn proportionally to the fraction of the hour during which the core operates,
and the start energy $\estart \pgrossel\,\si{\hour}$ (where \si{\hour} is one hour) is a fixed quantity required for each core start.
The start power level $\rstart \pgrossel$ models a brief peak power draw from outside the plant. 
It does not enter into the recirculating power calculation; rather, spare power capacity must be available in the same zone during the hour that the core starts.

A final parameter for the core is $\corevom$, the variable operations and maintenance (OM) cost of generating a quantity of heat from the core.
This represents the cost of replacements for the blanket and divertor, which are assumed to need replacement after being bombarded with a certain quantity of neutrons; neutron exposure is proportional to heat generated in the core.
Damage due to accumulation of thermal cyclic fatigue is not explicitly modeled.
However, since the fusion plants generally operate near their peak capacity, costs of thermal cyclic fatigue accumulated during successively-pulsed operation could be incorporated into the variable operational cost.

The PCS has five main parameters: the thermal efficiency $\dischargeEfficiency$;
the capital cost $\geninv$; the fixed OM cost $\genfom$; and the variable OM cost $\genvom$, which is accrued in proportion to the gross electric power generated;
and the minimum output level $\rho^{min}$.
The thermal storage system (TSS) has only one parameter, the storage capacity investment cost, which is varied between specific cases.
For simplicity, we assume no efficiency losses associated with the thermal storage, nor costs associated specifically with maximum inflow or outflow rates.

From the parameters of the core and PCS, one can calculate several derived quantities: $f_\mathrm{active}$, the fraction of the time that the core can be active; 
$f_\mathrm{netavgcap}$, the ratio of the time-averaged net electric power produced to the gross
electric power generation capacity;
$f_\mathrm{recirc}$, the fraction of the gross electric power which is used by the plant itself;
$Q_\mathrm{eng}$, the ratio of the net power output to the recirculating power;
and $\totvom$, the total cost of generating a unit of net electrical energy.
Formulas for these are given in the Methods section.

\begin{table}
\centering
\caption{Reference pulsed tokamak models used for this study.}\label{tab:fusmodels}
\begin{tabular}{@{}lllll@{}} \toprule
& Pessimistic & Mid-range & Optimistic &  \\ \midrule
\multicolumn{3}{@{}l}{Core parameters} & \\
$\tpulse{}$ & 2 & 4 & 1  & h \\
$\tdwell{}$ & 0.15 & 0.15 & 0.063 & h\\
$\ract{}$ & 0.2 & 0.1 & 0.014 & \\
$\rpass{}$ & 0.2 & 0.1 & 0.027 & \\
$\rstart{}$ & 0.2 & 0.1 & 0 & \\
$\estart{}$  & 0.05 & 0.025 & 0 & \\
$\corevom$ & 5 & 3 & 1  & $\si{\$/\mega\watt\hour}_\mathrm{th}$ \\ \midrule
\multicolumn{3}{@{}l}{Power conversion system parameters} & \\
$\thermeff$ & \multicolumn{3}{c}{0.4} &  \\
$\genMinStableOutput$ & \multicolumn{3}{c}{0.4} & \\ 
$\geninv$ & \multicolumn{3}{c}{750} & $\si{\$/\kilo\watt}_\mathrm{e}$  \\
$\genfom$  & \multicolumn{3}{c}{18.75\ } & $\si{\$/\kilo\watt}_\mathrm{e}\si{\year}$ \\
$\genvom$ & \multicolumn{3}{c}{1.74} & $\si{\$/\mega\watt\hour}_\mathrm{e}$ \\ \midrule
\multicolumn{3}{@{}l}{Derived quantities} & \\
$f_\mathrm{active}$ & 0.925 & 0.9625 & 0.9375 \\
$f_\mathrm{netavgcap}$ & 0.515 & 0.76 & 0.897 \\
$f_\mathrm{recirc}$ & 0.44 & 0.21 & 0.043 & \\
$Q_\mathrm{eng}$ & 1.26 & 3.76 & 22.4 & \\
$\totvom$ & 26 & 12 & 4.4  & $\si{\$/\mega\watt\hour}_\mathrm{e}$ \\ \bottomrule
\end{tabular}
\end{table}
The fusion plants in this study are based on one of three reference designs, listed in Table~\ref{tab:fusmodels}.
These are labeled pessimistic, mid-range, and optimistic, based on their core parameters and especially their resulting VOM cost $\totvom$.
The pessimistic plant requires a large dwell time between pulses, about 44\% of the gross electric power generated is required to operate the plant\cite{zohm_minimum_2010, minucci_electrical_2020}, and the marginal cost of net energy generation $\totvom$ is \SI{26}[\$]{\per\mega\watt\hour}, closer to that of a natural gas plant with carbon capture and storage (NG-CCS) than that of a fission plant.

The optimistic plant has a shorter dwell time, recirculating power levels tenfold lower, and has less costly operation at \SI{4.4}[\$]{\per\mega\watt\hour}, half that of fission plants.
A mid-range design has pulse cycle parameters, recirculating power, and marginal costs of net generation roughly halfway between the optimistic and pessimistic designs.

All plants share the same PCS design. 
The assumed capital cost of \SI{750}[\$]{\per\kilo\watt} is a 28\% reduction, based on an economy of scale, from the `Power cycle' cost of the Molten Salt Power Tower in the 2018 NREL System Advisor Model\cite{blair_system_2018}: see Table~3 of Turchi\cite{turchi_csp_2019}.
The thermal conversion efficiency is $\thermeff = 0.4$, and the variable operations and maintenance cost $\genvom = \SI{1.74}[\$]{\per\mega\watt\hour}$.

The annual fixed OM cost is assumed to be 2.5\% of the capital cost; the same assumption is made for the TSS and the core itself.

\section{Cost targets for fusion plants without integrated thermal storage}
We determined cost thresholds as function of capacity penetration for the three reference plants in three main scenarios. 
The three scenarios, termed low, medium, and high fusion market opportunity, differ in the cost of the available resources, which are: solar photovoltaics (PV), on- and off-shore wind, fission, natural gas plants with 100\% carbon capture and storage (NG-CCS),
%combined-cycle and combustion turbine plants burning a zero-carbon fuel (ZCF-CC and ZCF-CT, respectively),
plants burning a zero-carbon fuel such as hydrogen or biomethane in a combined cycle or combustion turbine (ZCF-CC or ZCF-CT, respectively),
lithium-ion batteries and metal-air batteries.
The investment costs and operational costs of these resources, listed in Tables~\ref{tab:capex} and~\ref{tab:varcosts}, respectively, are lowest in the low market opportunity scenario and highest in the high market opportunity scenario.
All the scenarios have identical nominal loads, with average and peak values of \SI{600}{\giga\watt} and \SI{1100}{\giga\watt}, respectively. 
One difference is in the quantity of certain loads, representing electric vehicle charging and residential hot water-heaters, which can be shifted in time by a few hours: 
for example, in the low, medium, and high market opportunity scenarios, 0.9, 0.75, and 0.6, respectively, of the vehicle charging loads can be delayed by up to 5 hours. 
See Table~S2 for full details.
The three scenarios do not differ in the costs or maximum procurable quantities of inter-regional transmission.
In the high market opportunity scenario, fission is not built because it is too expensive, and nearly the maximum amount of transmission is required for the resulting renewables-dominated grid.
In cases in medium and low market opportunity scenarios without fusion, there is about \SI{100}{\giga\watt} of fission (see also Figs S35 and S36) and the grid is less limited by transmission constraints.

\begin{figure}
\centering
\includegraphics[width=\textwidth]{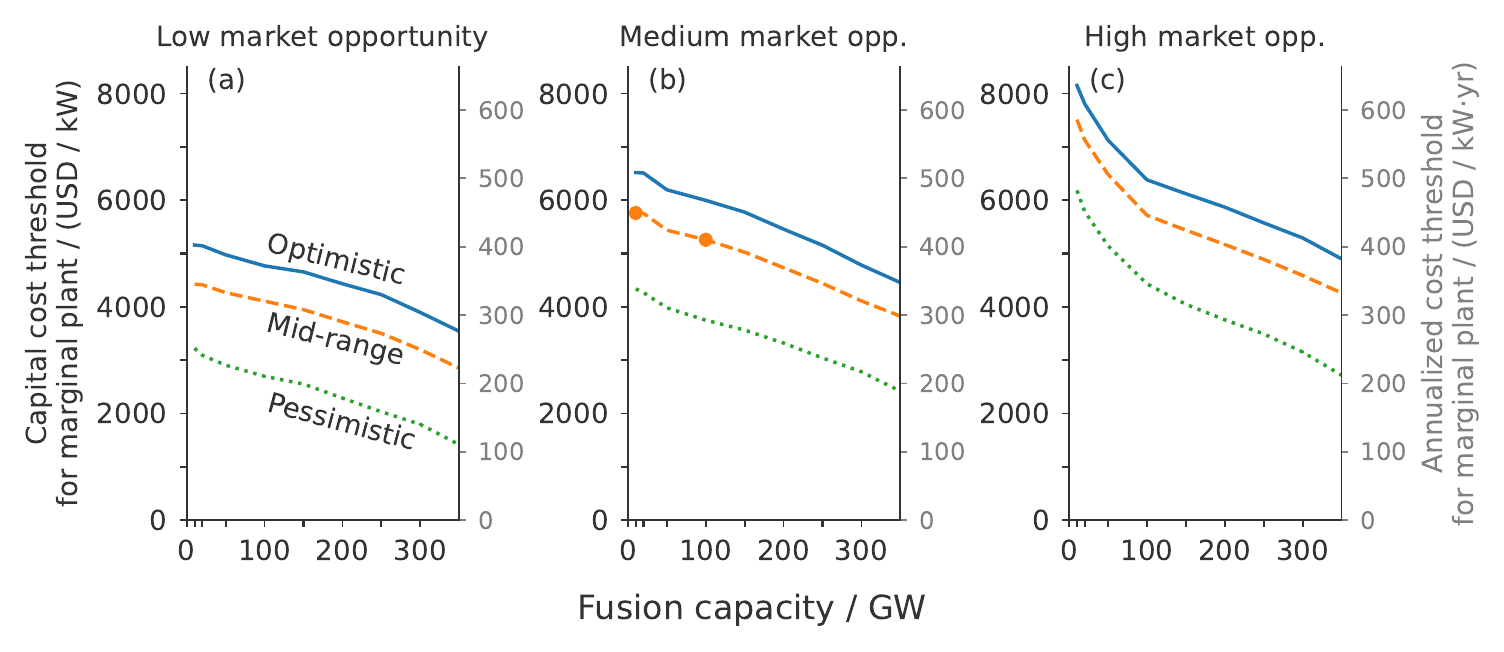}
\caption{Cost thresholds for a marginal unit of fusion capacity as a function of the total installed fusion capacity.
This corresponds to the maximum break-even capital cost for the final fusion plant built to reach a specified total installed capacity.
Results are presented for three reference reactors without intermediate thermal storage and in the three market opportunity scenarios.
Annualized costs are 7.8\% of the capital cost per year: see Note S2 for details.
The two bold points on the mid-range core, medium market opportunity scenario curve highlight how a decrease in capital cost of \$500/kW can increase the capacity penetration from \SI{10}{\giga\watt} to \SI{100}{\giga\watt}.
%Also see Fig.~SX for cost thresholds in additional scenarios.
}\label{fig:buildout}
\end{figure}
Figure~\ref{fig:buildout} shows the cost thresholds for a marginal plant
for each plant design in each scenario, as the fusion capacity penetration is set from \SI{10}{\giga\watt} to \SI{350}{\giga\watt}.
In equilibrium, the capital cost of a plant built to reach a specified total installed capacity must be equal to or lower than this respective curve.
For a fixed fusion capacity, the cost targets differ between plant designs as much as they differ between the three market opportunity scenarios. % duh
For many of the curves a small cost decrease leads to much wider adoption:
for the mid-range reactor in the medium opportunity scenario, the two bold points show where a cost decrease of \SI{500}[\$]{\per\kilo\watt} increases the equilibrium fusion capacity from \SIrange{10}{100}{\giga\watt}.
This suggests that if the initial cost targets can be met, even shallow learning curves could lead to a significant fusion capacity.

\section{Internal and external drivers of value}\label{sec:intext}
Within a given market opportunity scenario, the difference in value between these pulsed tokamak designs is driven by the difference in the marginal cost of net generation $\totvom$, and between market opportunity scenarios, by differences in the costs of competitor resources.

\begin{figure}
\centering
\includegraphics[width=0.90\textwidth]{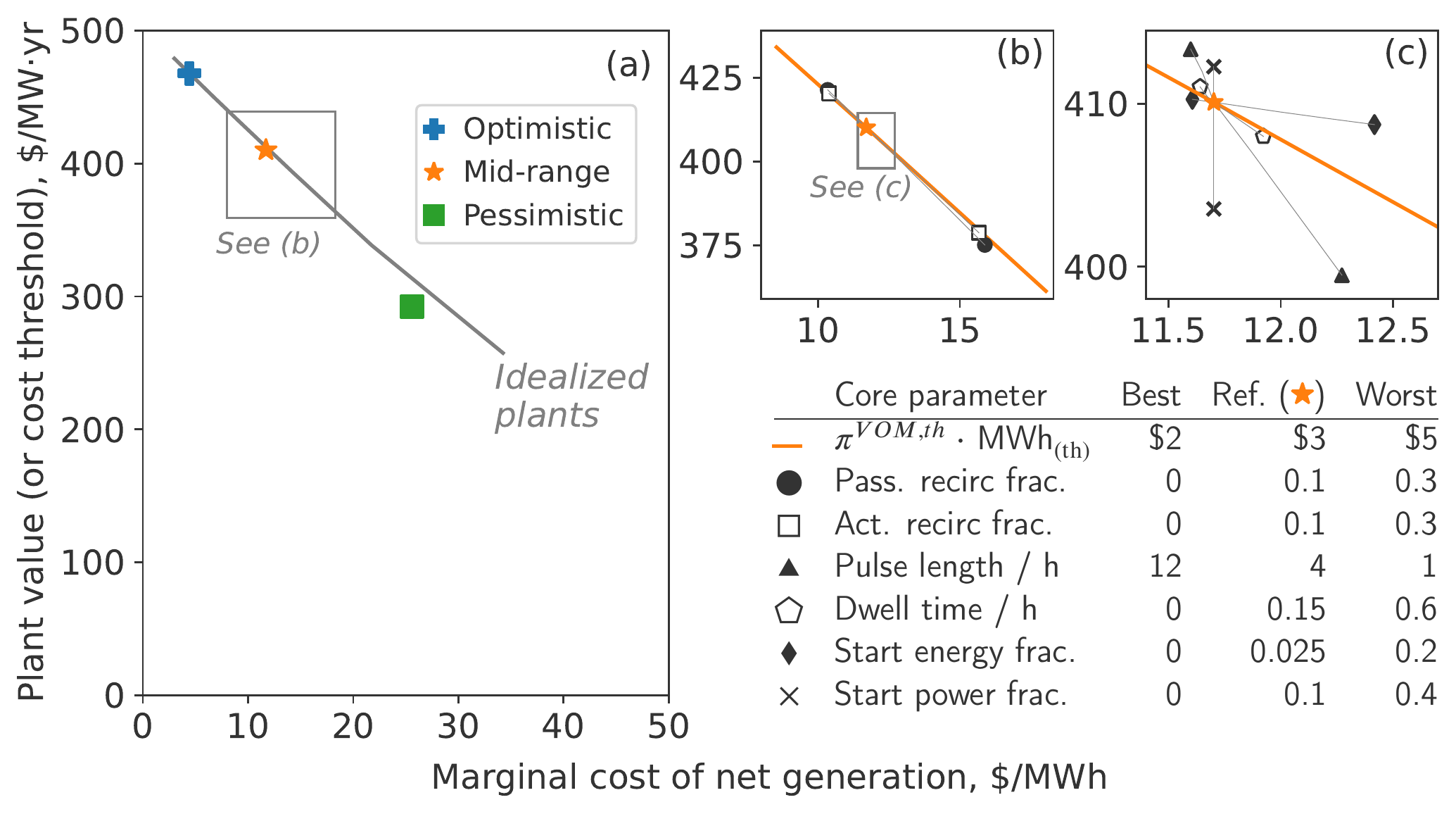}
\caption{Part (a): The marginal value of fusion plants at \SI{100}{\giga\watt} of capacity in the medium market opportunity scenario are shown as a function of their marginal cost of net generation.
The gray line shows the values of a set of plants with an idealized fusion core, which has no pulse constraints or recirculating power, but finite VOM costs.
Parts (b) and (c): as the parameters of a plant with a mid-range fusion core are altered from their reference values, one at a time, variation in plant value is largely due to the variation in the marginal cost of net generation, shown by the orange line.
The table lists the ranges over which the parameters were varied.
The unitless parameters are specified as fractions of the plant's peak gross electric power; see the Section~\ref{sec:fusionmodel}.
}\label{fig:variations}
\end{figure}

Figure~\ref{fig:variations}(a) plots the three reference plants, along with a gray line representing a set of idealized plants with no recirculating power or pulse constraints, but finite VOM costs.
All three reference plants lie close to this curve, which demonstrates that the penalty of pulsed operation is small.
The mid-range plant value (orange star) is just 0.6\% below that of the idealized plant with the same marginal cost of net generation; the pessimistic plant value (green square) is about 6\% lower than that of the equivalent.
Its deviation from the curve is driven largely by the core start power constraint.

Parts (b) and (c) examine the variation in the value of the mid-range plant as its operational parameters are modified, one at a time, over ranges shown in the table.
After $\corevom$, the two parameters with the largest effect are the passive and active recirculating power fractions.
The slopes of these are nearly the same as the slope of the curve formed by modifying $\corevom$.
This indicates that these parameters have altered the plant value primarily through changing the marginal cost of net generation.
Modifying the other quantities yields variations in plant value of less than 3\%.
When the pulse length is changed from 4 hours to 1 hour the value decreases due to the increased marginal cost of net generation, and also due to the increased quantity of power which must be drawn from the grid; the same value decrease is observed when quadrupling the core start power at the original pulse length.
Since within a given scenario the plant value is predicted almost entirely by the cost of marginal generation, our studies should be applicable to assess the value of a wide range of devices, not only pulsed tokamaks, as part of similar future electricity systems.
Given this finding, much of the further study focuses on plants like the mid-range reference plant with a modified $\corevom$.

\begin{figure}%
\centering
\includegraphics[width=3.5in]{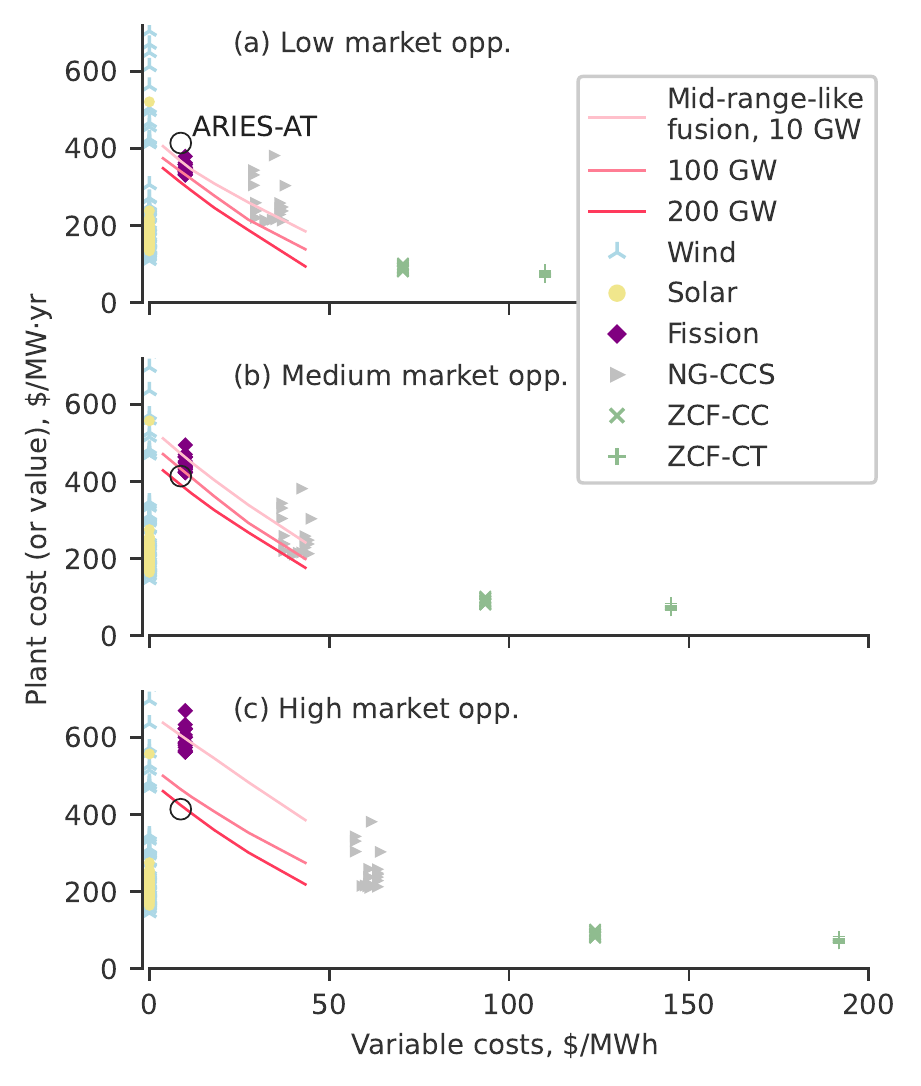}
\caption{For a given fusion plant design, differences in its value between scenarios depends largely on the costs of competitor resources.
Cost thresholds to reach three capacity penetration levels
for mid-range-like fusion plants are shown alongside the costs of other resources.
These plants have the operational constraints of the mid-range reference plant but altered variable costs.
Resource costs are plotted separately for each geographic zone and costs of solar and wind are normalized by their annual availability.
NG-CCS are plants burning natural gas with carbon capture and storage.
ZCF-CC and -CT are combined cycle and combustion turbine plants, respectively, burning zero-carbon fuels.
%Some competitors with high variable costs are not shown; see Table~SX.
A marker for the ARIES-AT fusion plant study\cite{najmabadi_aries-at_2006} is provided for comparison.
%Contours of constant cost of electricity show that fusion capacity penetration is not determined strictly by this metric.
Fig.~S14 shows cost and value data in the variant scenarios.
%\todo{mark opt, pes, and mid-range vom costs}.
}\label{fig:scenario_competition}
\end{figure}

The marginal value of fusion is determined by the resources that it competes with or complements, % this is kind of trivially true
and the composition of this set depends on the $\totvom$ of a given fusion plant.
Figure~\ref{fig:scenario_competition} shows how the cost thresholds for fusion at various capacity penetrations vary between the three main scenarios with the cost of other resources.
Particularly at low fusion capacity penetrations, the cost of fission strongly affects the potential value of fusion:
in order for a fusion plant with a similar $\totvom$ to that of fission to be built, it must have a lower capacity cost.
Note that a significant penetration of fusion is not guaranteed even if neither fission nor natural gas with carbon capture and storage is available: in additional scenarios with neither (see Fig.\ S14) fusion competes with a combination of renewables, storage, and peaker plants burning zero-carbon fuels, with the last acting as a firm generator.

\begin{figure}%
\centering
\includegraphics[width=0.95\textwidth]{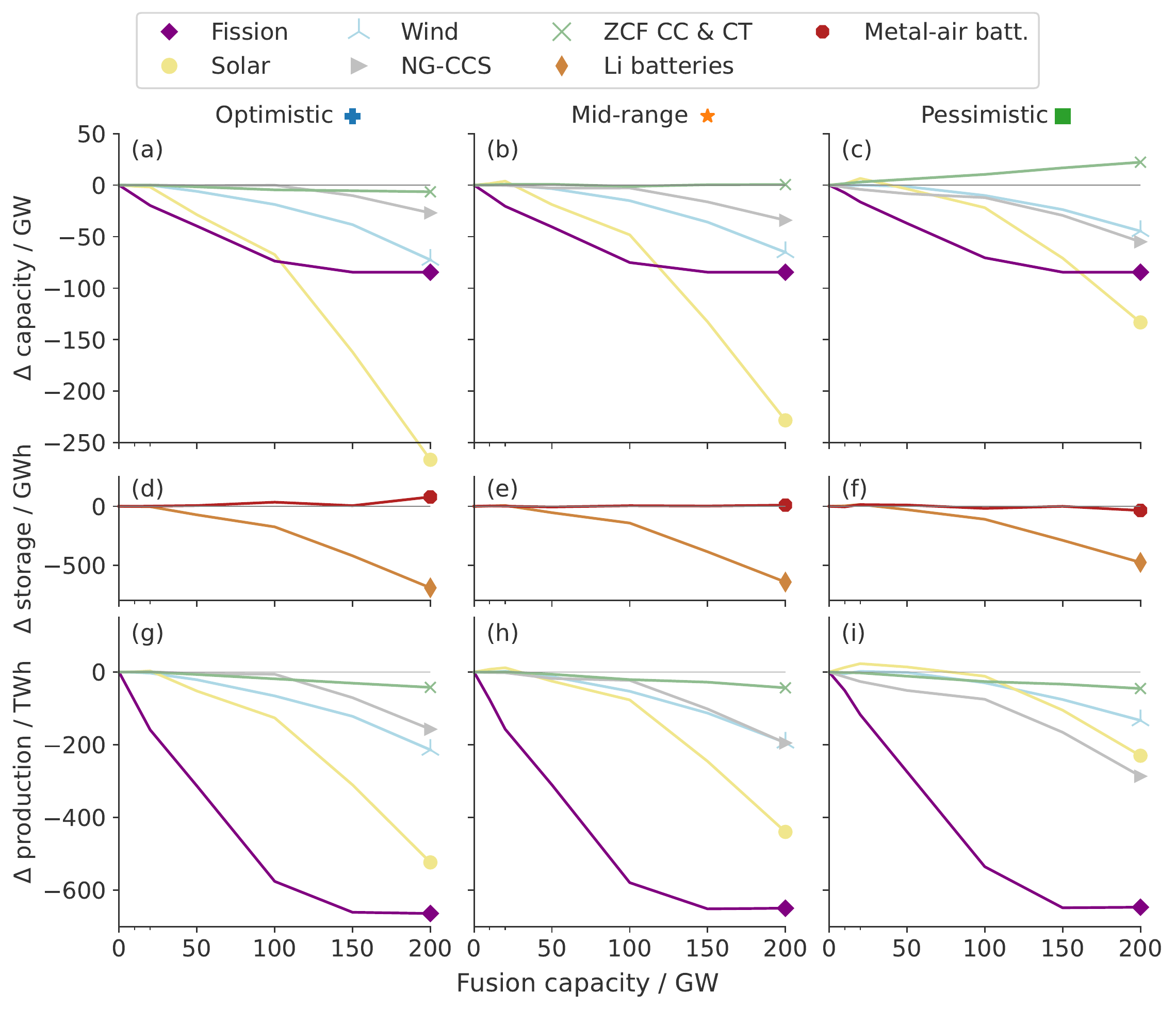}
\caption{Altered generation capacity, storage capacity, and annual energy production of competitor resources as a function of fusion's capacity penetration, in sets of cases for each of the three reference fusion plants, for the medium market opportunity scenario.
See also Fig.~S19 for the absolute quantities of each resource in this scenario, Figs.~S15--S24 for those in other scenarios, and Figs.~S25--S34 for altered quantities in other scenarios.
% Fusion primarily competes with fission at until the latter is displaced. At higher capacities it must compete with gas with CCS, and at still higher capacities with variable renewables and batteries. Peaker plants running on zero-carbon fuels are affected least.
}\label{fig:competition_replacement}
\end{figure}

Figure~\ref{fig:competition_replacement} shows explicitly that (in the medium market opportunity scenario) fission is the first competitor for any of the three reference plants.
In terms of annual energy production, all three primarily substitute for fission until \SI{100}{\giga\watt} when the latter is fully displaced.
Afterward, plants with the optimistic design displace mainly solar, wind, and batteries, while the pessimistic plant substitutes for more NG-CCS than solar.
In systems with fusion plants of the pessimistic design, slightly more solar is built at low fusion capacity penetrations, as recirculating power can be drawn from solar that would have been curtailed.
%See Figs.~S15--S24 for charts of the energy mix in several scenarios, and Figs.~S25--S34 for charts similar to Figure~\ref{fig:competition_replacement} which highlight the changes in the energy mix with fusion.
Integrated thermal storage for fusion plants, discussed in the next section, further increases the value of solar and decreases the value of other resources---see also Figs.\ S10--S13.

\section{Value of plants with thermal storage}
\label{sec:storage}
Pulsed tokamak designs may require\cite{barucca_pre-conceptual_2021} an intermediate thermal storage system (TSS) to supply the power conversion system (PCS) with heat during the dwell period; the PCS typically cannot handle the sudden decline in heat production associated with the end of the fusion pulse.
However, these systems only store a few core-minutes of heat.
We studied the value of adding an inter-hourly TSS with energy capacity costs similar to those of molten salt between the fusion core and PCS.
We independently optimize the core capacity, storage energy capacity, and PCS generation capacity in each model zone.
This allows for generators to be oversized relative to their fusion cores, in order to generate more energy during the most valuable time periods, such as periods of peak demand and/or minimal wind and solar availability.
This section describes the increase in the value of the core with the option to build storage, and Section~\ref{sec:opcycles} describes changes in operational patterns for a plant with storage.

\begin{figure}
\centering
\includegraphics[width=\textwidth]{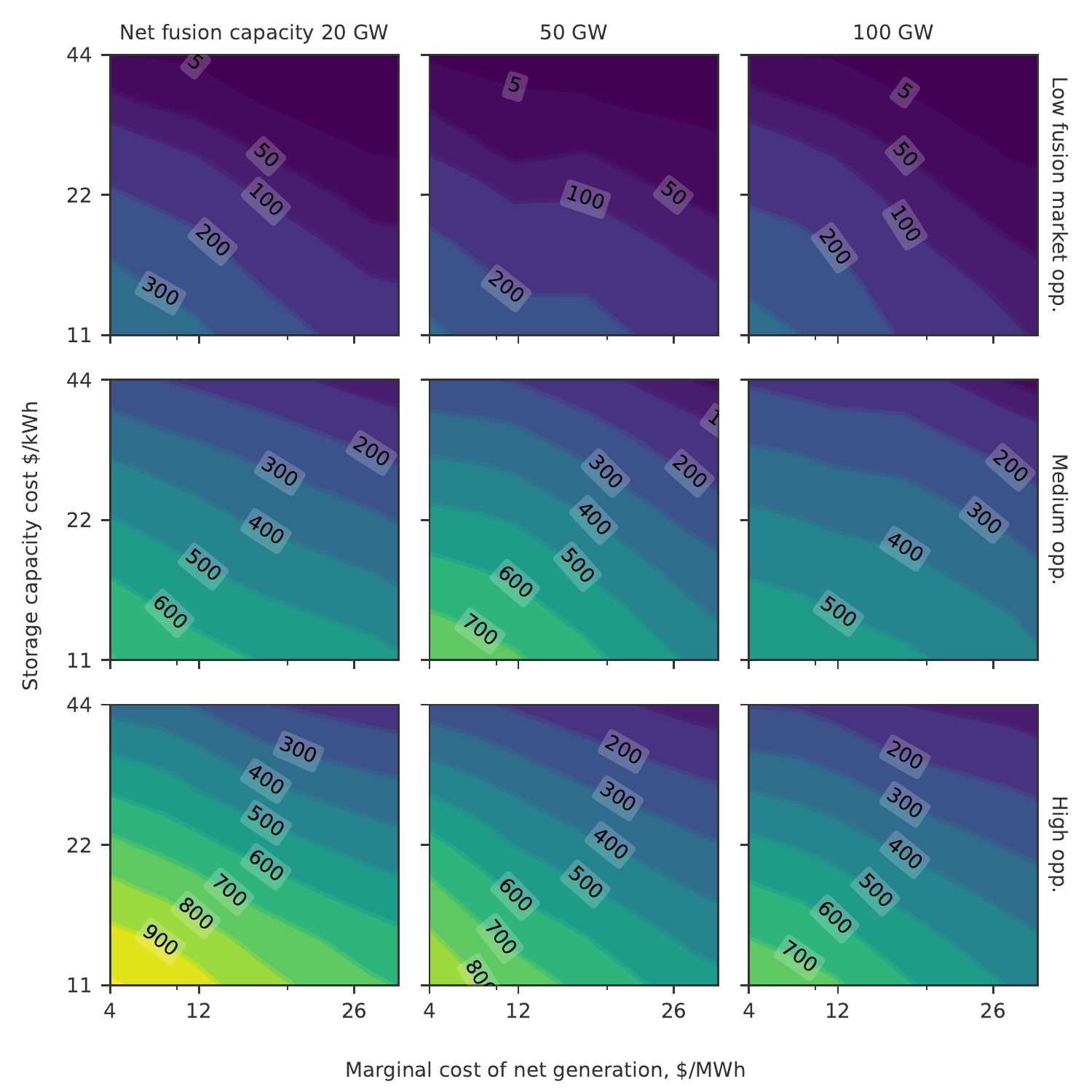}
\caption{Increase in the break-even price per unit of capacity for a mid-range-like fusion core with the option to build intermediate thermal storage to reach the given equilibrium penetration level, in \$/kW.
The plant's marginal cost of net generation $\totvom$ is varied as a proxy for variation between the three reference cores. See Figs.~S10--S13 for data on the total core value, optimal storage duration, and optimal PCS size.
}\label{fig:corevalincrease}
\end{figure}
Figure~\ref{fig:corevalincrease} shows the increase in the equilibrium value of the fusion core
with operational parameters like that of the mid-range reference plant, per unit of capacity,
as functions of the marginal cost of net generation $\totvom$ and the storage capacity cost.
Especially for plants in the medium and high fusion opportunity scenarios,
this is a substantial increase in the threshold capital cost for the plant core. For example, adding TSS at a cost of \$22/kW$_\textrm{th}$ increases the value of a mid-range fusion plant's core at \SI{50}{\giga\watt} of penetration by \SI{490}[\$]{\per\kilo\watt}, or about 10\%.
The option to build storage is more valuable at lower fusion capacity penetrations because the optimal storage quantity per plant is larger.
This suggests that a TSS could be especially valuable for the first generation of fusion plants.
As fusion penetration increases and the total thermal storage capacity along with it, the marginal value per unit of additional storage capacity declines.

The optimal thermal storage system (TSS) duration (Fig.\ S10) generally ranges from \SIrange{2}{8}{\hour}, depending foremost on the storage capacity cost, and the optimized PCS capacity (Fig.\ S11) generally ranges from 1.1 to 1.35 times the amount needed to serve the fusion cores without storage.
The durations are suitable for diurnal storage and allow the PCS to supply increased power during the night, when the lack of solar power production and overnight demand from EV charging and electric heat pumps makes electricity more valuable in this deeply decarbonized energy system.

\section{Capacity factors and the value of flexibility}
\begin{figure}
\centering
\includegraphics[width=0.95\textwidth]{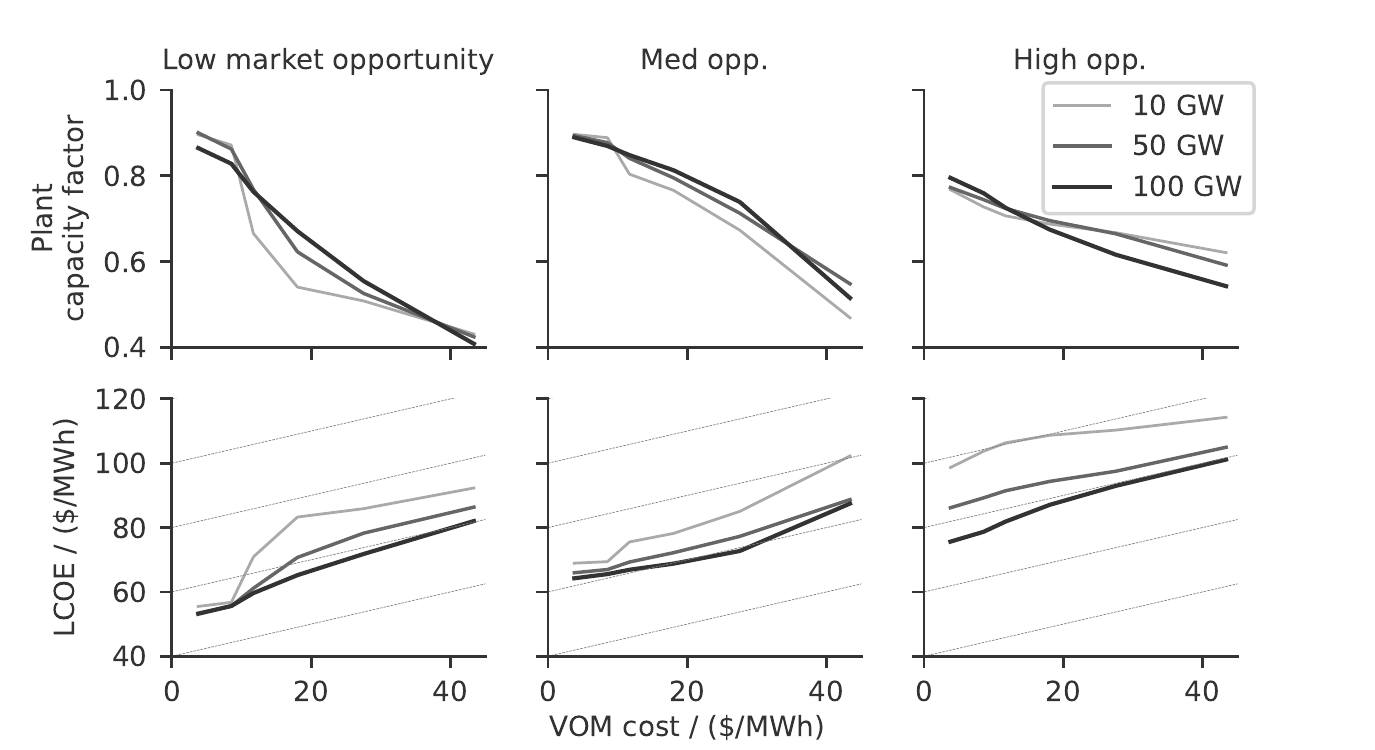}
\caption{Top row: capacity factors for mid-range-like plants with varying VOM costs in the three market opportunity scenarios, for three levels of capacity penetration, and without a TSS.
Bottom row: computed `cost of electricity' metrics for the same plants, taking into account their capacity factor. Dashed lines with a slope of 0.5 guide the eye.
}\label{fig:capacityandcoe}
\end{figure}

In grids dominated by variable renewables like those in this study,
the electricity price is often zero (in the base cases without fusion, 10-50\% of the year, depending on the scenario and geographic zone); 
generating electricity from fusion during these hours is not profitable.
As shown in Section~\ref{sec:intext}, an increased variable operations and maintenance (VOM) cost decreases the marginal value of a plant.
Plants with a higher VOM cost are run less frequently, as under economic dispatch they are typically only called on to generate when electricity demand is high and plants with lower variable operating costs are already generating at maximum capacity.
Figure~\ref{fig:capacityandcoe}, parts (a) through (c), show the capacity factor---the ratio of the fusion plants' annual net output to their maximum possible net output---for mid-range-like plants without thermal storage systems in the three market opportunity scenarios, and for three levels of capacity penetration.
For the \SI{10}{\giga\watt} level in the low and medium market opportunity scenarios, a stepwise decrease in capacity factor occurs if the VOM cost exceeds the variable operating cost of the fission plants.
The capacity factor for the fleet can increase with penetration, as fusion displaces fission.
In plants with passive recirculating power, the capacity factor of the core is higher than that of the overall plant.
In Medium market opportunity cases with \SI{50}{\giga\watt} of fusion, the annual capacity factors of the fusion plants (cores) are 90\% (90\%), 84\% (86\%), or 73\% (81\%), for plants of optimistic, mid-range, or pessimistic designs, respectively.

Although fusion plants have generally been considered as `baseload' plants that run continuously, dispatchable operation (the ability to turn on and off, or otherwise modulate their power output) adds to their value.
In scenarios where we force fusion plants of the three reference designs to operate at full capacity year-round, their marginal break-even cost decreases by
\SI{50}[\$]{\per\kilo\watt},
\SI{130}[\$]{\per\kilo\watt}, and
\SI{340}[\$]{\per\kilo\watt}, respectively, as they accumulate operational costs during periods of low prices.
This loss of value is especially important for plants with high VOM costs which already have a lower capital cost threshold.

Thermal storage increases the capacity factor of the plants and modifies the operational patterns of the cores and PCSs.
In equivalent cases in the same scenario, adding the option to build thermal storage for \$22/kWh$_\textrm{th}$ increases the capacity factors of the plants (core) to 98.4\% (98.6\%), 94.7\% (95.5\%), and 86\% (90\%) for the three reference reactor designs.
Correspondingly, this mitigates the penalties of inflexibility: forcing the core to operate at full capacity decreases the marginal value by just
\SI{12}[\$]{\per\kilo\watt},
\SI{67}[\$]{\per\kilo\watt}, and
\SI{150}[\$]{\per\kilo\watt}, for the optimistic, mid-range, and pessimistic cores, respectively.
See Fig.~S5 for the value of dispatchable operation as a function of capacity penetration.

Figure~\ref{fig:capacityandcoe} also shows the levelized cost of electricity (\slcoe{}) for these plants.
If the capital cost of a plant $\plantcapex$ and fixed charge rate $\fixedchargerate$ are known as well as the variable cost $\totvom$ and annual capacity factor $\capacityfactor$, one can compute
\begin{equation}\label{eq:simplecoe}
\slcoe{} = \frac{\plantcapex \fixedchargerate}{8760 \; \capacityfactor} + \totvom.
\end{equation}
Here we take the capital cost of the plants to be equal to the marginal value at the given capacity penetration.
As long as the plants are dispatchable, the maximum allowable \slcoe{} \textit{increases} with the variable cost, even as the maximum allowable capacity investment cost (Figure~\ref{fig:scenario_competition}) falls.
This effect is due to the distribution of electricity prices: plants make most of their revenues in relatively few hours of the year.
It also shows that the equilibrium capacity penetration, and more generally, the merits of a plant for an electricity system, cannot be determined solely by the \slcoe{}.

\section{Annual and daily operation cycles}\label{sec:opcycles}

\begin{figure}
\centering
\includegraphics[width=0.95\textwidth]{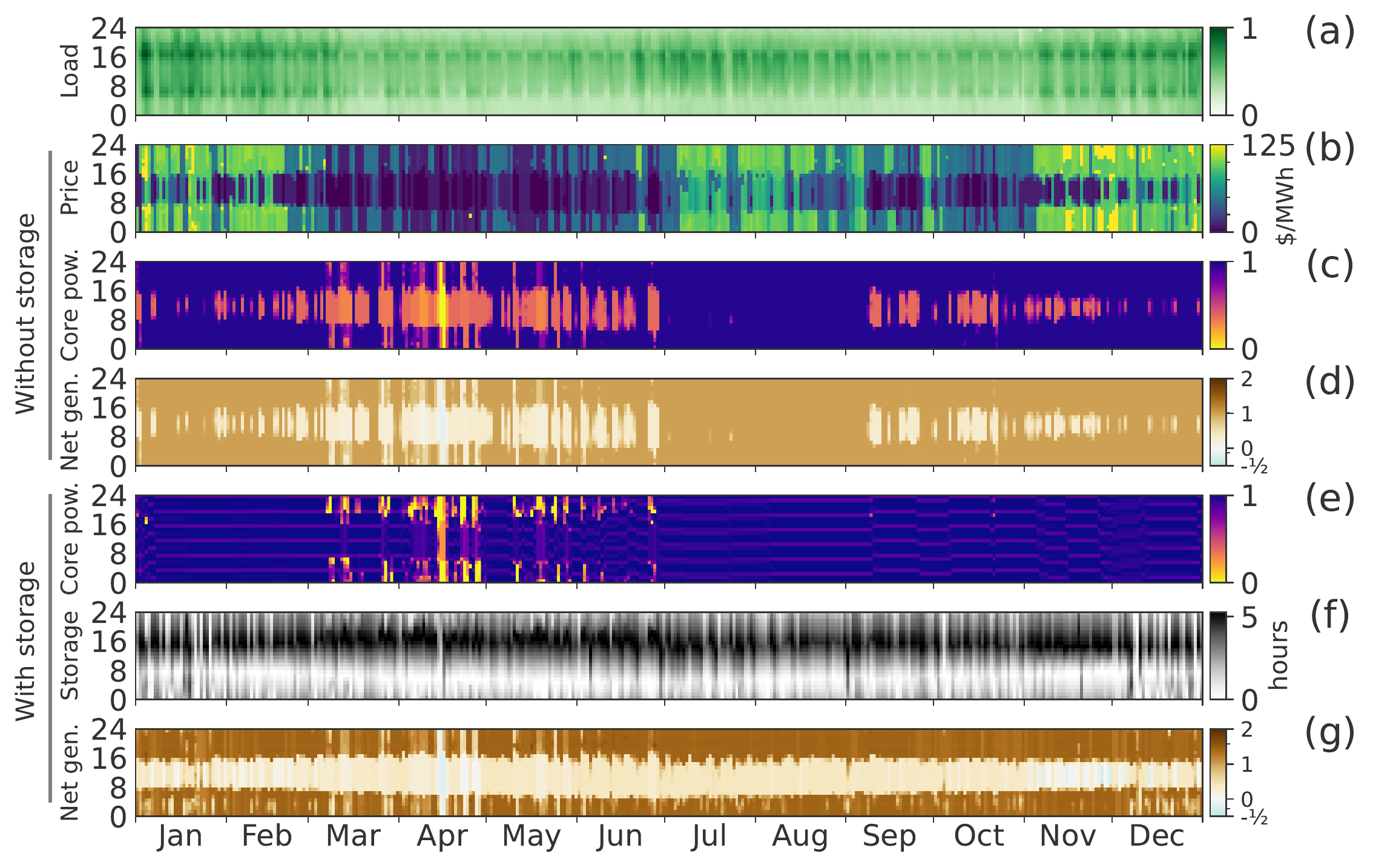}
\caption{Optimal hourly operational behavior of the fusion cores and PCSs for plants with mid-range cores with and without a TSS option, in one typical zone of a medium market opportunity scenario with a total system fusion capacity of \SI{50}{\giga\watt}. Part (a) shows the load in the zone, (b) shows the price, parts (c) and (e) shows thermal power output of the core normalized by its peak power, parts (d) and (g) show the net generation of the plant normalized by its long-run capacity, and part (f) shows the state of energy storage in the TSS measured in hours of the peak thermal capacity of the core.
See Figs.~S39 and S40 for operation of the other reference plants with and without thermal storage.
}\label{fig:hourlychart}
\end{figure}

Figure~\ref{fig:hourlychart} shows the operation of mid-range fusion plants in the medium market opportunity scenario with \SI{50}{\giga\watt} of fusion capacity.
Parts (c) and (e) show the operation of the cores and the normalized net output of the plants.
In times such as the spring and fall where loads (a) and prices (b) are lower and solar power is available during the day, the plants follow a diurnal cycle, decreasing their output during the solar peak.
During the periods of decreased output the plant is generally not shut off; rather, the power conversion system (PCS) is operated at levels approaching the minimum power level of 40\%.

With a thermal storage system (TSS), the core runs during the solar peak, drawing electricity from the grid and storing heat while the PCS operates at its minimum capacity.
In the evening, the oversized PCS draws power from the core and from storage, allowing plant to export power at 145\% of its nominal (long-run) capacity.

\begin{figure}
\centering
\includegraphics[width=\textwidth]{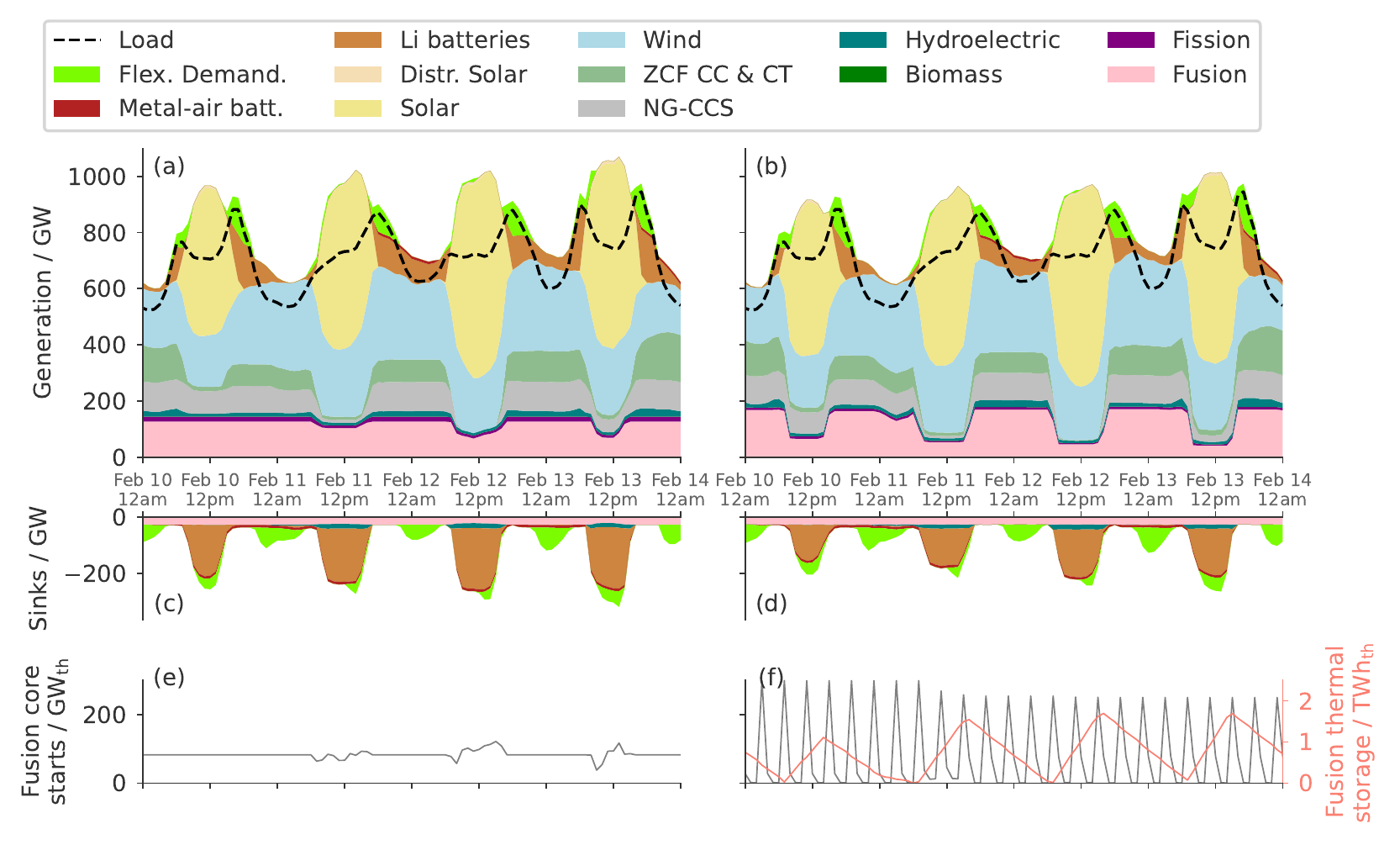}
\caption{Operation of the whole electricity system during four days in February, for cases in the medium market opportunity scenario with \SI{100}{\giga\watt} of mid-range fusion plants.
At left, a case where the plants do not have a TSS; at right, a medium-cost thermal storage (\$22/kW$_\textrm{th}$) can be built.
Parts (a) and (b) show the generation by resources in the system, as well the nominal load; parts (c) and (d) show the sinks of power including energy storage and the fusion recirculating power, and parts (e) and (f) quantify the total thermal power of the fusion cores for which a pulse begins in each hour.
Part (f) also shows the state of charge in the fusion thermal storage systems.
In both cases, heat is produced by the cores at a constant rate, but with storage, the output of the plants are modulated daily by storing and discharging heat.
Without the TSS, the fusion plants decrease their output when the variable renewables are sufficient.
With the TSS, the fusion plants store their heat output during the day and release it at night.
}\label{fig:systemops}
\end{figure}

The impact of a thermal storage system on operation can be seen further in Figure~\ref{fig:systemops},
which compares the behavior of plants with and without a TSS.
Without a TSS, the pulses of the fusion fleet (shown in Part (e)) are staggered so that together they produce nearly constant power output (a), other than during daylight hours when their output is reduced.
With a TSS, the precise timing of the pulses (shown in part (f)) is less important because the TSS buffers the PCS.
Within the model, at least at this time, there is not a strong incentive to stagger the pulses.
The TSS is filled and emptied on a diurnal cycle (part (f)) in order to generate power preferentially during the afternoon and night (part (b)). 
Incorporating a TSS with fusion also reduces the role of lithium battery storage.
Less total battery capacity is required (Fig.~S28) and as seen here, less charging and discharging takes place.

\section{Regional opportunities for fusion}

\begin{figure}
\centering
\includegraphics[width=1\textwidth]{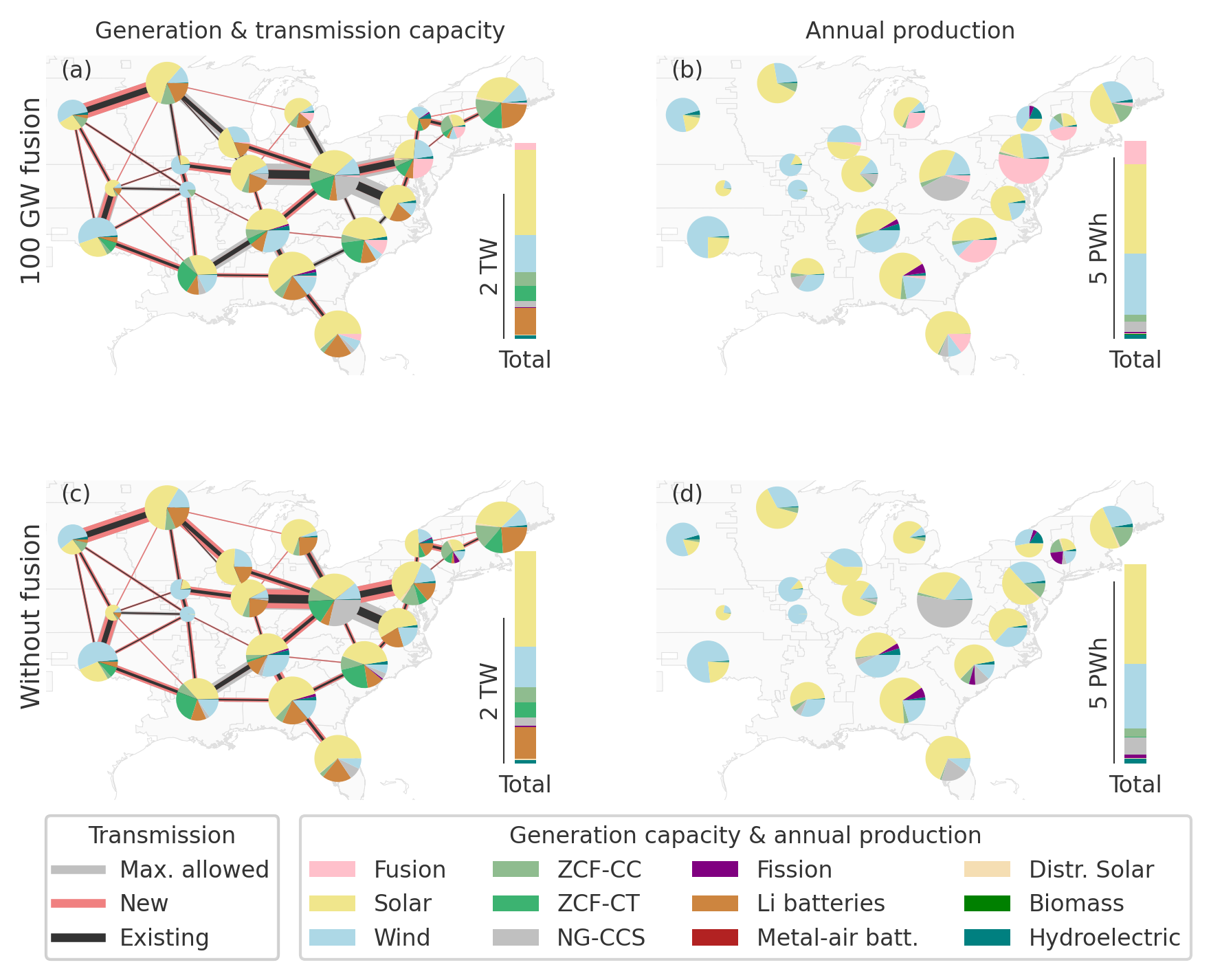}
\caption{Generation capacity, transmission capacity, and annual production of each resource type in the high market opportunity scenario, for cases with \SI{100}{\giga\watt} of mid-range fusion reactors (top) and without fusion (bottom).
Fusion capacity is mostly along the eastern seaboard, where it is built (parts (a) versus (c)) instead of solar, wind, NG-CCS, and batteries and (parts (b) versus (d)) displaces a portion of their electricity production.
The quantity of new transmission built (parts (a) versus (c)) is significantly smaller with the addition of fusion as a firm resource.
}\label{fig:map}
\end{figure}

Fig~\ref{fig:map} shows the most valuable places for \SI{100}{\giga\watt} of fusion energy to be built (a) and produce energy (b) in the high market opportunity scenario.
It is built preferentially in eastern and northern regions, likely due to the higher population density and lower resource potential of solar and wind there compared to regions further west.
Compared to the equivalent case without fusion, it leads to fewer new transmission lines (c) needing to be constructed and (d) less energy generated from solar, NG-CCS and wind.

In the low and medium market opportunity scenarios, fusion is sited in similar regions.
Less new transmission is required in these scenarios, 
and as the first \SI{100}{\giga\watt} of fusion mostly substitute for new-build fission, 
the direct effect on transmission is limited; see also Figure~S6.

\section{Discussion}\label{sec12}
There are six major implications of this study.
First, fusion could be a major firm resource for the US Eastern Interconnection, providing an annual net value of tens of billions of dollars, especially if renewables and storage, nuclear fission, or gas with carbon capture and storage fail to reach their cost targets.
As an example, in the medium and high market opportunity scenarios, a plant with costs like that of ARIES-AT\cite{najmabadi_aries-at_2006} (see Figure~\ref{fig:scenario_competition}) would have equilibrium capacities of more than \SI{100}{\giga\watt}, but zero in the low market opportunity scenario.
In particular, fission is the primary competitor to fusion, so fusion stakeholders should closely monitor its development as well as trends in public acceptance of each.
The United States may wish to consider fusion as a hedge for its energy portfolio\cite{united_states_of_america_long-term_2021} in case fission and other competitor resources fail to emerge.
Alternately, it could develop the technology for export, especially for locations where land availability or safety concerns limit the viability of renewables and fission.

Second, the value of a fusion plant depends strongly on its marginal cost of net power generation (or variable operations and maintenance cost),
so fusion developers must take into account the costs of operating and maintaining future reactors, not only the capital cost.
However, if fusion plants are dispatchable, then even a simple levelized cost of energy (\slcoe{}) calculation which incorporates both fixed and variable costs does not capture all the complexities.
%It ignores the dependence of a plant's capacity factor on its variable cost and ignores the widely-varying distribution of electricity prices throughout the year.
For two plant designs with the same \slcoe{}, the one with the higher variable cost (and lower capital and fixed costs) would reach a larger equilibrium capacity penetration.
This suggests some changes to how plants should be designed, if one's goal is to maximize fusion capacity.
If designs were optimized for \slcoe{} in the past,
then on the margin, more weight should be given to reducing capital costs at the expense of higher variable costs.
This could mean using less durable components or subjecting components to higher loads, even if they must be replaced more often.
This argument, along with the sustained periods of low prices in the models during spring and fall (Figure~\ref{fig:hourlychart}(b)),
suggests that rather than the typical proposed maintenance cycles of, for example, four to six months followed by two years of operation\cite{crofts_maintenance_2014},
 a plant with annual maintenance which is limited to the periods of low electricity prices could be an attractive option.
While the model in this study neglects the needs for maintenance, a detailed assessment of the lost value due to these scheduled outages will be the subject of a future paper.

The above argument to trade capital costs for variable costs depends on fusion plants being dispatchable.
The value attributable to dispatchable operation suggests that 
even fundamentally steady-state plants should either be designed with the ability to throttle their power output (once per day, on roughly half the days in the year, would be sufficient)
or coupled to a more flexible power conversion system though a thermal storage system.

Third, plant value depends only weakly on the particulars of an hourly-scale pulse cycle when examined with an hourly resolution.
If the technical challenges of pulsed tokamaks can be overcome, their varying output would not impose a significant penalty on their value,
though future work should examine their integration with the grid on finer spatial and temporal scales\cite{gaio_status_2022}.

Fourth, while not studied explicitly, the strong dependence of plant value on variable cost suggests that multiple classes of fusion plants at different locations on the capital--variable cost frontier could coexist.

Fifth, our study finds that the equilibrium capacity penetration of fusion increases significantly with relatively small decreases in the cost of a marginal plant.
This suggests that if cost targets for an initial market penetration can be met, further cost decreases driven by incremental improvements and learning-by-doing could allow fusion to reach a much higher capacity.

Finally, integrated thermal storage such as molten salt increases the plant value by a modest amount by better serving daily demand cycles.
It would be especially valuable while the total fusion capacity is small, and would further be valuable if early plants need to operate in a quasi-steady-state mode (for example due to a limited tolerance for thermal cycling).
For an initial market penetration, as long as the power conversion system remains dispatchable, the core need not be.

This study has several limitations due to the nature of the GenX capacity expansion model and our method of using it.
First, our study cannot address the question of the optimum unit size for fusion plants as all costs are assumed to scale linearly with capacity.
While magnetic fusion concepts like tokamaks and stellarators tend to multi-gigawatt scales when optimizing for the minimum cost-per-watt,
the procuring utilities have finite ability to finance project costs, and maximum scales may be limited by concerns about the stability of the grid\cite{takeda_dynamic_2015} should a plant suffer an unplanned outage.
However, the question of the optimum unit size is distinct from the cost thresholds we have determined.
For our study of plants with thermal storage systems, economies of scale would affect the optimal capacity ratios for the core, power conversion system, and storage.

Second, the temporal and spatial resolution of the simulations are likely somewhat favorable to fusion.
The hourly time basis used here does not resolve the details of a tokamak pulse cycle.
Future work could use finer steps to resolve any start-up power flows from the grid, which could lead to price spikes.
Similarly, the coarse spatial resolution underestimates the challenges of integrating pulsed reactors with the grid\cite{minucci_electrical_2020}.

Third, GenX optimizes capacity expansion for a long-run equilibrium, but the energy transition is dynamic.
The two modeled capacity expansion cycles, which represent \SI{15}{\year}, each use a single representative year with perfect foresight and rational decision-making.
We ignore cost-decreasing learning effects, which could lead to ``technology lock-in'' if, for example, small modular fission reactors are successfully mass-produced and adopted.
We implicitly assume that fusion becomes available just as electricity demand grows, as opposed to adoption being driven by retirements from the existing fleet\cite{spangher_characterizing_2019}, and ignore limitations to fusion's growth rate due to finite tritium stocks\cite{pearson_tritium_2018}.

Fourth, in our studies, the electricity system is not coupled with wider energy markets, and fuel costs are both fixed throughout the year and independent of demand.
Conversely, applications of fusion energy other than electricity generation, such as co-production of industrial heat or hydrogen, could add value to fusion plants, but these opportunities are not studied.
Neither is the use of fusion to re-power existing thermal plants\cite{handley_potential_2021}, which could reduce capital costs and make fusion easier to site.

Finally, we assume perfect availability for fusion with no unscheduled downtime and no need for scheduled maintenance periods.
The latter is a planned topic of study.

\section{Experimental Procedures}\label{sec11}
\subsection*{Resource availability}
\subsubsection*{Lead contact}
Further information and requests for resources and materials should be directed to and will be fulfilled by the lead contact, J.A.~Schwartz (\url{jacobas@princeton.edu}).
\subsubsection*{Materials availability}
This study did not generate new unique materials.
\subsubsection*{Data and code availability}\label{sec:dataark}
Data for this paper, as well as the code used for analyses, will be archived at \url{http://arks.princeton.edu/ark:/88435/dsp012j62s808w}.
% (For purposes of review, it is available through Globus at  \url{https://app.globus.org/file-manager?destination_id=dc43f461-0ca7-4203-848c-33a9fc00a464&destination_path=\%2Ff8em-3c49\%2F} )

A branch of GenX v0.2.0 was used for the model runs.
It is available at \url{https://github.com/cfe316/GenX/tree/fusion}.
GenX is available at \url{https://github.com/GenXProject/GenX}.

\subsection{Modeling technique}

We studied the value of fusion plants of various designs in a United States Eastern Interconnect electricity system with net-zero \ce{CO2} emissions during a period representing 2036--2050.
While this period is somewhat early for a significant build-out of commercial fusion plants,
it was chosen as a compromise due to the increasing uncertainty associated with periods further into the future.

We used the GenX capacity expansion and operations code\cite{genx_software}.
Given a representation of the electricity system, described as a set of geographic zones, transmission networks between zones, hourly loads, hourly availability profiles for variable renewable resources and time-shiftable loads, sets of existing resources, and resources which could be built, it determines the sets of resources to be built or retired and the hourly operation of each resource in order to minimize the annual cost of the electricity system.
The model contains 20 geographic zones, each describing one or more of the regions based on the EPA Integrated Planning Model (IPM \verb|v6|) regions\cite{united_states_environmental_protection_agency_epas_2021} in the Eastern Interconnection.
A full year (\SI{8760}{\hour}) of hourly operation is modeled.

We examined three main scenarios with differing costs of resources other than fusion; we refer to these as the low, medium, and high fusion market opportunity scenarios.
Each scenario has the same load profiles, but different quantities of time-shiftable loads.
Input data are from a variety of sources, including PUDL\cite{selvans_pudl_2021} and the NREL Annual Technology Baseline (ATB) 2021\cite{nrel_national_renewable_energy_laboratory_annual_2021}.
Eleven variant scenarios are also described in the SI.

For each of the three main scenarios, in order to determine the resources existing at the start of the 2036--2050 period, we performed a simulation representing the years 2021--2035, without the net-zero \ce{CO2} constraint; the generator capacity expansion results were used as inputs for the subsequent model runs.

\subsection{Resource cost and operational assumptions}

Table~\ref{tab:capex} lists the capital costs of the resources which can be built in the 2036--2050 simulations. 
% this is true for Utility-Scale PV
% this is true for Li batteries. It can be checked by looking at the 4 hour and 8 hour time frames: 
% the 4 hour batt uses 1 power and 4 storages, the 8 hour batt uses 1 power and 8 storages. Look at the year 2043. It is utility-scale battery storage.
% It's true for onshore wind.
% Unsure about offshore wind, though the numbers are close, they're not exact. Maybe we are tracking the base and siting and hookup differently.
%
Capital costs, fixed operational costs, variable operational costs, and heat rates for new-build technologies in the medium and high market opportunity scenarios generally follow the `Moderate' cost assumptions from the ATB, while the low market opportunity scenario generally follows the `Advanced' cost assumptions.
Capital costs are taken to be the average of those in the period 2036--2050.

\begin{table}[ht]
\begin{center}
\caption{Median capital costs of generation and storage in \si{\$ \per \kilo\watt} and \si{\$/\kWh} in 2036--2050 for the three market scenarios, the real weighted average cost of capital (WACC) in \% for each technology, and the assumed lifetime in years. Low-cost fission is used only in the so-named variant scenarios.}\label{tab:capex}
%\rowcolors{2}{gray!10}{gray!30}
\begin{tabular}{@{}ld{3.0}d{3.0}d{3.0}cc@{}} \toprule
& \multicolumn{1}{r}{Low} & \multicolumn{1}{r}{Medium} & \multicolumn{1}{r}{High} & \multicolumn{1}{r}{Real WACC} & \multicolumn{1}{r@{}}{Lifetime}  \\
\midrule
Utility-scale Solar PV & 536 & 686 & 686 & 2.57 & 30 \\
Onshore wind & 586 & 826 & 826 & 3.00 & 30 \\
Offshore wind & 1603 & 1918 & 1918 & 3.21 & 30 \\
%Conventional hydroelectric & 29 & 29 & 29 & \\
%Natural gas without carbon capture* & 199 & 164 & 120 & \\
ZCF-CT & 787 & 787 & 787 & 3.34 & 30 \\
ZCF-CC & 942 & 942 & 942 & 3.34 & 30 \\
NG-CCS & 2318 & 2318 & 2318 & 3.34 & 30 \\
Fission & 4176 & 6233 & 9348 & 3.34 & 40 \\
\textcolor{gray}{Fission (low-cost)} & \textcolor{gray}{3740} & \textcolor{gray}{4986} & \textcolor{gray}{6233} & \textcolor{gray}{3.34} &  \textcolor{gray}{40}  \\ \midrule
Li batteries - power & 80 & 187 & 187 & 2.57 & 15 \\
Li batteries - storage & 86 & 117 & 117 & 2.57 & 15\\
Metal-air batteries - power & 800 & 1200 & 2000 & 2.57 & 25 \\
Metal-air batteries - storage & 8 & 12 & 20 & 2.57 & 25 \\ \bottomrule
\end{tabular}
\end{center}
\end{table}

Exceptions are metal-air batteries, which are not listed in the ATB and for which we use cost and performance assumptions from \cite{baik_what_2021}, and NG-CCS plants.
Because only the `Conservative' cost and performance assumptions for NG-CCS in the ATB assume a conventional combined cycle plant (rather than a fuel cell or an interpolation between the two), we use the conservative case as our baseline for this category of plant.
NG-CCS cost and performance parameters are adjusted further to reflect the requirement of 100\% carbon capture efficiency in our system, an increase from the 90\% efficiency assumed in the ATB: capital cost is increased by \SI{116}[\$]{\per\kilo\watt}, heat rate by 0.365, fixed OM by \SI{9.67}[\$]{\per\kilo\watt\year}, and variable OM by 7.6\%. We further consider the need for \ce{CO2} transport and storage infrastructure, the cost of which varies by model zone.
\ce{CO2} pipeline construction costs are added to NG-CCS plant investment costs, and are calculated using methodology developed in \cite{larson_net-zero_2021}, assuming an average plant size of \SI{500}{MW}, a 100\% utilization rate, and a length equal to the distance between the largest major metro area in each GenX zone and the edge of the nearest \ce{CO2} injection basin.
Variable injection costs per ton of \ce{CO2} are added to NG-CCS plant variable operational costs, and vary by injection basin.
\ce{CO2} pipeline costs by GenX zone are listed in Table~S3. 

Costs for the ZCF combustion turbine (CT) and closed cycle gas turbine (CC) correspond to those of the ``Natural Gas FE CT'' and ``Natural Gas FE CC'', respectively.
%Costs for gas with CCS are based off those from the ``moderate'' scenario, but adjusted \todo{How?} to account for 100\% CCS rather than the 90\% assumed by the ATB.
%Further regional adjustments account for costs of constructing pipelines for \ce{CO2} sequestration.
Capital costs of these combustion plants do not vary by scenario, but the fuel costs do; see Table~\ref{tab:varcosts}.

Fission costs in the medium fusion opportunity scenario are from the ``moderate'' ATB scenario, which are from the EIA Annual Energy Outlook 2021\cite{energy_information_administration_annual_2021}.
The low and high market opportunity scenarios use \sfrac{2}{3} and \sfrac{3}{2} of this cost, respectively, and the three low-cost nuclear scenarios use \sfrac{3}{5}, \sfrac{4}{5}, and \sfrac{5}{5} of this cost, respectively.

Lithium ion batteries have charge and discharge efficiencies of 0.92 each and metal-air storage has charge and discharge efficiencies of 0.65 each, for round-trip efficiencies of 0.85 and 0.42, respectively.

Resources have an additional cost to account for transmission spur lines, with regional costs from \$3686 per \si{\mega\watt}-mile to \$6320 per \si{\mega\watt}-mile.
The length of the spur lines is shown in Table~S4; wind and solar have variable spur line lengths and costs from a method developed for the Net-Zero America study\cite{larson_net-zero_2021}.

Table~\ref{tab:varcosts} lists the fuel costs and total variable costs of the resources in the 2036--2050 cases in the fusion market opportunity scenario classes.
The total variable cost includes the cost of the fuel and the variable OM cost of the plant itself.
The two ZCF fuel resources nominally use the same type of fuel, but the ZCF-CC plant is more efficient, so its total variable cost is lower.
For fusion we list the variable OM cost of the power conversion system (PCS) only.
This cost is incurred proportional to the gross generation, so for a fusion plant with 50\% recirculating power the PCS variable OM cost would be twice as high per \textit{net} \si{MWh}$_{e}$. 
The fusion plant also has a VOM cost for the core operation, but these vary by the reactor design: see Table 1 of the main paper.

Zonal costs for conventional fuel types are from the EIA Annual Energy Outlook\cite{energy_information_administration_annual_2021}.
Natural gas costs for the Low, Medium, and High market opportunity cases are taken from the AEO's high oil and gas supply, reference, and low oil and gas supply cases, respectively.
The nature of the zero-carbon fuel is not explicit, but the cost of ZCF in the Medium market opportunity scenario is set equal to the average \ce{H2} cost in the three high-electrification scenarios in the Net-Zero America report\cite{larson_net-zero_2021}, \SI{15.20}[\$]{\per\giga\joule}.
ZCF costs in the Low and High market opportunity cases are \sfrac{2}{3} and \sfrac{3}{2} of this value, respectively.

\begin{table}
\caption{\label{tab:varcosts} Fuel costs and total variable costs in \si{\$/MMBTU} and \si{\$/\mega\watt\hour}, respectively, in 2036--2050, for the three market opportunity scenario classes.}
\centering
\begin{tabular}{@{}ld{3.2}d{3.2}|d{3.2}d{3.2}|d{3.2}d{3.2}@{}} \toprule
 & \multicolumn{2}{c}{Low} & \multicolumn{2}{c}{Medium} & \multicolumn{2}{c@{}}{High} \\ \midrule
ZCF-CT         & 10.81  & 110.01  & 14.41  & 145.00 & 19.21 & 191.66 \\
ZCF-CC   & 10.81  & 70.49  & 14.41  & 93.39 & 19.21 & 123.92 \\
NG-CCS  & 2.75   & 33.20 & 3.75 & 40.72 & 6.50 & 61.39 \\
Fission             & 0.73   & 9.96   & 0.73   & 9.96 & 0.73 & 9.96 \\%  \bottomrule
Li batteries        &    & 0.15   &    & 0.15 &  & 0.15 \\
Metal-air storage   &   & 0   &   & 0 &  & 0 \\ % \bottomrule
Fusion: PCS operation & & 1.74 & & 1.74 & & 1.74 \\ \bottomrule
\end{tabular}
\end{table}

Power flow between resources and loads inside zones is considered to be lossless; this is sometimes called the copper-plate assumption.
Power flow between zones is limited by the inter-regional transmission capacity.
These start at 2020 values.
The capacity of each route can be expanded by up to 150\% or \SI{1.5}{\giga\watt}, whichever is larger.
This assumption is somewhat arbitrary, but even either forbidding new transmission or allowing unlimited transmission does not affect the value of fusion by more than \SI{150}[\$]{\per\kilo\watt} in the medium market opportunity scenario or more than \SI{900}{\per\kilo\watt} in the high market opportunity scenario; see Figure~S6.
Costs per GW-mile for transmission are dependent on the regions connected.
The existing transmission capacity and maximum possible capacities are listed in Table S6.
Transmission losses are linear with the power transmitted between zones.

\subsection{Pulsed tokamak plant model}\label{sec:fusimp}

When assessing costs, such as the value of the core (as in Figure~\ref{fig:corevalincrease}), the core represents not just the fusion chamber itself,
but all parts of the plant other than the TSS and PCS. 
The structure of the implementation motivates the component definitions.
The core represents most of the plant: the reactor itself, maintenance facilities, land, parking lots, waste storage, and so on.
The model PCS includes some heat exchangers, the turbines, generators, and heat rejection systems.
Its costs are assumed to scale linearly with its rated power capacity.
Any heat exchangers upstream of the TSS are part of the core.
The model TSS includes tanks required for inter-hourly energy storage, but does not include additional piping, heat exchangers, or other systems which affect the input or output power capacity of the storage system; those are part of the core.
The TSS costs scale linearly with the energy storage capacity.
The core also includes any intra-hourly energy storage systems which are required for interfacing with the PCS.

In a real reactor, certain components may need to be replaced due to cyclic fatigue rather than damage from cumulative neutron exposure; this could be modeled by adding a finite cost to start a pulse, but it is not implemented here.
In this study the fusion plant is operated at its maximum cadence throughout most of the year, so pricing the pulse start in addition to the heat generated would lead to little difference in operation.

\subsubsection*{Reference reactors}

The pessimistic core loosely models a plant similar to DEMO (which is not a competitive commercial power plant\cite{federici_demo_2018-1, kembleton_prospective_2022}). 
The two-hour total cycle time is a rounding down of the $\sim\SI{8000}{\second}$ total cycle time including a \SI{2}{\hour} flat-top, and the \SI{0.15}{\hour} dwell time is a rounding down of the $\lt 10$ minute length, both listed in Table~3 of the overview paper by Federici et al\cite{federici_demo_2018-1}.
The active and recirculating power fractions are both set to 0.2. 
Together these represent an improvement on the \SI{475}{\mega\watt} and \SI{392}{\mega\watt} of \textit{total} recirculating power demands required by the steady state electrical systems during the burn flat-top and dwell period, respectively, as estimated as the ``active power'' for the HCPB design (Figure 15 and Table 6) by Gaio et al\cite{gaio_status_2022}. Note that here ``active'' is in contrast to reactive power.
For the ``indirect'' coupling configuration between the fusion core and PCS, the gross electric power is \SI{900}{\mega\watt}, so this would translate to a $\rpass = 0.435$ and $\ract = 0.092$.
In our model moving a fixed quantity from $\rpass$ to $\ract$ is always beneficial, so $\ract=\rpass=0.2$ is strictly an improvement.
The start power and start energy fractions are roughly estimated from the size of the loads of the DEMO pulsed power electrical network (PPEN) \cite{minucci_electrical_2020}.
The variable operations and maintenance cost is estimated by applying the Sheffield costing method\cite{sheffield_generic_2016} to a 2015 version of DEMO\cite{reux_demo_2015} and computing the costs of blanket and divertor replacement; this came to \$7.3$/\mathrm{MWh}_\mathrm{th}$ and was rounded down to \$5$/\mathrm{MWh}_\mathrm{th}$.

The mid-range core is based off the pessimistic core, but with a pulse cycle duration of \SI{4}{\hour} and with recirculating power parameters half as large as those of the pessimistic core.
Its $\corevom$ is the average of those of the two other cores.

The optimistic core is based on an optimistic view of the ARC reactor\cite{sorbom_arc_2015, sorbom_recent_2020}.
The 2020 ARC design point\cite{sorbom_recent_2020} describes it as having roughly 30 minute pulses; this is changed to \SI{1}{\hour} in order to fit in our hour-based model; the dwell time was increased proportionally.
It has a total recirculating power of 4.3\%\cite{sorbom_recent_2020}, no external start power, and a variable operations and maintenance cost arbitrarily set at 1/5th that of the pessimistic core.
Though this model must start a pulse every hour, because there is zero start energy and start power this is not a significant burden; we call it ``optimistic'' due to its low total VOM cost.

Table~\ref{tab:fusmodels} lists parameters for the fusion plant which can be derived from the core and power conversion system parameters, assuming that the plant is operated at its maximum capacity.
The core active fraction
\begin{equation}
    f_\mathrm{act} = \left(1 - \frac{\tdwell}{\tpulse}\right).
\end{equation}
is the fraction of the time that the core is producing heat, and
when the active recirculating power is required.
The core net capacity factor
\begin{equation}
    f_\mathrm{netavgcap} = f_\mathrm{act}\left(1 - \ract\right) - \rpass - \frac{\estart}{\tpulse}.
\end{equation}
is the ratio of the time-averaged net electric power produced to the gross electric power generation capacity.
The core peak thermal capacity $\peakthcap$ is related to the plant time-average net electrical capacity $\netavgcap$,
\begin{equation}
    \netavgcap = \thermeff\,f_\mathrm{netavgcap}\,\peakthcap.
\end{equation}
In this paper, the fusion plant capacity penetration is specified in terms of $\netavgcap$.
This allows a comparison between cores with different operational characteristics.
The marginal cost of net generation
\begin{equation}
    \totvom =  \frac{f_\mathrm{act} \left(\corevom + \thermeff \genvom \right)}{\thermeff f_\mathrm{netavgcap}}
\end{equation}
reflects the variable operations and maintenance cost for the core and generator, taking into account the recirculating power and dwell times.
As shown in Figure~\ref{fig:variations}, for plants without a thermal storage system, the required total plant cost is largely determined by this quantity.
Finally, the recirculating power fraction
\begin{equation}
    f_\mathrm{recirc} = \frac{f_\mathrm{act} \ract + \rpass + \estart/\tpulse}{f_\mathrm{act}}.
\end{equation}
is the fraction of the gross power generated which must be used to operate the device itself.
This can be related to the plant's time-averaged engineering gain $Q_\mathrm{eng}$\cite{wurzel_progress_2021}:
\begin{equation}
    Q_\mathrm{eng} = (1 - f_\mathrm{recirc})/f_\mathrm{recirc}.
\end{equation}

\subsection{Plant cost threshold determinations}
In this paper we determine the relationship between the cost of fusion and its equilibrium capacity penetration in the electricity system.
While perhaps the most straightforward method would be to find the capacity penetration as a function of cost,
here we find the cost as a function of the capacity penetration.
The method employed is not entirely straightforward, but has some advantages in practice, and ultimately achieves the same results.

The most obvious method would have been to set a trial cost for the fusion core and solve the capacity expansion problem to determine the total fusion capacity in the optimized electricity system.
While this method is conceptually straightforward, it has a minor disadvantage: if the cost of fusion core was too high, no fusion plants would be built, wasting a several-hour computation.
Additionally, finding the cost threshold for an initial penetration into the market---a key metric for fusion commercialization---becomes a root-finding problem.

Instead we use a method\cite{de_sisternes_value_2016, mallapragada_long-run_2020, heuberger_systems_2017} that determines the marginal value of a plant as a function of its capacity penetration.
Unless the user requests an extremely large fusion capacity penetration (beyond \SI{350}{\giga\watt} in the scenarios described here), this method always results in a nonzero marginal value for fusion energy.
It is thus arguably more efficient in its use of compute time.
There are two additional advantages.
First, differences in the operational parameters of the plant (such as those explored in Fig.~\ref{fig:variations}) directly translate into differences in value, rather than differences in the equilibrium capacity penetration. We believe this to be a more relevant metric for fusion developers and policymakers.
Second, the value-based approach perhaps has a conceptual advantage, in that it disentangles the question of the cost of a plant from the notion of its value to the system as a whole.
The method is explained below.

The plant is composed of the core, PCS, and TSS.
%The PCS and its costs are modeled after the power conversion system of a concentrated solar plant and the TSS costs, based on the energy capacity cost of molten salt storage, are varied case by case.
In each case, the investment cost and fixed operations and maintenance costs of the fusion core are set to zero,
and the total net fusion capacity in the system is fixed by a constraint.
The value of the fusion core is calculated as the \textit{dual value} of this constraint: the amount by which the model's objective function would decrease given a relaxation of the constraint by one unit.
Because system costs would decline by this amount were an additional unit of fusion core capacity deployed in the system, this can equivalently be interpreted as the minimum cost that the fusion core could have for this deployment to be profitable.
For equilibrium market conditions in GenX the profit of a marginal plant is exactly zero, and so we can interpret the core value at a specified total fusion penetration as the exact core \textit{cost} at which fusion would naturally achieve that penetration in a competitive market.
For plants without a TSS the ratio of fusion core capacity to PCS capacity is fixed.
The sum of the (known) annual PCS cost and the core value is the maximum annual cost of the whole fusion plant,
which includes the annualized investment cost and the fixed operations and maintenance costs.
We refer to this as the plant value, or equivalently, the threshold cost for a marginal plant.
In cases where a TSS is allowed, a precise equivalent to this quantity cannot be determined by simply adding its costs, since the ratios of PCS and TSS capacities to fusion core capacities vary strongly as a function of the fusion capacity penetration.
Instead we refer solely to the value of the core itself.

As a check of this system value method we performed a run (using the ``straightforward'' method) with the cost of the fusion core set equal to the marginal value previously determined at \SI{100}{\giga\watt} of capacity penetration.
The optimal fusion capacity was \SI{99.77}{\giga\watt}. 
The imperfection is due to finite tolerances in the optimizer.

\backmatter

\section*{Supplemental information description}
The supplemental information (SI) describes the model configuration, economic parameters, and other input data for the main scenarios, and for eleven variant scenarios not discussed in the main text.
It contains a formal description of the fusion plant implementation and a full set of input parameters.
The SI also includes additional outputs: plots of added value and relative component sizes of fusion plants with thermal storage,
capacity and energy production mixes for the various scenarios, maps showing where fusion plants are built, and charts showing the operational behaviors of the other reference plants.

\section*{Acknowledgments}
Thanks to Greg Schively for help with PowerGenome.
%Thank you to C.\ Swanson, A.\ Rutkowski, and R.\ Goldston for interesting discussions.

\section*{Author contributions}
JAS: Conceptualization, Methodology, Software, Formal analysis, Investigation, Data curation, Writing, Visualization.
WR: Methodology, Software, Resources, Writing-original draft.
EK: Conceptualization, Methodology, Writing-review and editing, Supervision, Funding acquisition.
JJ: Conceptualization, Methodology, Software, Writing-review and editing, Supervision, Funding acquisition.

\section*{Funding}
J.\ A.\ Schwartz and E.\ Kolemen were supported through the U.S. Department of Energy under contract number DE-AC02-09CH11466.
The United States Government retains a non-exclusive, paid-up, irrevocable, world-wide license to publish or reproduce the published form of this manuscript, or allow others to do so, for United States Government purposes.

W. Ricks received support for this work from the Princeton Zero-Carbon Technology Consortium, which is supported by unrestricted gifts from General Electric, Google, and ClearPath.

\bigskip
\begin{flushleft}%

\end{flushleft}

\bibliography{PlasmaControl}

\end{document}

\title[Valuing fusion for the US : Supplemental]{Supplemental information for: The value of fusion energy to a decarbonized United States electric grid}

%%=============================================================%%
%% Prefix	-> \pfx{Dr}
%% GivenName	-> \fnm{Joergen W.}
%% Particle	-> \spfx{van der} -> surname prefix
%% FamilyName	-> \sur{Ploeg}
%% Suffix	-> \sfx{IV}
%% NatureName	-> \tanm{Poet Laureate} -> Title after name
%% Degrees	-> \dgr{MSc, PhD}
%% \author*[1,2]{\pfx{Dr} \fnm{Joergen W.} \spfx{van der} \sur{Ploeg} \sfx{IV} \tanm{Poet Laureate}
%%                 \dgr{MSc, PhD}}\email{iauthor@gmail.com}
%%=============================================================%%

\author*[1,2]{\fnm{Jacob} \sur{Schwartz}}\email{jacobas@princeton.edu}
%\equalcont{These authors contributed equally to this work.}

\author[1,2]{\fnm{Wilson} \sur{Ricks}}\email{wricks@princeton.edu}
%\equalcont{These authors contributed equally to this work.}

\author*[1,2]{\fnm{Egemen} \sur{Kolemen}}\email{ekolemen@princeton.edu}
%\equalcont{These authors contributed equally to this work.}

\author[1,2]{\fnm{Jesse} \sur{Jenkins}}\email{jdj2@princeton.edu}
%\equalcont{These authors contributed equally to this work.}

\affil*[1]{\orgdiv{Department of Mechanical and Aerospace Engineering}, \orgname{Princeton University}, \orgaddress{\street{Olden St.}, \city{Princeton}, \postcode{08540}, \state{NJ}, \country{USA}}}

\affil[2]{\orgdiv{Andlinger Center for Energy and the Environment}, \orgname{Princeton University}, \orgaddress{\street{Olden St.}, \city{Princeton}, \postcode{08540}, \state{NJ}, \country{USA}}}

%%==================================%%
%% sample for unstructured abstract %%
%%==================================%%

\keywords{nuclear fusion, capacity expansion, technology assessment}

%%\pacs[JEL Classification]{D8, H51}
%%\pacs[MSC Classification]{35A01, 65L10, 65L12, 65L20, 65L70}

% \maketitle

\beginsupplement

\section*{Note S1: Model configuration}\label{sec:modelconfig}

This note describes the configuration of the model and construction of the three main scenarios which are discussed in the main text, as well as eleven additional scenarios which are variants of the main three.

This work uses a branch of the open-source electricity system capacity expansion and operations code \verb|GenX 0.2.0|\cite{genx_software}.
It optimizes the expansion of generation, storage, and transmission capacities, as well as hourly operations for groups of generators in a model electricity system.
This branch implements the fusion plant model described in Section~\ref{sec:fusimp}.

This work studies the value of fusion in a future representing the period 2036--2050, using a two-stage myopic brownfield optimization.
The first step starts with the existing fleet of generators and models the period 2021--2035 with no federal decarbonization mandate.
The second step, representing 2036--2050, introduces the option to build fusion plants and enforces a requirement for net-zero carbon electricity.

The model input data, including load profiles, transmission network specifications, existing generation resources, technology costs, and operational parameters, were compiled using PowerGenome \cite{powergenome}.
Time-shiftable electrified load profiles were generated using the \verb|efs_demand| branch of PowerGenome, which has since been merged into the \verb|master| branch (but see commit \verb|06fcb3cb|).
These represent loads due to light-duty vehicle charging and residential water heating, which have some flexibility in when exactly they are served.
The total energy to be served to these categories is fixed---they do not represent ``demand-response'' programs.
All other inputs were compiled using the \verb|master| branch \verb|v0.5.3|.
All cost data are in 2019 dollars, and are primarily taken from the 2021 NREL Annual Technology Baseline (ATB) \cite{nrel_national_renewable_energy_laboratory_annual_2021}.

Full sets of input data for both PowerGenome and GenX are available via the Princeton DataSpace, \url{http://arks.princeton.edu/ark:/88435/dsp012j62s808w}.
(For purposes of review, data is available through Globus at  \url{https://app.globus.org/file-manager?destination_id=dc43f461-0ca7-4203-848c-33a9fc00a464&destination_path=\%2Ff8em-3c49\%2F} )

\subsection*{Geographic zones and aggregation}
\begin{figure}[htb]
\centering
\includegraphics[width=0.95\textwidth]{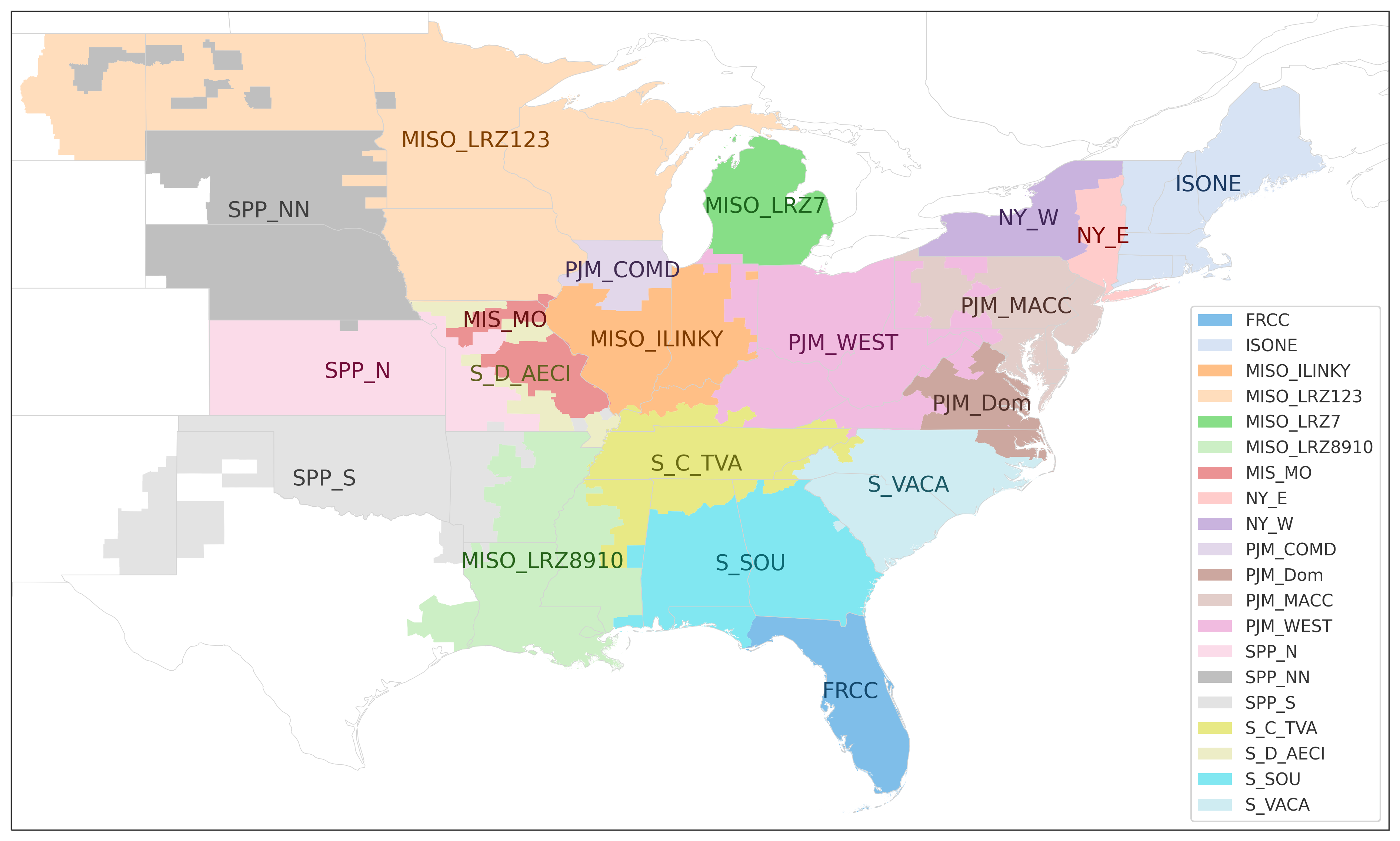}
\caption{Map of the twenty model zones. Some are identical to IPM regions, and some are conglomerates---see Table \ref{tab:IPMZoneAggregation}.}\label{fig:zonelegendmap}
\end{figure}

All the scenarios model the same electrical system: the United States (US) portion of the Eastern Interconnection.
The model uses 20 geographic zones, which are constructed from the US Integrated Planning Model (IPM \verb|v6|) regions\cite{united_states_environmental_protection_agency_epas_2021}. While the Eastern Interconnection does extend into Canada, these regions are not considered.
Figure~\ref{fig:zonelegendmap} shows the geographic regions covered by each zone, and Table~\ref{tab:IPMZoneAggregation} lists the IPM regions composing each zone.

A `copper-plate' model is used within each zone, so transmission within a zone is lossless.
Between zones, maximum transmission capacities and losses are considered.
Interregional transmission expansion is also possible; costs are calculated via a transmission routing methodology developed in \cite{REPEAT} that finds cost-optimal transmission routes between and within model zones based on a cost surface map.

\begin{table}[htb]
\caption{\label{tab:IPMZoneAggregation} Formation of the twenty model zones from Integrated Planning Model (IPM) regions. Some zones are identical to IPM regions; others are conglomerates. See also Fig.~\ref{fig:zonelegendmap}.}
\centering
\begin{tabular}{@{}ll@{}} \toprule
Zone & is composed of \\ \midrule
\verb|FRCC| & - \\
\verb|ISONE| & \verb|NENG_ME, NENG_CT, NENGREST| \\
\verb|MISO_ILINKY| & \verb|MIS_IL, MIS_INKY| \\
\verb|MISO_LRZ123| & \verb|MIS_MAPP, MIS_MNWI, MIS_WUMS, MIS_MIDA, MIS_IA| \\
\verb|MISO_LRZ7 | & \verb|MIS_LMI| \\ 
\verb|MISO_LRZ8910| & \verb|MIS_AR, MIS_LA, MIS_AMSO, MIS_WOTA, MIS_D_MS| \\
\verb|MIS_MO| & - \\
\verb|NY_E| & \verb|NY_Z_F, NY_Z_G-I, NY_Z_J, NY_Z_K| \\
\verb|NY_W| & \verb|NY_Z_A, NY_Z_B, NY_Z_C&E, NY_Z_D| \\
\verb|PJM-COMD| & - \\
\verb|PJM-Dom| & - \\
\verb|PJM_MACC| & \verb|PJM_WMAC, PJM_EMAC, PJM_SMAC, PJM_PENE| \\
\verb|PJM_WEST| & \verb|PJM_West, S_C_KY, PJM_AP, PJM_ATSI| \\ 
\verb|SPP-N| & - \\
\verb|SPP_NN| & \verb|SPP_WAUE, SPP_NEBR| \\
\verb|SPP_S| & \verb|SPP_SPS, SPP_WEST| \\ 
\verb|S_C_TVA| & - \\
\verb|S_D_AECI| & - \\
\verb|S_SOU| & - \\
\verb|S_VACA| & - \\
\bottomrule \end{tabular}
\end{table}

\subsection*{Time basis and loads}

The simulations model a single full year on an hourly time basis, with a total of 8760 hours.
Inter-temporal constraints couple the first hour of the year to the last.

The hourly load profiles in each zone are the same in each scenario.
Table~\ref{tab:tfloads} lists the peak hourly load in each zone, and Figure~\ref{fig:hourlyloadchart} shows the hourly load profiles.
The three market opportunity scenarios differ in the quantity of the loads which are shiftable---able to be advanced or delayed for a few hours, due to a less-time-sensitive nature.
Hourly load profiles are constructed from base load profiles, based on the 2012 weather year and scaled to account for incremental load growth, plus additional loads representing  electrified residential water heaters (RWH) and light-duty vehicles charging (LDV).
Hourly electrified load profiles are from the NREL Electrification Futures Study\cite{mai_electrification_2018}, with the magnitude of electrified load scaled to reflect results from the high electrification scenario in the Net-Zero America study \cite{larson_net-zero_2021}.
A fraction of the LDV loads can be shifted later by up to \SI{5}{\hour}: 0.9, 0.75, and 0.6, for the Low, Medium, and High fusion market opportunity scenarios, respectively.
A fraction of the RWH loads can be shifted earlier or later by \SI{2}{\hour}: 0.2, 0.1, and 0, for the Low, Medium, and High fusion market opportunity scenarios, respectively.
Zero shiftable RWH load in the High fusion market opportunity scenario means this load is not shiftable in time.

Table~\ref{tab:tfloads} lists the peak hourly shiftable load for the two categories for each zone in the three scenarios.
The shiftable fraction changes hourly and is often much smaller, with an average of about 25\%.

\begin{table}[b]
    \caption{\label{tab:tfloads}Peak loads and shiftable loads in the system. For each zone, peak hourly loads and peak values for the quantity of the load which is shiftable.
    LDV is light duty vehicle charging and RWH is residential water heating. 
    LDV loads can be delayed by up to \SI{5}{\hour} and RWH loads can be advanced or delayed by \SI{2}{\hour}.
    All quantities are in MW.
    Zeros in the last column indicate that in the High fusion market opportunity scenario, RWH loads are \textit{not} shiftable.}
    \centering
    \begin{tabular}{@{}rrrrrrrr@{}}
    \toprule
    &&\multicolumn{2}{c}{Low Opp.} & \multicolumn{2}{c}{Medium Opp.} & \multicolumn{2}{c}{High Opp.} \\
    Zone & Peak load & LDV & RWH & LDV & RWH & LDV & RWH \\ \midrule
\verb|FRCC| & 82067 & 28663 & 578 & 23886 & 289 & 19109 & 0 \\
\verb|ISONE| & 126985 & 17233 & 567 & 14361 & 284 & 11489 & 0 \\
\verb|MISO_ILINKY| & 48318 & 15261 & 546 & 12717 & 273 & 10174 & 0 \\
\verb|MISO_LRZ123| & 89346 & 29550 & 806 & 24625 & 403 & 19700 & 0 \\
\verb|MISO_LRZ7| & 43162 & 12420 & 386 & 10350 & 193 & 8280 & 0 \\
\verb|MISO_LRZ8910| & 57326 & 17427 & 381 & 14523 & 190 & 11618 & 0 \\
\verb|MIS_MO| & 18282 & 6260 & 161 & 5217 & 80 & 4174 & 0 \\
\verb|NY_E| & 48391 & 8275 & 453 & 6896 & 227 & 5517 & 0 \\
\verb|NY_W| & 15205 & 2613 & 143 & 2178 & 72 & 1742 & 0 \\
\verb|PJM_COMD| & 46727 & 11187 & 457 & 9322 & 228 & 7458 & 0 \\
\verb|PJM_Dom| & 38844 & 12544 & 326 & 10453 & 163 & 8363 & 0 \\
\verb|PJM_MACC| & 128107 & 31388 & 1264 & 26157 & 632 & 20926 & 0 \\
\verb|PJM_WEST| & 130713 & 40055 & 1330 & 33379 & 665 & 26703 & 0 \\
\verb|SPP_N| & 29951 & 9771 & 244 & 8142 & 122 & 6514 & 0 \\
\verb|SPP_NN| & 19920 & 7153 & 197 & 5961 & 99 & 4769 & 0 \\
\verb|SPP_S| & 42827 & 13890 & 238 & 11575 & 119 & 9260 & 0 \\
\verb|S_C_TVA| & 59311 & 21454 & 467 & 17878 & 234 & 14303 & 0 \\
\verb|S_D_AECI| & 3019 & 1033 & 26 & 861 & 13 & 689 & 0 \\
\verb|S_SOU| & 93706 & 34575 & 585 & 28813 & 293 & 23050 & 0 \\
\verb|S_VACA| & 84275 & 28300 & 622 & 23583 & 311 & 18867 & 0 \\ \bottomrule
    \end{tabular}
    \label{tab:my_label}
\end{table}

\subsection*{Generators}

PowerGenome compiled data on existing generators from PUDL\cite{selvans_pudl_2021} \verb|v0.5.0|.
This includes nuclear fission, hydroelectric, coal, natural gas fired combined cycle (NG-CC), natural gas fired combustion turbine (NG-CT), natural gas steam turbines (NG-ST), additional peaker plants, plants combusting biomass, solar photovoltaics, onshore and offshore wind, and hydroelectric pumped storage. Existing generators are clustered by technology using a k-means process, resulting in 1--2 clusters per technology per zone. Hydropower reservoirs in the eastern US are assumed to have a storage capacity equal to two times the average annual reservoir inflow, and pumped hydro facilities are assumed to have storage durations of 15.5 hours \cite{phs}. Existing distributed solar capacities are gathered by state from the Energy Information Administration (EIA) Form 861M \cite{861M}. States with mandated distributed generation targets as a percentage of total load have capacities calculated directly based on these requirements \cite{NCSL}.
The existing fleet is used for the set of initial generators for 2021--2035 model period.
Generators that are not retired during that period are assumed to be available at the start of the 2036--2050 period, with some exceptions.
First the decarbonization constraint forces most existing fossil-fuel burning plants to retire. Only newly-built natural gas plants, which are allowed to convert to burn zero-carbon fuel (ZCF), i.e. hydrogen or biomethane, at no additional cost, are able to remain in service.
Second, lithium battery systems are assumed to have a lifetime of 15 years, so any batteries built in the first period do not count as existing generators at the start of the second period.
Third, existing plants, especially fission plants, that would reach their scheduled retirement before 2050 are also unavailable during the second period.

In the 2021--2035 period the model has the option to deploy new generation and storage technologies, including solar, onshore and offshore wind, lithium-ion batteries, long-duration metal-air batteries, nuclear, natural gas combined cylce (CC) and combustion turbine (CT) plants, and natural gas plants with 100\% carbon capture (NG-CCS).
The 2036--2050 period adds the option to build new  zero-carbon fuel combined cycle plants (ZCF-CC) and zero-carbon fuel combustion turbine plants (ZCF-CT), and eliminates the option to build new unabated natural gas CCs and CTs.

Hourly variability profiles for utility-scale wind and solar in each zone are calculated within PowerGenome for each resource cluster based on data from the 2012 weather year.
Distributed solar profiles for the largest urban area in each GenX zone are downloaded from Renewables Ninja \cite{reninja}, and assume fixed-angle panels with 10\% losses.
All technology cost and performance input data for each market opportunity case are available in the archived data for this work.

\subsection*{Policies}
The 2021--2035 period includes state clean energy standards and minimum capacity requirements for certain generation and storage resources \cite{NCSL}.

The simulations for 2036--2050 permit no \ce{CO2} emission from the electricity sector, so any coal or standard natural gas plants are forced to retire at the start of the period.

All simulations incorporate capacity reserve margin constraints based on regional NERC reference reserve margins \cite{nerc}.
This ensures that each of the six NERC assessment area subregions within the US Eastern Interconnection has, at all times, a prescribed level of available firm generating capacity. Inclusion of this constraint ensures reliability by accounting for potential unavailability of generation resources resulting from unscheduled outages, abnormal weather conditions, or other non-modeled causes. % language?

\subsection*{Model solution method}
We use the commercial optimization solver Gurobi.
Models are solved using a barrier method with a convergence tolerance of \num{e-4}.
Doubling the tolerance for a typical case increased the value of the objective function by \$12.5\,M, which is less than 0.006\% of the total.
Cases require 3-6 hours of wall clock time on 12 cores and a peak memory usage of \SI{180}{GB}.

We use a linearized unit commitment method to model operational constraints for fusion cores, fusion power conversion systems, and thermal generators more broadly.
A brief summary of unit commitment follows.
Devices have a `commitment' state: they are off or on. 
After turning a device off, there is a minimum time before it can be turned on, and visa-versa. 
Starting a device has an associated fuel cost and an additional monetary cost.
Devices have a minimum stable power level.
Devices have a maximum ramp rate at which they can increase or decrease their power level, which is expressed in fractions of the maximum power level per hour.
% Unit commitment treats

Integer unit commitment uses an integer variable to represent the number of committed devices for each resource in a zone; this option also means that only an integer number of devices can exist or be constructed. 
However, introduction of integer variables makes the problem a Mixed Integer Linear Program and greatly increases the solution time.
The integer constraints can be `relaxed' into linear constraints.
Resources in the linear unit commitment formalism can be considered to be composed of a set of infinitesimal devices, each with the same operational constraints as a unit device. 
This leads to LP problems where the solutions for capacities are within one half the nominal device unit size of the MILP solution.
Since the electricity system studied here has a typical scale much larger than that of a single fusion plant, this is deemed to be acceptable.

\subsection*{Additional scenarios}
We studied eleven scenarios in addition to the main three market opportunity scenarios discussed in the main text. % main main
Each of them is based on one of the main three scenarios.

\sisetup{
per-mode=symbol,
}%

% In the low-cost nuclear scenarios, fission plants have capital costs of
% \SI{3739.599}[\$]{\per\kilo\watt},
% \SI{4986.132}[\$]{\per\kilo\watt}, and
% \SI{6232.665}[\$]{\per\kilo\watt}, respectively, rather than the values shown in Table~\ref{tab:capex}.
The three low-cost fission scenarios are otherwise identical to the three main scenarios.
Fission plant costs are given in Table~2 of the main paper.
These scenarios demonstrate the sensitivity of the value of fusion to the cost of fission.

In the three constrained-renewables scenarios, each new-build solar and wind resource in each zone is constrained to 2/3 of the capacity which would be built without this constraint in the case with zero fusion.
These scenarios model cases in which land-use restrictions or other policies prevent building as much solar and wind as would be built in a system optimized for cost alone.

Three scenarios forbid new-build fission, and two forbid both new-build fission and gas with CCS, in order to model futures where one or both of these firm resources does not become widely available.
Note, however, that fusion is not the only clean firm resource.
The zero-carbon-fuel combined cycle (ZCF-CC) and combustion turbine (ZCF-CT) plants are still available in these scenarios.
As in the main scenarios, their variable cost is much higher than that of fusion plants, so they occupy a different niche in the electricity market.

\subsection*{Additional sensitivity cases}\label{sec:additionalsensitivitycases}
Three sets of additional \textit{cases} are studied, using the medium and high fusion market opportunity scenarios.

The first explores the value of dispatchable operation for fusion plants.
In this, we set the fusion core to constantly operate at its maximum capacity, and compare the value to that in standard cases where the core is dispatchable.
Figure~\ref{fig:dispatchsens} displays the results for the value of dispatchability.

The second two sets test the sensitivity of our results to the constraints on transmission expansion.
In one set, new transmission was forbidden, and in the other, there is no restriction on the amount of new transmission (the cost per gigawatt-mile is the same).
Figure~\ref{fig:transmissionsens} shows the influence of the two alternate transmission policies.

\clearpage

\section*{Note S2: Economic and operational parameter tables}\label{sec:econop}
Tables~2 and~3 of the main paper list the capital and variable costs of the various resources.
This section lists additional data for the GenX cases.

Capital costs for the NG-CCS plants (Table~2) and their operational costs (Table~3) include these costs for \ce{CO2} pipelines to a geological storage basin.
\ce{CO2} pipeline costs by GenX zone are listed in Table~\ref{tab:pipelines}. 

\begin{table}[htb]
\caption{\label{tab:pipelines} \ce{CO2} pipeline and injection cost adders for NG-CCS plants by GenX zone.}
\centering
\begin{tabular}{@{}ld{7.0}d{1.2}@{}} \toprule
Zone &
\multicolumn{1}{c}{Annuity (\$/MW)} &
\multicolumn{1}{c}{Variable OM (\$/MWh)} \\ \midrule
\verb|FRCC| & 4691 & 5.85 \\
\verb|ISONE| & 157793 & 6.6 \\
\verb|MISO_ILINKY| & 4691 & 6.6 \\
\verb|MISO_LRZ123| & 22350 & 6.6\\
\verb|MISO_LRZ7 | & 41655 & 6.6\\ 
\verb|MISO_LRZ8910| & 4691 & 5.65\\
\verb|MIS_MO| & 4691 & 6.6\\
\verb|NY_E| & 122972 & 6.6\\
\verb|NY_W| & 83678 & 6.6\\
\verb|PJM-COMD| & 9189 & 6.6\\
\verb|PJM-Dom| & 86798 & 6.6\\
\verb|PJM_MACC| & 108548 & 6.6\\
\verb|PJM_WEST| & 4691 & 6.6\\ 
\verb|SPP-N| & 35259 & 6.6\\
\verb|SPP_NN| & 54360 & 6.6\\
\verb|SPP_S| & 4691 & 6.2\\ 
\verb|S_C_TVA| & 6952 & 5.65\\
\verb|S_D_AECI| & 11410 & 6.6\\
\verb|S_SOU| & 30976 & 5.85\\
\verb|S_VACA| & 42718 & 5.85\\
\bottomrule \end{tabular}
\end{table}

Resources have an additional cost to account for transmission spur lines, with regional costs from \$3686 per \si{\mega\watt}-mile to \$6320 per \si{\mega\watt}-mile. The length of the spur lines is shown in Table~\ref{tab:spurlines}; wind and solar have variable spur line lengths and costs from a method developed for the Net-Zero America study\cite{larson_net-zero_2021}.

\begin{table}[ht]
\begin{center}
\caption{Spur line lengths for new-build plants.}\label{tab:spurlines}
\begin{tabular}{@{}ld{2.0}@{}} \toprule
& \multicolumn{1}{r@{}}{Spur length / miles} \\
\midrule
Utility-scale Solar PV & var. \\
Onshore wind & var. \\
Offshore wind & var.  \\
ZCF-CT & 20  \\
ZCF-CC & 20  \\
NG-CCS & 20  \\
Fission & 50  \\
Li batteries  & 10  \\
Metal-air batteries & 10 \\
\end{tabular}
\end{center}
\end{table}

Startup costs for thermal plants are from the Western Wind and Solar Integration Study\cite{lew_western_2013}.
These and other thermal plant operational parameters, such as ramp rates and minimum up and down (commitment) times, are listed in Table~\ref{tab:thermoperational}.

\begin{table}
\caption{\label{tab:thermoperational} Operational parameters for the thermal generators.
Note (a): start `fuel' is implemented somewhat differently for fusion plants. Instead of requiring a larger fuel supply during the start hour, the fusion PCS wastes heat equivalent to 0.2 of the peak core thermal output. }
\centering
\begin{tabular}{@{}ld{2.0}d{1.2}d{1.1}d{3.0}d{1.1}@{}} \toprule
 & \multicolumn{1}{c}{Min.\ commit} 
 & \multicolumn{1}{c}{Ramp} 
 & \multicolumn{1}{c}{Min}
 & \multicolumn{1}{c}{Start cost / }
 & \multicolumn{1}{c}{Start fuel / } \\
 & \multicolumn{1}{c}{time / h} 
 & \multicolumn{1}{c}{rate $\cdot$ h} 
 & \multicolumn{1}{c}{power}
 & \multicolumn{1}{c}{(\$/MW)}
 & \multicolumn{1}{c}{(MMBTU / MW)} \\ \midrule
ZCF-CT      & 1  & 1    & 0.2 & 134 & 3.5 \\
ZCF-CC      & 6  & 0.64 & 0.3 & 103 & 2 \\
NG-CCS      & 6  & 0.64 & 0.6 & 103 & 2 \\
Fission     & 24 & 0.25 & 0.5 & 278 & 0 \\
Fusion: PCS & 1  & 1    & 0.4 & 100 & \multicolumn{1}{c}{0.2$^a$} \\ \bottomrule
\end{tabular}
\end{table}

\subsection*{Transmission lines}
Table~\ref{tab:powerlines} lists the inter-zonal transmission lines in the system, their existing capacity, the potential for expansion, the cost of expansion, and their length.
The existing capacity is equal to that in 2020.
The maximum capacity expansion is 150\% of the existing capacity or \SI{1.5}{\giga\watt}, whichever is greater.
Expansion costs depend on the length and regional factors.
Transmission losses are proportional to the power transmitted and the length of the line, at a rate of 1\% per hundred miles: for example, the first line loses 3.35\% of power transmitted.
In the main scenarios, transmission expansion is typically about 10 times larger in the 2036--2050 period than in the 2021--2035 period.

\begin{table}[htb]
\caption{\label{tab:powerlines} Existing transmission lines and their potential for expansion.}
\centering
\begin{tabular}{@{}d{2.0}d{2.0}d{2.2}d{2.2}d{2.1}d{3.0}@{}} \toprule
\multicolumn{2}{c}{} & \multicolumn{1}{c}{Capacity/}& \multicolumn{1}{c}{Max. expansion/} & \multicolumn{1}{c}{Cost/} & \multicolumn{1}{c}{Length/} \\
\multicolumn{2}{c}{Zones} & 
\multicolumn{1}{c}{GW} & \multicolumn{1}{c}{GW} &
\multicolumn{1}{c}{(M\$/\si{\giga\watt.yr})} & 
\multicolumn{1}{c}{mile} \\ \midrule
 1& 19&             3.6&            5.4&        64.6&      335 \\
 2&  8&             2.2&            3.2&        88.3&      199 \\
 2&  9&             0.0&            1.5&        48.6&      290 \\
 3&  4&             1.3&            1.9&        63.7&      578 \\
 3&  5&             0.1&            1.5&        34.8&      339 \\
 3&  7&             4.5&           6.75&        21.8&      204 \\
 3& 10&             7.1&           10.6&        20.0&      160 \\
 3& 13&            11.9&           17.9&        29.9&      312 \\
 3& 17&             2.0&            3.0&        50.0&      276 \\
 4&  5&             0.1&            1.5&        80.3&      555 \\
 4&  8&             2.3&            3.4&        60.7&      512 \\
 4& 10&             5.8&            8.7&        41.5&      429 \\
 4& 14&             0.0&            1.5&        62.2&      510 \\
 4& 15&             8.5&           12.7&        57.5&      276 \\
 4& 18&             0.1&            1.5&        65.4&      524 \\
 5& 13&             4.8&            7.3&        51.6&      322 \\
 6& 14&            0.17&            1.5&        44.4&      527 \\
 6& 16&             3.9&            5.8&        41.6&      435 \\
 6& 17&             6.3&            9.5&        49.3&      349 \\
 6& 18&             2.4&            3.6&        38.8&      421 \\
 6& 19&             2.2&            3.2&        50.3&      382 \\
 7& 14&             1.0&            1.5&        20.2&      305 \\
 7& 18&             3.2&            4.9&        25.2&       34 \\
 8&  9&             3.6&            5.4&        71.8&      126 \\
 8& 12&             0.6&            1.5&        37.2&      206 \\
 9& 12&             2.5&           3.75&        46.8&      191 \\
10& 13&             4.0&            6.0&        35.2&      396 \\
11& 12&             2.8&            4.2&        21.1&      219 \\
11& 13&            11.8&           17.7&        50.6&      271 \\
11& 20&             2.6&            3.9&        56.8&      218 \\
12& 13&             9.4&           14.1&        48.8&      300 \\
13& 17&             4.5&           6.75&        63.0&      364 \\
13& 20&             2.2&            3.3&        66.5&      327 \\
14& 15&             2.3&            3.5&        15.9&      371 \\
14& 16&             7.5&           11.3&        26.9&      239 \\
14& 18&             2.1&            3.1&        17.8&      279 \\
15& 16&             1.5&            2.3&        29.4&      581 \\
15& 18&             1.4&            2.1&        27.5&      533 \\
16& 18&             1.7&            2.6&        28.0&      418 \\
17& 18&             0.6&            1.5&        45.4&      364 \\
17& 19&             5.1&            7.6&        58.8&      237 \\
17& 20&            0.28&            1.5&        75.0&      379 \\
18& 20&             3.0&            4.5&        65.1&      331 \\

\bottomrule
\end{tabular}
\end{table}

\subsection*{Capacity reserve margin policy}
The capacity reserve margin (CRM) policy ensures that during each hour there is sufficient spare capacity in each of six regions, listed in Table~\ref{tab:crmOverview}.

\begin{table}[htb]
\caption{\label{tab:crmOverview} Capacity reserve margin constraint regions and included zones}
\centering
\begin{tabular}{@{}lll@{}} \toprule
CRM region & Margin &  Zones \\
1 & 0.183 & \verb|ISONE| \\
2 & 0.15 & \verb|NY_E|, \verb|NY_W| \\
3 & 0.155 & \verb|PJM_COMD|, \verb|PJM_Dom|, \\
  &       & \verb|PJM_MACC|, \verb|PJM_WEST| \\
4 & 0.12 & \verb|SPP_N|, \verb|SPP_NN|, \verb|SPP_S| \\
5 & 0.12 & \verb|MISO_ILINKY|, \verb|MISO_LRZ123|, \verb|MISO_LRZ7| \\
  &  & \verb|MISO_LRZ8910|, \verb|MIS_MO| \\
6 & 0.15 & \verb|FRCC|, \verb|S_C_TVA|, \verb|S_D_AECI| \\
  &      & \verb|S_SOU|, \verb|S_VACA| \\
\bottomrule \end{tabular}
\end{table}

Wind and solar provide 80\% of their unused capacity to the CRM.
Batteries provide 95\% of the power they generate to the CRM.
Thermal generators, including fusion, provide 90\% of their unused capacity to the CRM.
See the GenX documentation for details of this policy.

\subsection*{Economic assumptions for fusion}

This study assumes a 30 year asset life and 3.34\% real WACC for all fusion plants; this is the same real weighed average cost of capital (WACC) as for fission in the NREL ATB Market + Policies case\cite{nrel_national_renewable_energy_laboratory_annual_2021}, and leads to a capital recovery factor of 5.3\%.
We also assume a fixed operations and maintenance (FOM) cost of 2.5\% of the capital cost, leading to annual cost of 7.8\% of the capital cost.
To translate the plant capital costs in the paper to a scenario with different assumptions, 
multiply them by the entry in Table~\ref{tab:CAPEXconv} for the chosen combination of asset life and real WACC.

\begin{table}
\caption{\label{tab:CAPEXconv} Capital cost conversion ratios between different asset life and real weighted average cost
of capital (WACC) assumptions.}
\centering
\begin{tabular}{@{}d{1.2}|d{1.2}d{1.2}d{1.2}d{1.2}d{1.2}d{1.2}d{1.2}d{1.2}@{}} \toprule
\multicolumn{1}{c}{} & \multicolumn{8}{c}{Asset life / years} \\
\multicolumn{1}{@{}l}{WACC} & \multicolumn{1}{c}{25} & \multicolumn{1}{c}{30} & \multicolumn{1}{c}{35} & \multicolumn{1}{c}{40} & \multicolumn{1}{c}{45} & \multicolumn{1}{c}{50} & \multicolumn{1}{c}{55} & \multicolumn{1}{c}{60} \\ \midrule
1.00\% & 1.11 & 1.23 & 1.33 & 1.41 & 1.49 & 1.55 & 1.61 & 1.66 \\
2.00\% & 1.03 & 1.12 & 1.20 & 1.27 & 1.33 & 1.38 & 1.42 & 1.46 \\
3.00\% & 0.95 & 1.03 & 1.09 & 1.15 & 1.19 & 1.23 & 1.26 & 1.28 \\
3.34\% & 0.93 & 1.00 & 1.06 & 1.11 & 1.15 & 1.18 & 1.21 & 1.23 \\
4.00\% & 0.88 & 0.95 & 1.00 & 1.04 & 1.07 & 1.09 & 1.11 & 1.13 \\
5.00\% & 0.82 & 0.87 & 0.91 & 0.94 & 0.96 & 0.98 & 1.00 & 1.01 \\
6.00\% & 0.76 & 0.80 & 0.83 & 0.86 & 0.87 & 0.89 & 0.89 & 0.90 \\
7.00\% & 0.71 & 0.74 & 0.77 & 0.78 & 0.79 & 0.80 & 0.81 & 0.81 \\ \bottomrule
\end{tabular}
\end{table}

\subsection*{Base case loads and prices in the scenarios}

\begin{figure}%
\centering
\includegraphics[width=0.95\textwidth]{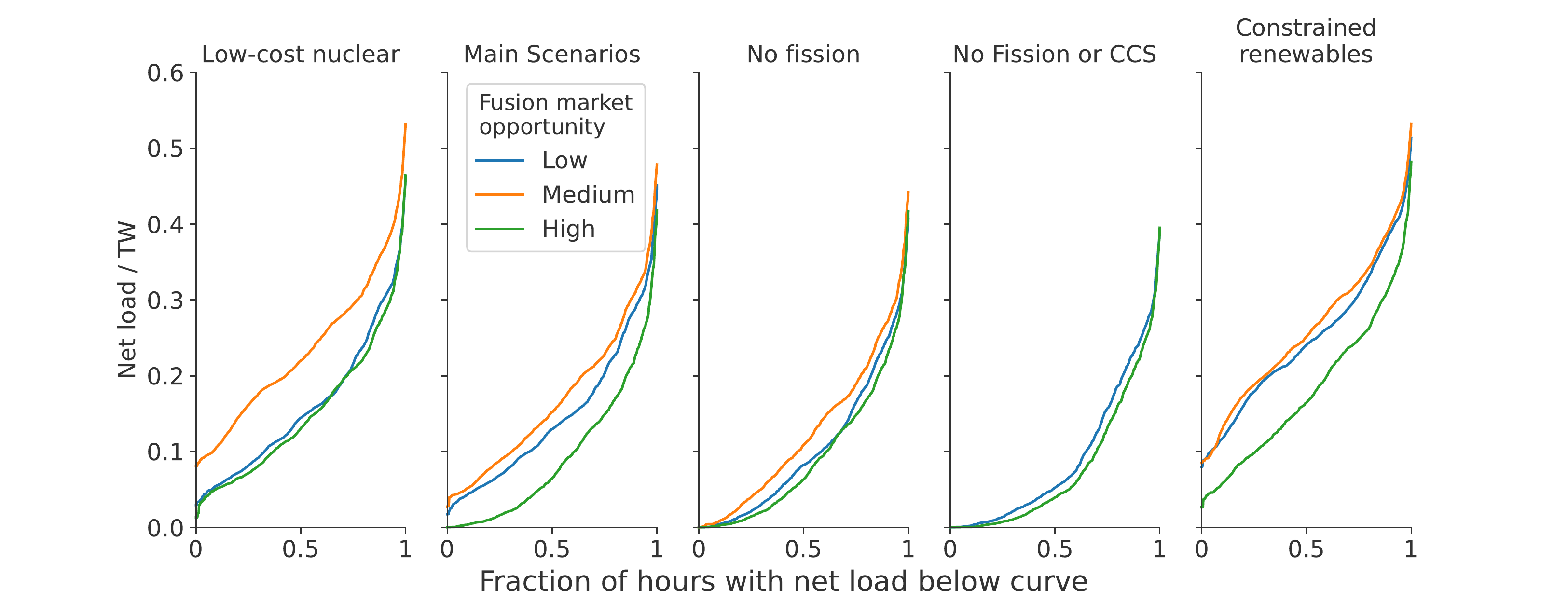}
\caption{Net load duration curves---the fraction of hours that the total system load supplied by thermal generators is above a certain level---for base cases (those with zero fusion capacity) in the fourteen scenarios.
}\label{fig:netloadduration10scenarios}
\end{figure}
Figure~\ref{fig:netloadduration10scenarios} shows the net load duration curves for the base cases of the fourteen scenarios. Cases with a shallower curve typically have more power supplied by renewables. A steeper curve means that thermal generators are used more often.

\begin{figure}%
\centering
\includegraphics[width=0.95\textwidth]{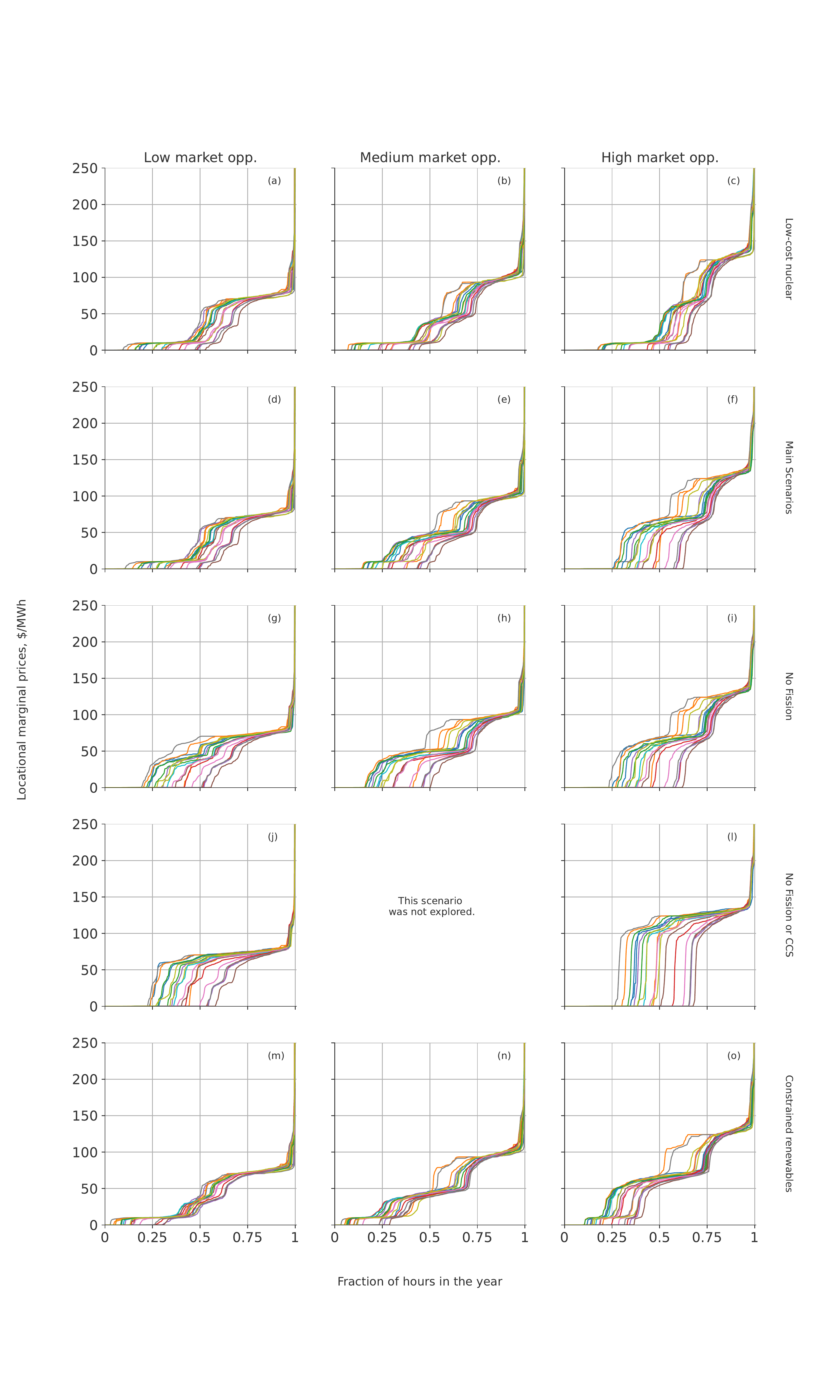}
\caption{Sorted hourly locational marginal prices for each of the twenty zones in the base cases (those without fusion) for each of the ten scenarios.
}\label{fig:base_case_prices}
\end{figure}
Figure~\ref{fig:base_case_prices} shows the sorted hourly price of electricity each zone in the base case of each of the fourteen scenarios.
Prices are zero during hours when the load can be met entirely with variable renewables.
Plateaus in price occur at the marginal cost of generation for the various types of plants---fission, gas with CCS, and combined cycle or combustion turbine plants burning zero carbon fuels.

\begin{figure}%
\centering
\includegraphics[width=0.95\textwidth]{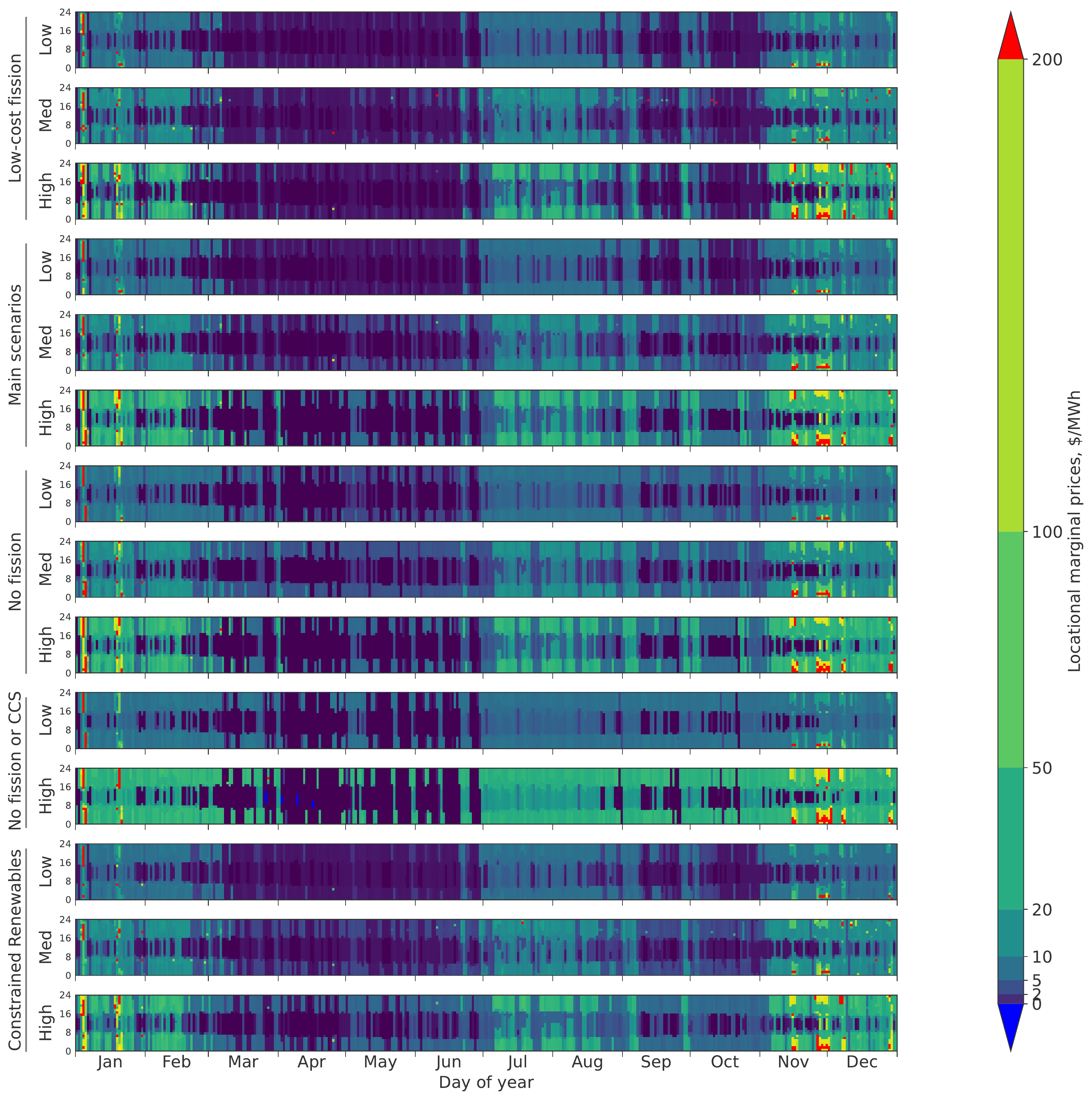}
\caption{Hourly price charts for the \texttt{PJM\_MACC} (Mid-Atlantic) zone in the base cases (those without fusion) in each scenario.
This zone is where fusion is built first in the main scenarios, and the price patterns are typical of most of the zones.
Hours with prices higher than \$200/\si{\mega\watt\hour} are shown in red, and hours with negative prices are shown in blue.
The period of three to four months roughly from March through the end of June, has low prices.
This could be a good time to schedule fusion reactor maintenance.
}\label{fig:base_hourly_price_charts}
\end{figure}
Figure~\ref{fig:base_hourly_price_charts} shows the hourly price series in a particular region, \verb|PJM_MACC| (Mid-Atlantic), for the base case in each of the scenarios.
There is a sustained period of low prices in the spring, from March through June. 
This suggests that if fusion maintenance periods could be limited to this window, a fusion plant would not lose much of its value to the system.

%\FloatBarrier
\clearpage
\section*{Note S3: Fusion plant implementation}\label{sec:fusimp}
This Section describes the mathematical formulation of the fusion plant in GenX.

\subsection*{Fusion module indices, variables, and parameters}
\begin{table}[h]
\centering
\caption{Model indices and sets} 
\begin{tabular}{@{}l p{8cm}@{}}
\toprule
Notation & Description \\ \midrule
$t \in T$ & Where $T$ denotes the set of all hours in the modeled weather year and $t$ denotes a particular hour  \\
$z \in Z$ & Where $Z$ is the set of all model zones and $z$ denotes a particular zone (a geographical region) \\
$ y \in G$ & Where $G$ denotes the set of available technologies and $y$ is a particular technology \\
$FUS \subseteq G$ & The set of fusion plants \\
\bottomrule
\end{tabular}
\end{table}

\renewcommand{\arraystretch}{1.4}

\begin{table}[h]
\centering
\caption{Model variables} 
\begin{tabular}{@{}l p{8cm}@{}}
\toprule
Notation & Description \\ \midrule
$\vCCAP$ & Core thermal capacity \\
$\vTSCAP$ & Energy capacity \\
$\vCAP$ & Power conversion system capacity \\

$\vCP$ & Thermal energy generated by the core \\
$\vTS$ & Level of stored thermal energy \\
$\vP$ & Power injected into the grid \\
% $\Pi_{z,t}$ & Energy withdrawn from the grid \\
$\vFCOMMIT$ & Commitment state of the core \\
$\vFSTART$ & Startup decisions of the core \\
$\vFSHUT$ & Shutdown decisions of the core \\

$\vCOMMIT$ & Commitment state of the power conversion system \\
$\vSTART$ & Startup decisions of the power conversion system \\
$\vSHUT$ & Shutdown decisions of the power conversion system \\

\bottomrule
\end{tabular}
\end{table}

\begin{table}[h]
\centering
\caption{Model expressions} 
\begin{tabular}{@{}l p{8cm}@{}}
\toprule
Notation & Description \\ \midrule
$\eRecircPass$ & Thermal core passive recirculating power \\
$\eRecircAct$ & Thermal core active recirculating power \\
$\eRecircStart$ & Thermal core startup energy \\
$\eRecircTot$ & Thermal core total recirculating power \\
\bottomrule
\end{tabular}
\end{table}

% \begin{table}[h]
% \centering
% \caption{Model expressions} 
% \begin{tabular}{@{}l p{8cm}@{}}
% \toprule
% Notation & Description \\ \midrule
% $e^{\mathrm{CVar}, \mCore}$ & Variable cost of core operation in an hour \\
% $e^{\mathrm{totalCVar}, \mCore}$ & Sum of the variable cost of core  operation over the model period \\ 
% $e^{\mathrm{CFixed}, \mCore}$ & Fixed costs of thermal cores \\
% $e^{\mathrm{CFixed}, \mStorage}$ & Fixed costs of storage \\

% \bottomrule
% \end{tabular}
% \end{table}

\begin{table}[h]
\centering
\caption{Model parameters.} 
\begin{tabular}{@{}l p{8cm}@{}}
\toprule
Notation & Description \\ \midrule
$\hourweight$ & Weight of the hour $t$ \\
$\coreUnitSize$ & Unit size of the thermal core \\
$\genUnitSize $ & Unit size of the power conversion system \\

$\maxCoreCap$ & Maximum thermal core capacity \\
$\maxStorageCap $ & Maximum energy storage capacity \\
$\maxGenCap$ & Maximum power conversion system discharge capacity \\

$\coreInvCost$ & Annualized investment cost of the thermal core \\ 
$\storageInvCost$ & Annualized investment cost of the thermal storage system \\ 
$\genInvCost$ & Annualized investment cost of the power conversion system \\

$\coreFOMCost$ & Fixed OM cost of the thermal core \\ 
$\storageFOMCost$ & Fixed OM cost of the thermal storage system \\ 
$\genFOMCost$ & Fixed OM cost of the power conversion system \\

$\coreVOMCost$ & Variable OM cost of the thermal core \\ 
$\genVOMCost$ & Variable OM cost of the power conversion system \\

$\genStartCost$ & Start-up cost (of the power conversion system) \\
$\genStartHeat$ & Start-up heat for power conversion system \\
% $\eta^{charge}_{z}$ & Efficiency of storing heat rather than using it immediately \\
$\dischargeEfficiency$ & Efficiency of converting thermal energy to electrical energy \\ \midrule
%$\eta^{loss}_z$ & Self-discharge rate per time step per unit of installed capacity \\
% $\rho^{max}_{y,z,t}$ & Maximum fraction of discharge capacity available during time step $t$ \\
$\tpulse$ & Fusion reactor maximum pulse cycle length \\
$\tdwell$ & Fusion reactor dwell time \\
$\rpass$ & Fusion reactor passive recirculating power \\
$\ract$ & Fusion reactor active recirculating power \\ 
$\estart$ & Fusion reactor startup energy requirement\\
$\pstart$ & Fusion reactor startup peak power requirement\\
$\coreMinStableOutput$ & Minimum stable output level for the thermal core \\

$\Omega^{\mathrm{System\ total,\ FUS}}$ & Maximum fusion capacity in the system \\
\bottomrule
\end{tabular}
\end{table}
\newpage

\subsection*{Variable limits and constraints}

Capacities for the core, storage, and power conversion system (PCS) are non-negative.
In each hour, the thermal power produced by the core, the level of energy in the thermal storage system, and the amount of energy output to the grid by the PCS is non-negative.
The unit commitment variables for the core and PCS are also non-negative.
\begin{equation}
\begin{aligned}
    \vCCAP &\ge 0 \qquad &\vFCOMMIT & \ge 0   \\
    \vTSCAP &\ge 0 \qquad  &\vFSTART & \ge 0  \\
    \vCAP & \ge 0 \qquad &\vFSHUT & \ge 0 \\
    \vCP &\ge 0 \qquad     &\vCOMMIT & \ge 0 \\ 
    \vTS &\ge 0 \qquad  &\vSTART & \ge 0 \\
    \vP & \ge 0 \qquad  &\vSHUT&  \ge 0\\
\end{aligned}
\end{equation}

The core thermal power must be less than the installed core capacity,
\begin{equation}
    \vCP \le \vCCAP.
\end{equation}
The installed core capacity, storage capacity, and PCS capacity in a zone can optionally be limited:
\begin{equation}
\begin{aligned}
    \vCCAP & \le \maxCoreCap \\
    \vTSCAP & \le \maxStorageCap \\
    \vCAP & \le \maxGenCap, \\
    \end{aligned}
\end{equation}
though these constraints are not used in our simulations.
The variable cost of core operations during the year is the sum over the cost in each timestep,
\begin{equation}
    e^\mathrm{totalVarCore} = \hourweight \,\coreVOMCost\, \vCP.
\end{equation}
The annual fixed cost of the plants is the sum of the investment costs and fixed OM costs of their parts, 
\begin{equation}
\begin{aligned}
    e^\mathrm{\mInvest} = & \, \vCCAP \left(\coreInvCost + \coreFOMCost\right) \\ + & \, \vTSCAP \left(\storageInvCost + \storageFOMCost\right) \\ + & \, \vCAP \left(\genInvCost + \genFOMCost\right); \\
    \end{aligned}
\end{equation}
this expression is added to the objective.
The level of energy storage in the thermal storage system must be less than the installed capacity,
\begin{equation}
    \vTS \le \vTSCAP.
\end{equation}
The level of energy storage at the end of an hour equals the energy during the previous hour, plus any thermal power created by the core, less discharge to the PCS and any PCS startup energy usage.
\begin{equation}
\begin{aligned}
    \vTS{} = \vTSMinusOne{} +\; & \vCP{} \\
     -\;  & \vP{} / \dischargeEfficiency{}     \\
     -\; & \vSTART{} \, \genUnitSize{} \, \genStartHeat{} / \dischargeEfficiency{} \\
\end{aligned}
\end{equation}
The quantity of core committed in each hour is less than the total quantity of fusion cores.
\begin{equation}
\begin{aligned}
    \vFCOMMIT{} \le & \vCCAP{} / \coreUnitSize{} \\
    \vFSTART{} \le & \vCCAP{} / \coreUnitSize{} \\
    \vFSHUT{} \le & \vCCAP{} / \coreUnitSize{} \\
\end{aligned}
\end{equation}
The commitment state in a given hour is that of the previous hour, plus any that started, minus any that have shut down:
\begin{equation}
    \vFCOMMIT{} = \vFCOMMITMinusOne{} + \vFSTART{} - \vFSHUT{}.
\end{equation}
If the core is committed, its power is greater than the minimum power level:
\begin{equation}
    \vCP{} \ge \coreMinStableOutput{} \coreUnitSize{} \vFCOMMIT{}.
\end{equation}
The maximum thermal output is lower if the core is starting during this hour.
The dwell time between pulses is counted during the hour of a core starting:
\begin{equation}
    \vCP{} \le \coreUnitSize{} \left( \vFCOMMIT{} - \tdwell\,\vFSTART{} \right).
\end{equation}
Pulsed tokamaks have a maximum pulse cycle length $\tpulse$. In order for a quantity of the thermal core to be committed it must have started in the last $\tpulse$ hours:
\begin{equation}
    \vFCOMMIT{} \le \sum_{u = 0}^{\tpulse - 1} \vFSTARTHour{t - u}.
\end{equation}
The passive recirculating power required each hour,
\begin{equation}
    \eRecircPass{} =\rpass\,\dischargeEfficiency{} \vCCAP{},
\end{equation}
depends on the built core capacity.
The active recirculating power,
\begin{equation}
    \eRecircAct{} = \ract\,\dischargeEfficiency{}  \left(\vFCOMMIT{} - \tdwell \vFSTART{}\right) \coreUnitSize{}
\end{equation}
is proportional to the maximum power which can be produced during an hour. It does not decrease if the core is set at a lower power level.
The thermal core startup energy,
\begin{equation}
    \eRecircStart{} = \estart\,\dischargeEfficiency{} \vFSTART{} \coreUnitSize{}
\end{equation}
is proportional to the quantity of thermal core which starts. Note that although most units of output are given here in terms of power, they effectively represent total energy generated over an hour-long model timestep. Startup loads are assumed to have durations much shorter than an hour, so the total input energy required for each pulse is assumed to be smoothed over the model timestep in which the pulse occurs. Consequences of the large instantaneous power requirement are captured in fusion's contribution to the capacity reserve margin constraint, which is described below.
The total recirculating power is
\begin{equation}
    \eRecircTot{} = \eRecircPass{} + \eRecircAct{} + \eRecircStart{}.
\end{equation}
The net of recirculating power and generated power 

\begin{equation}
    \vP - \eRecircTot{}
\end{equation}
for each fusion generator is added to the power balance constraint in its respective zone.

Fusion also contributes to the capacity reserve margin constraint in GenX, which requires a certain level of net dispatchable capacity availability at each model timestep. Fusion's contribution to this constraint is equal to
\begin{equation}
    \vP - \eRecircPass{} - \eRecircAct{} - \pstart \dischargeEfficiency{} \vFSTART{} \coreUnitSize{}
\end{equation}
where the final term captures the impact of a fusion reactor's instantaneous startup power requirement on the system's capacity availability needs at a given timestep.

\subsection*{Fusion system-wide capacity constraint}
In this paper we use a constraint on the total fusion capacity in the system.
For plants that do not have a maximum pulse cycle length ($\tpulse > 0$),
\begin{equation}
\begin{aligned}
    f_\mathrm{active} & = 1 \\
    \left<f_\mathrm{start}\right> & = 0 \\
\end{aligned}
\end{equation}
and for those that do,
\begin{equation}
\begin{aligned}
    f_\mathrm{active} & = 1 - \tdwell / \tpulse \\
    \left<f_\mathrm{start}\right> & = \estart / \tpulse. \\
\end{aligned}
\end{equation}
The gross thermal capacity factor is
\begin{equation}
    f_\mathrm{netavgcap} = f_\mathrm{active}\left(1 - \ract\right) - \rpass - \left<f_\mathrm{start}\right>
\end{equation}
and the gross thermal to net electric capacity factor is
\begin{equation}
    f_\mathrm{netavgcap}^{el} = \dischargeEfficiency{} f_\mathrm{netavgcap}
\end{equation}
The sum of the thermal core capacity in each zone weighted by this factor must be less than the total system maximum fusion capacity.
\begin{equation}
    \sum_{y \in \mathrm{FUS}, z \in Z} \left(\vCCAP{} \; f_\mathrm{netavgcap}^{el}\right)_{y,z} \le \Omega^{\mathrm{System\ total,\ FUS}}
\end{equation}

\clearpage

\section*{Supplemental figures}\label{sec:suppfigs}

\begin{figure}[htb]
\centering
\includegraphics[width=\textwidth, center]{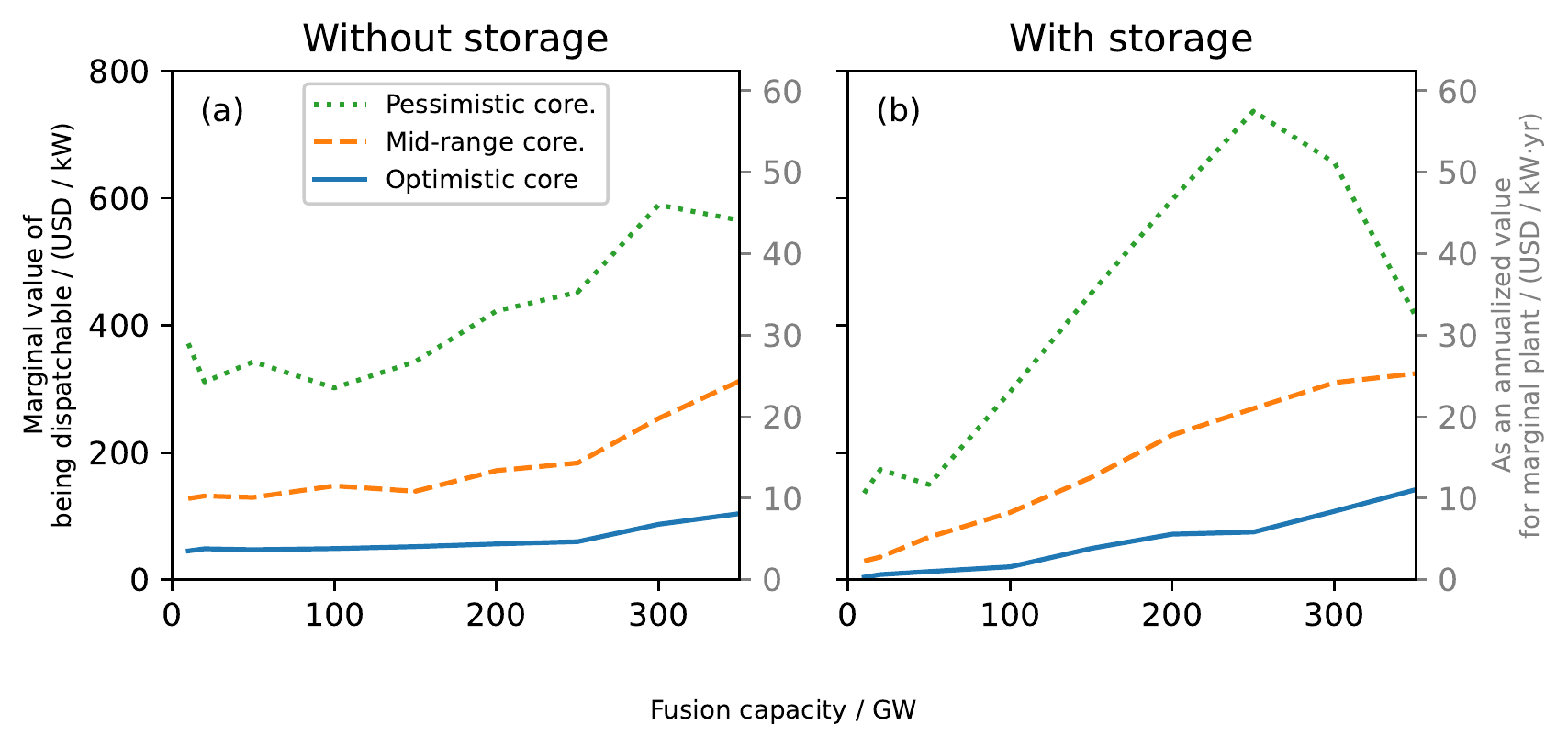}
\caption{The value of dispatchable operation for the core, as opposed to forcing the core to operate at its maximum capacity, for the three reference reactors with and without storage in the medium market opportunity scenario.}
\label{fig:dispatchsens}
\end{figure}
Figure~\ref{fig:dispatchsens} shows the marginal value provided by dispatchability for the fusion core, versus sensitivity cases where the core is forced to run at its maximum capacity. 
(The cores are dispatchable for the remainder of this work.)
Here, dispatchable means that the core can
\begin{itemize}
\item operate at any power level, from 0 to full power, while committed;
\item stop a pulse before the standard pulse length has ended;
\item lay idle any number of hours between pulses.
\end{itemize}
Dispatchability is much more important for plants with higher variable costs, like the pessimistic core, since they would waste more money on variable operations and maintenance (VOM) costs during times of low electricity prices.
Part (a) shows that the marginal value of dispatchability without thermal storage; it is nonzero and relatively insensitive to the fusion penetration.
Part (b) shows that with thermal storage, for plants with a low capacity penetration, being dispatchable is less important, and conversely more important at a high capacity penetration.
At low penetration, the non-dispatchable fusion cores can fill their thermal storage and discharge later.
At high penetration, the total appetite in the system for short-duration storage of this form may be saturated, hence more strongly diminishing returns for systems forced to constantly run.

\begin{figure}[htb]
\centering
\includegraphics[width=\textwidth, center]{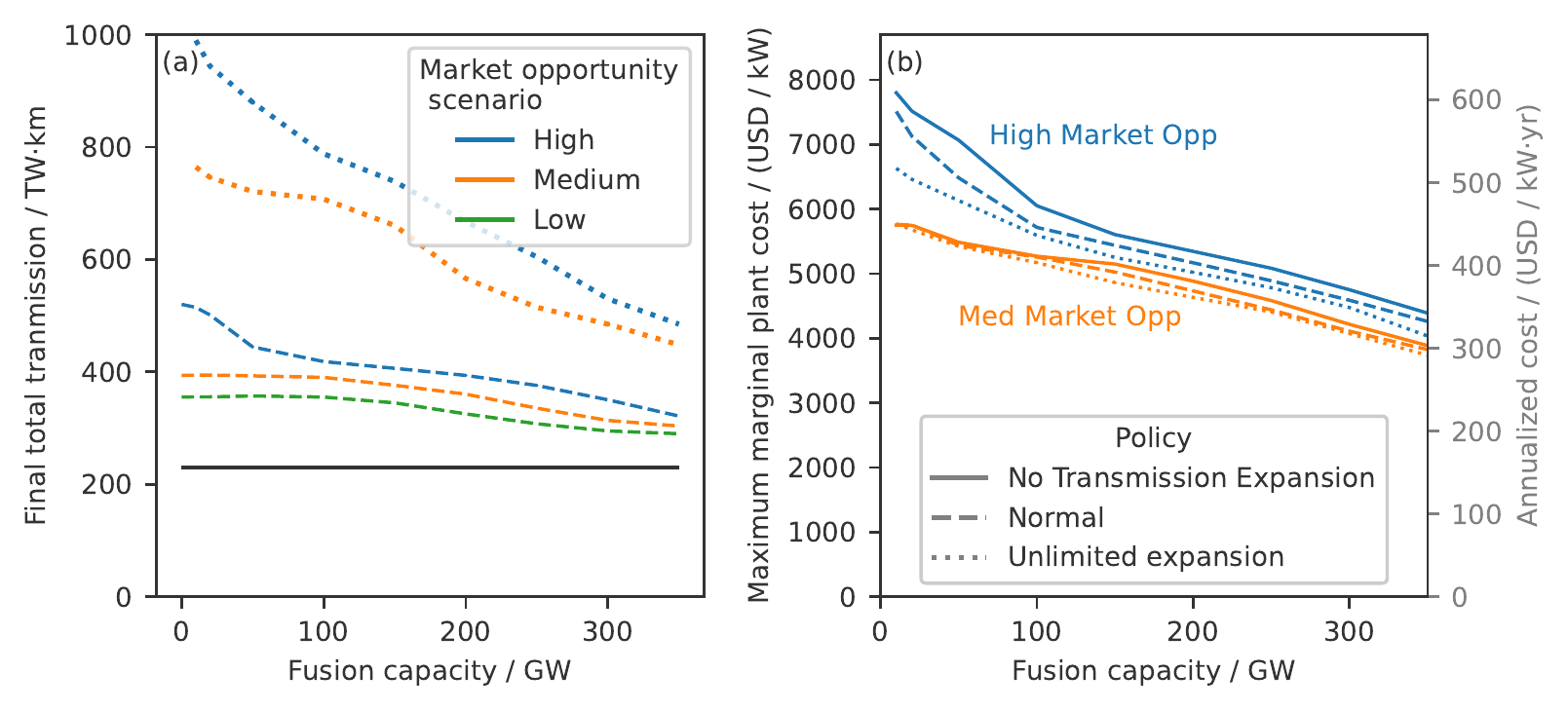}
\caption{Part (a): Transmission built in the standard cases (dashed) as well as additional sensitivity cases with unlimited transmission expansion (dotted), and with no transmission expansion (solid black), for mid-range fusion plants without thermal storage.
Part (b): Influence of the transmission expansion policy on the value of fusion.}
\label{fig:transmissionsens}
\end{figure}
Figure~\ref{fig:transmissionsens} shows the influence of the transmission policy on the amount of transmission built, and on the value of fusion.
In the medium market opportunity scenario, neither of these affected the value of fusion significantly, especially while the total fusion capacity was less than \SI{100}{\giga\watt}. Until this level of capacity penetration, fusion is largely substituting for new-build fission.
Above that capacity penetration, forbidding new transmission slightly increases the marginal value of fusion, by up to \SI{150}[\$]{\per\kilo\watt}, and unrestricted expansion slightly decreases the value, also by up to \SI{150}[\$]{\per\kilo\watt}.

In the base case of high market opportunity scenario (the case with zero fusion, and a finite allowed quantity of transmission expansion), many of the transmission constraints are binding, and an increasing penetration of fusion decreases the optimal quantity of new transmission.
Forbidding transmission expansion entirely increases the value of fusion by up to \SI{600}[\$]{\per\kilo\watt}, and allowing unlimited transmission expansion decreases its value by up to \SI{900}[\$]{\per\kilo\watt} (at the lowest capacity penetrations).

\begin{figure}[htb]
\centering
\includegraphics[width=1.25\textwidth, center]{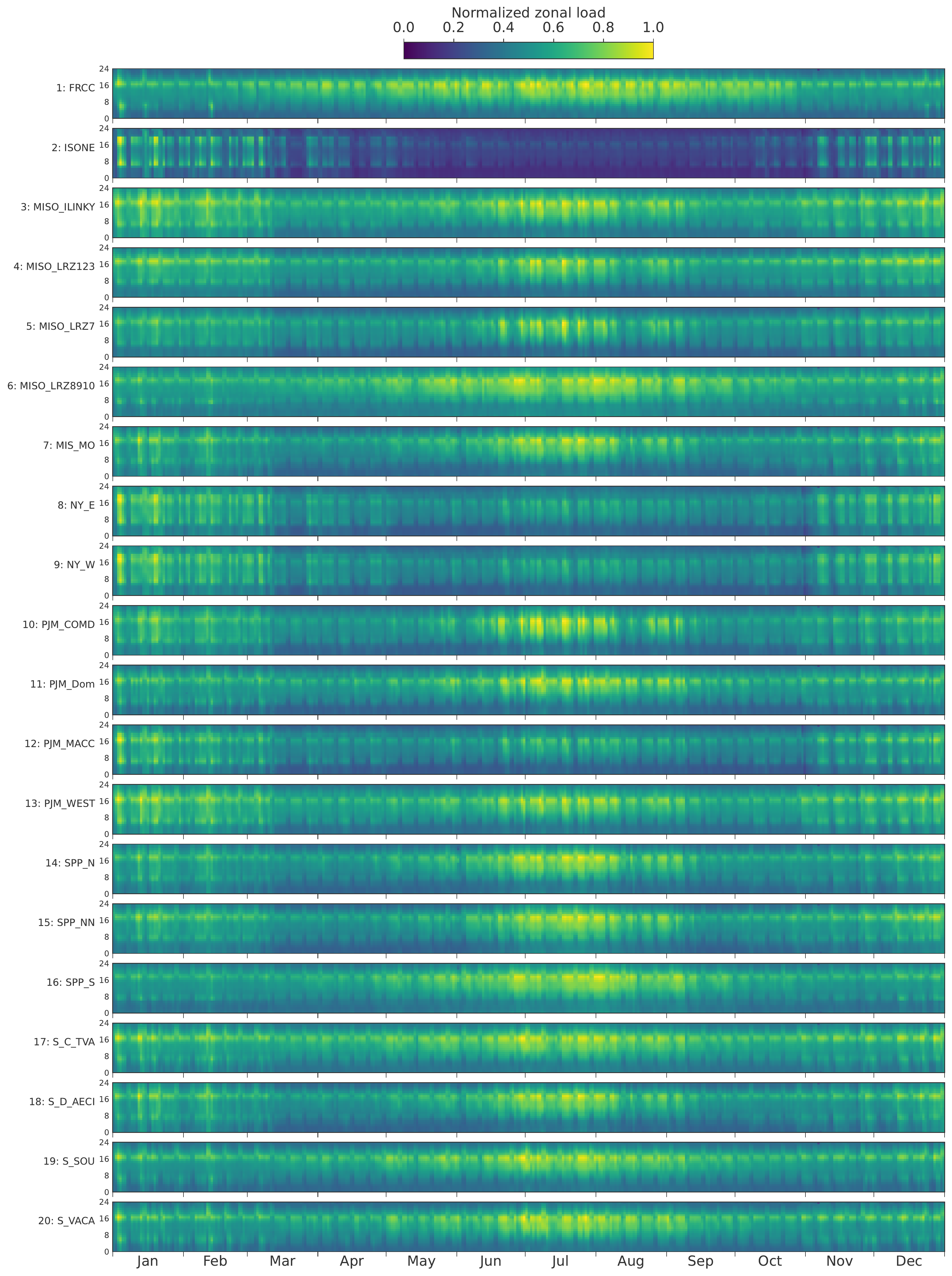}
\caption{Normalized hourly load profiles in each zone. Peak values are listed in Table \ref{tab:tfloads}.}
\label{fig:hourlyloadchart}
\end{figure}

\begin{figure}%
\centering
\includegraphics[width=0.95\textwidth]{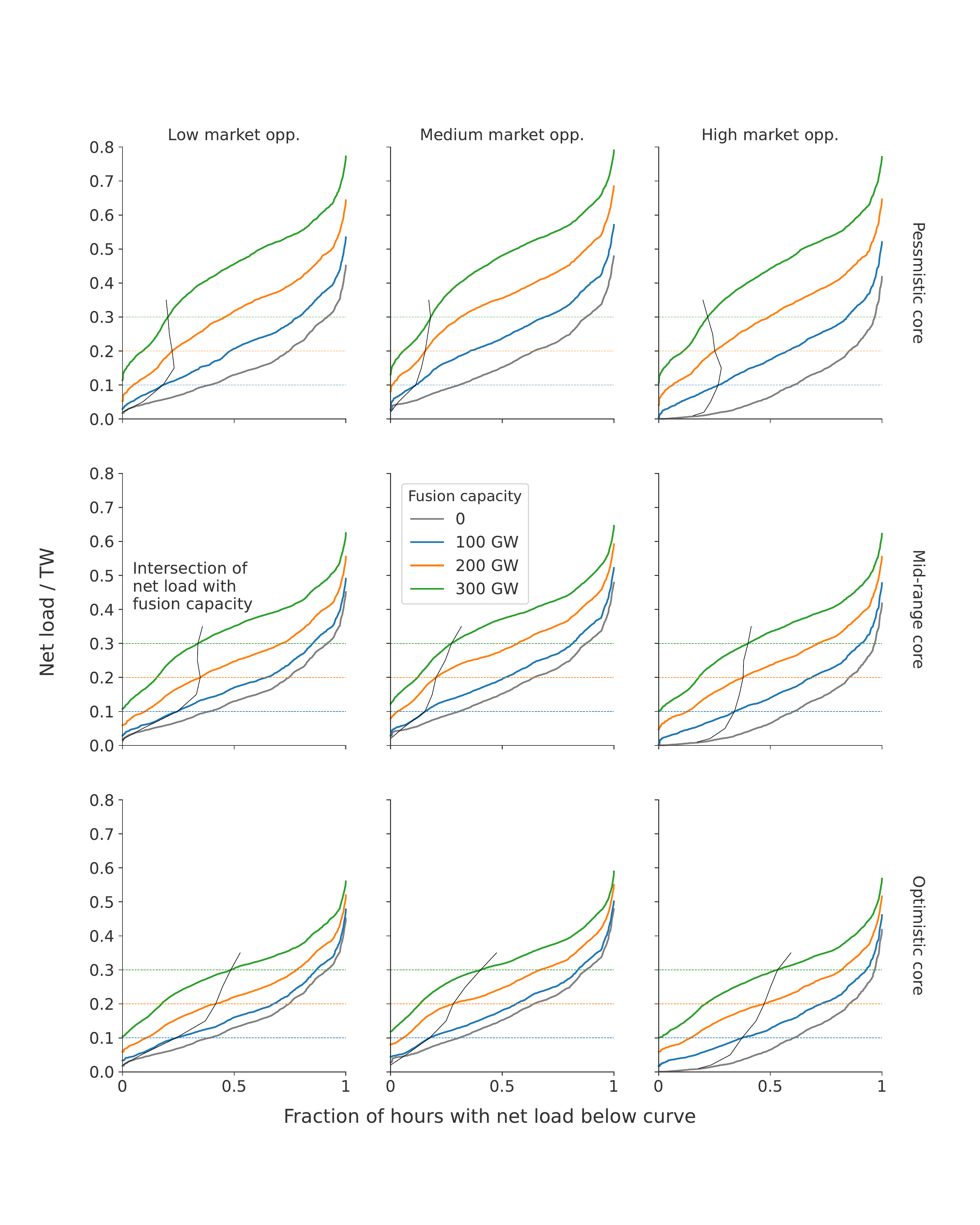}
\caption{Net load duration curves in scenarios with fusion, for each of the three reference reactors in the three reference scenarios, at fusion capacities of \SIrange{0}{300}{\giga\watt}.
These curves show the sorted hourly load which must be supplied by thermal generators, i.e. the load after subtracting power contributed by variable renewables, batteries, and that which was time-shifted.
Since fusion has a relatively low variable cost, when the net load is above the fusion capacity, it is likely that all the fusion reactors are turned on.
The black lines chart the intersections between the fusion capacity and the net load curves.
}\label{fig:netloaddurationwithfusion}
\end{figure}
Figure~\ref{fig:netloaddurationwithfusion} shows ``net load'' duration curves for each of the three main scenarios, for the three reference reactor designs at a range of fusion capacities.
The net load is that which must be supplied by thermal generators in the system: fusion, fission, NG-CCS, or ZCF plants.
These charts give a sense of the fraction of the year during which the net load is high enough that (all) the fusion reactors will be turned on.

\begin{figure}%
\centering
\includegraphics[width=0.95\textwidth]{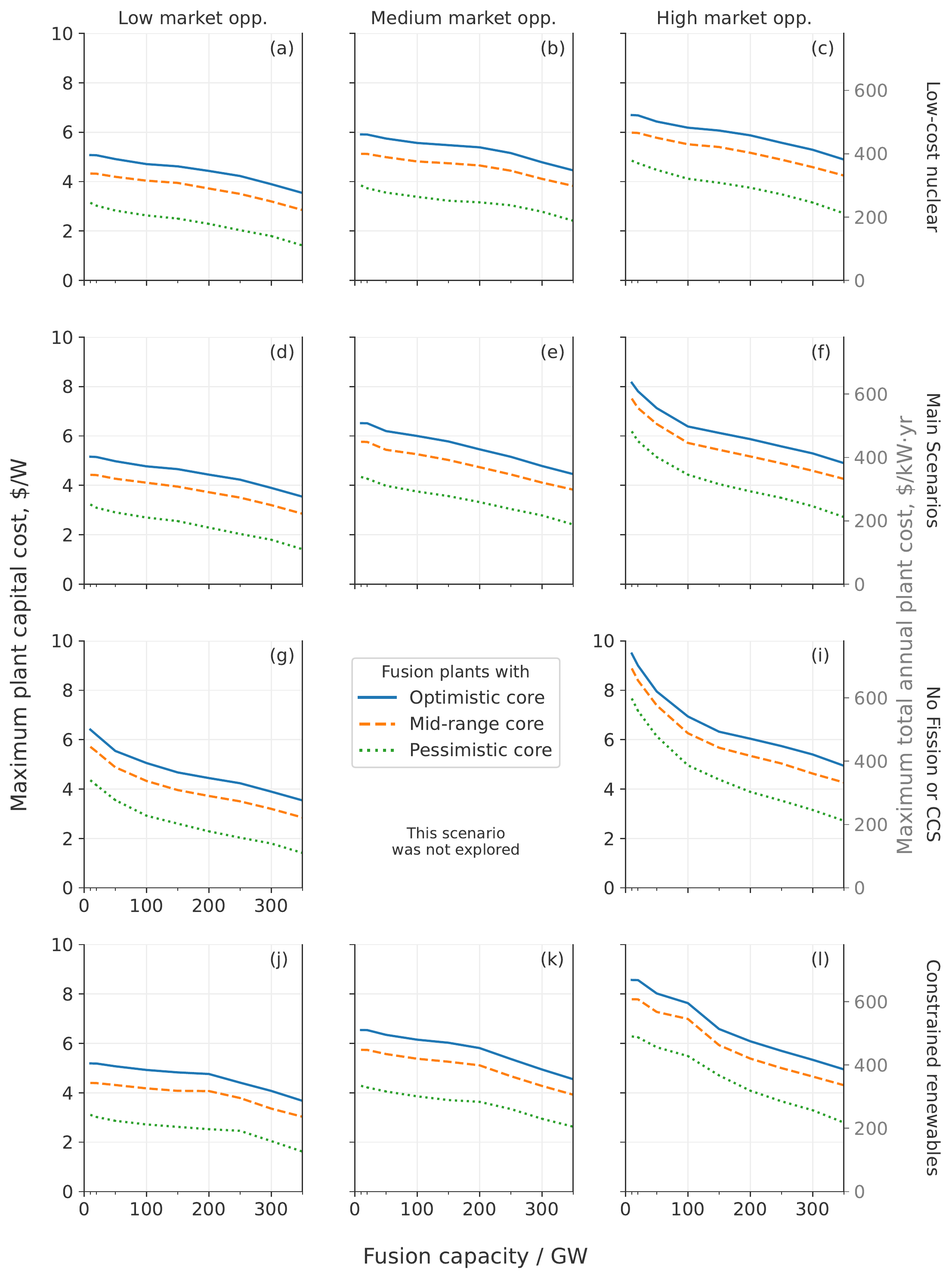}
\caption{Maximum plant capital costs for the three main plant designs
in three scenarios with lower-cost nuclear fission, (a), (b), (c),
in the three main market opportunity scenarios, (d), (e), (f); 
scenarios without new fission or NG-CCS, (g) and (i);
and the three constrained-renewables scenarios, (j), (k), (l).
}\label{fig:plant10scenarios}
\end{figure}
Figure~\ref{fig:plant10scenarios} shows the maximum plant capital costs and maximum total annual plant costs for plants of the three main designs in ten of the scenarios. Not all types of cases were run for all scenarios.
In the low-cost nuclear scenarios, more new fission plants are constructed and the value of fusion at low capacity penetrations is depressed.
The scenarios without fission or NG-CCS, the value of fusion at low capacity penetrations is increased, since fusion and ZCF-burning plants are now the only firm generators. However, the value of even the optimistic plant does not exceed \SI{10}[\$]{\per\watt}.
In the scenarios with constrained renewables, the value of fusion is somewhat increased relative to the base scenarios as fusion is built instead of fission and NG-CCS plants.

\begin{figure}
\centering
\includegraphics[width=0.9\textwidth]{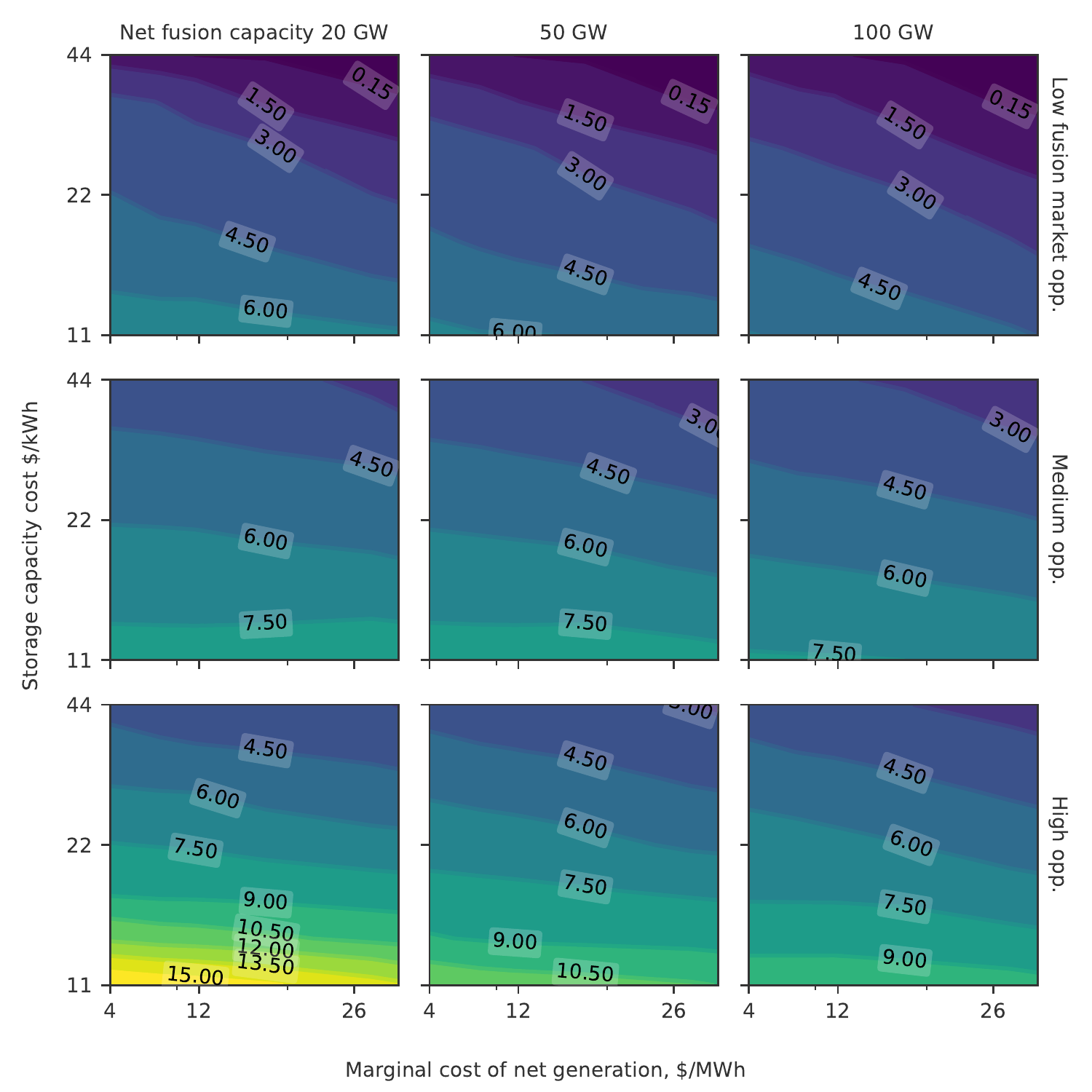}
\caption{Optimal storage capacities, measured in hours of the core's maximum instantaneous thermal output, generally range from 2 to 12 hours, and depend strongly on the storage capacity cost and the market opportunity scenario. These data are for plants with a mid-range core, with the $\corevom$ cost altered to produce different marginal costs of generation.
The data presented here correspond to those in Fig.~6.
}\label{fig:optimalduration}
\end{figure}

\begin{figure}%
\centering
\includegraphics[width=0.95\textwidth]{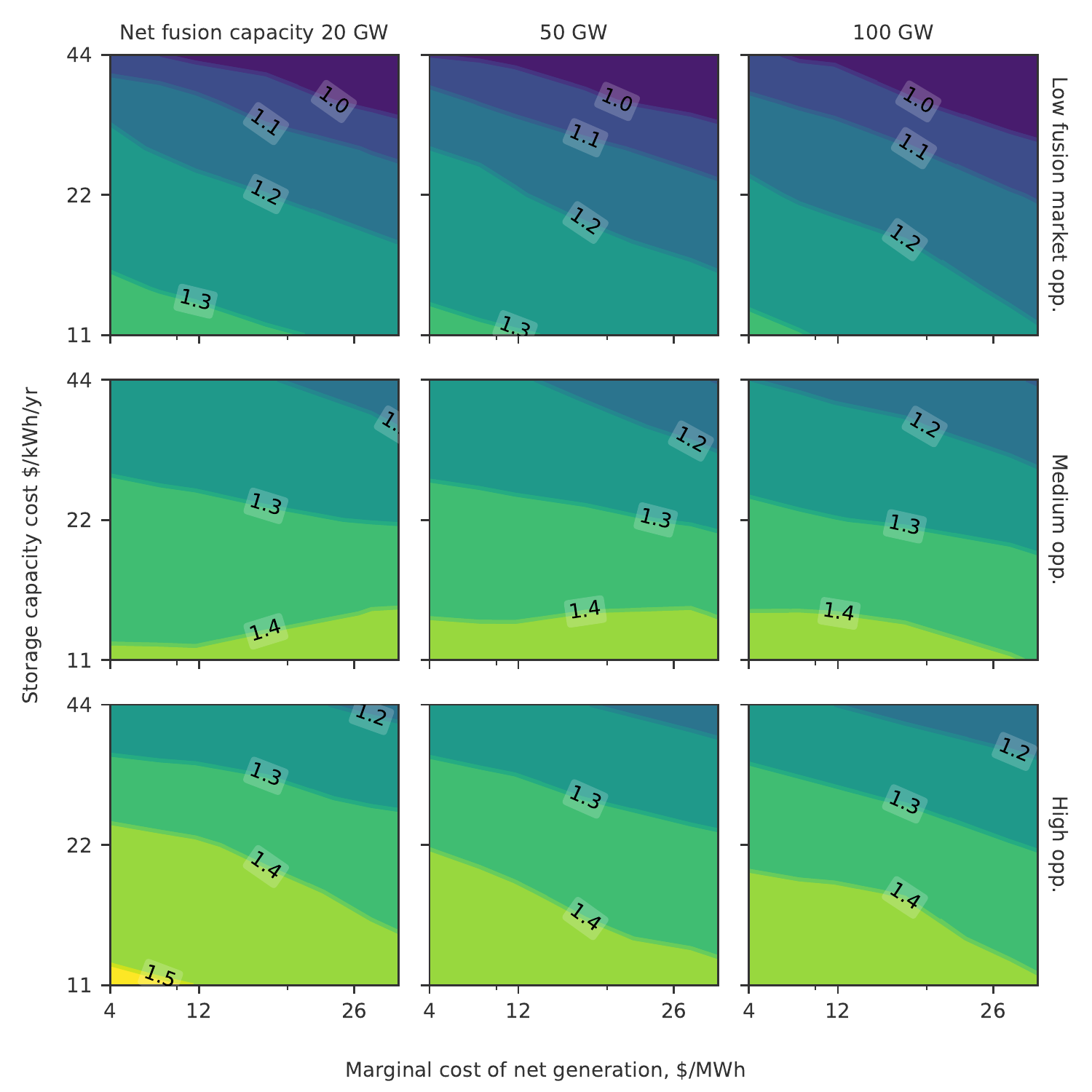}
\caption{Ratios of power conversion system (PCS) capacity for plants with thermal storage systems to the capacity of the PCS which would be needed without a thermal storage system, for mid-range plants. The data presented here correspond to those in Fig.~6.}\label{fig:midrangeGenRatios}
\end{figure}

\begin{figure}%
\centering
\includegraphics[width=0.95\textwidth]{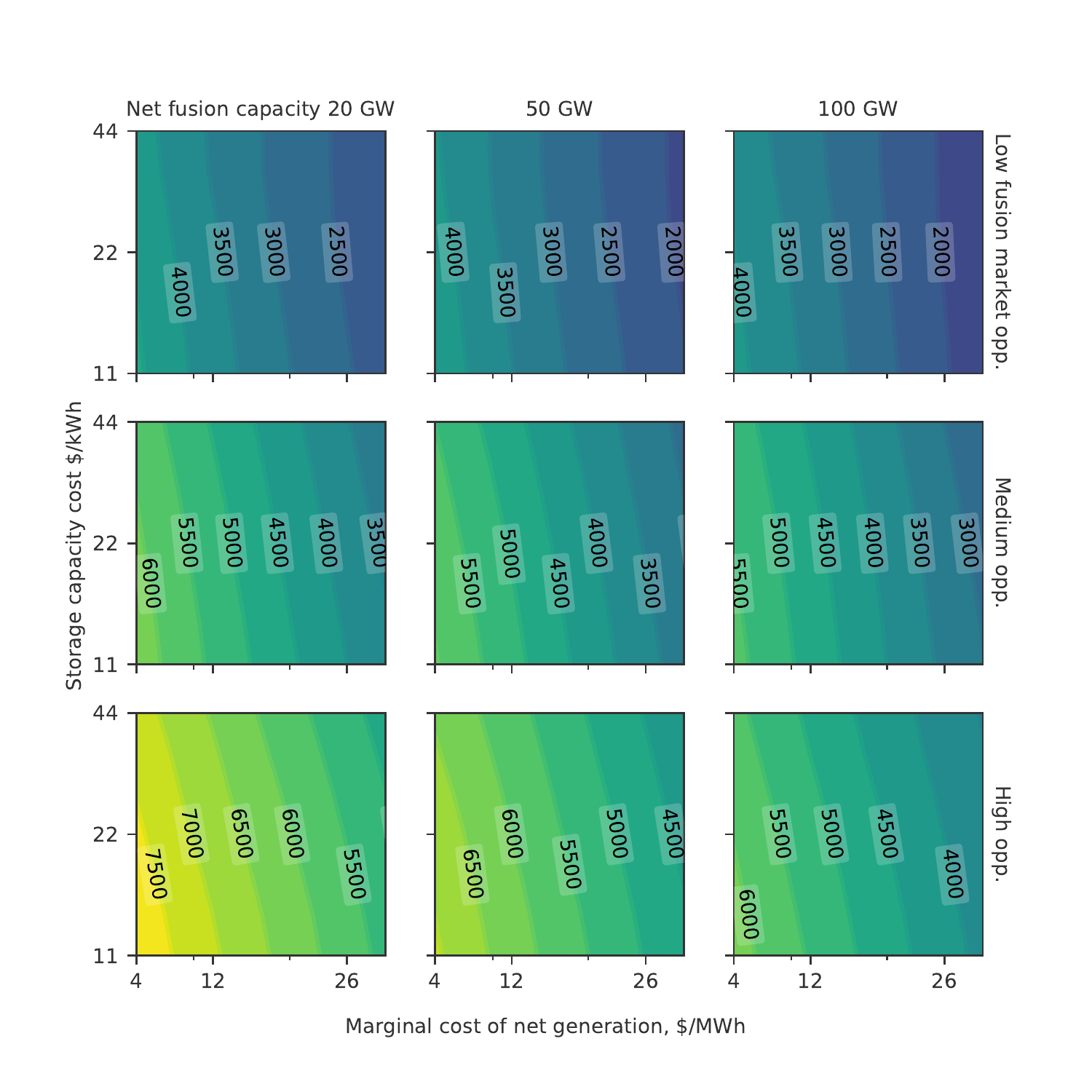}
\caption{Value of the fusion plant core, in \$/kW, when a thermal storage system is permitted.
The data presented here correspond to those in Fig.~6.
}\label{fig:storageCoreValue}
\end{figure}

\begin{figure}%
\centering
\includegraphics[width=0.95\textwidth]{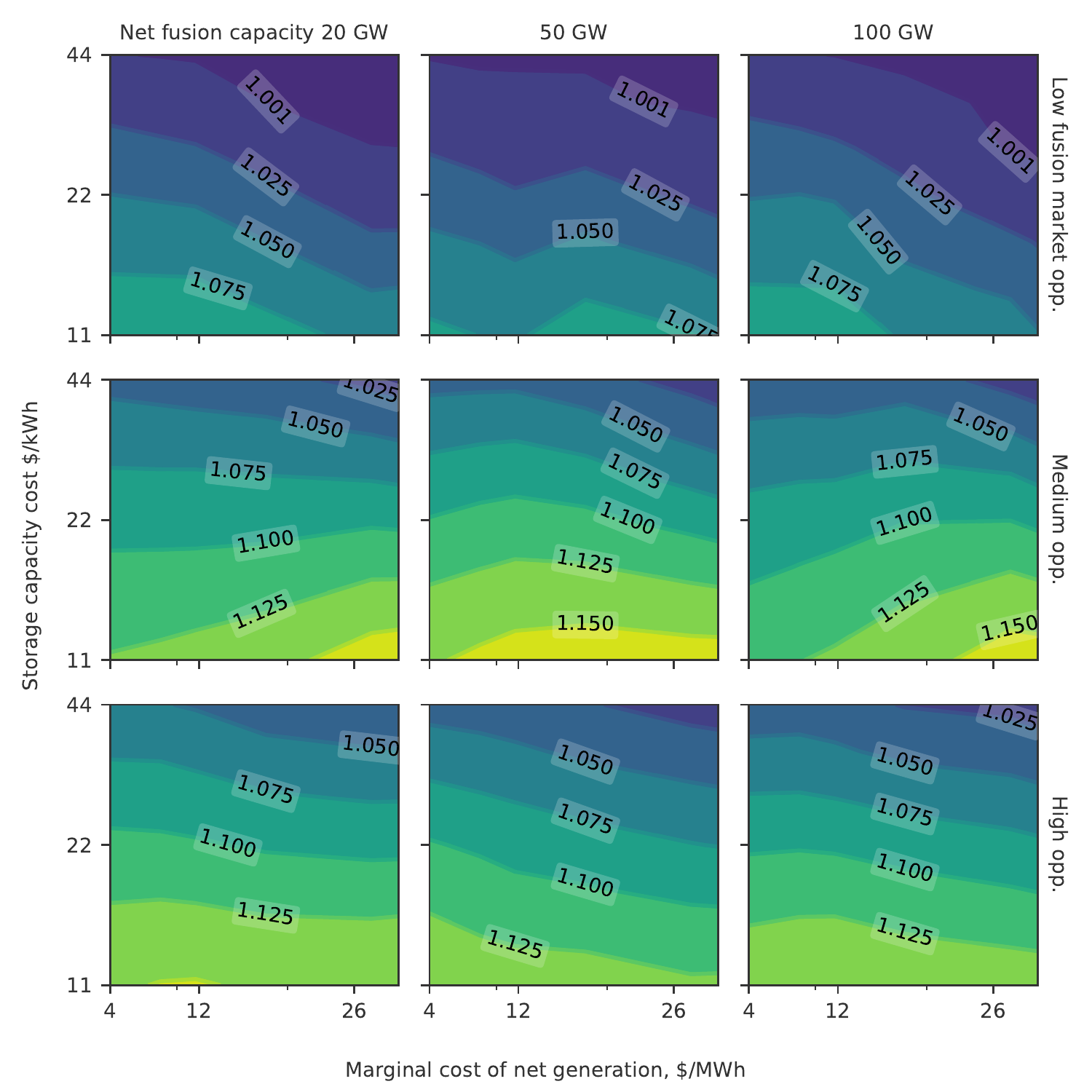}
\caption{Value of the fusion plant core when a thermal storage system (TSS) is permitted, relative to the value when a TSS is not permitted.
The data presented here correspond to those in Fig.~6.
}\label{fig:storageFractionalValueEnhancement}
\end{figure}

Figures~\ref{fig:optimalduration}--\ref{fig:storageFractionalValueEnhancement} display data from the same set of simulations as Figure~6 of the main paper.
The first shows the average storage duration, where duration is measured in hours of the core's peak thermal output.
The second shows the ratio of the total generation capacity in the system to the generation capacity that would be required without storage.
The third shows the value (or cost thresholds) for the fusion plant core (see Sec.~10.3) for the precise definition of the core), and the fourth shows the ratio of this value to the value of the core without storage.

\begin{figure}
\centering
\includegraphics[width=0.98\textwidth]{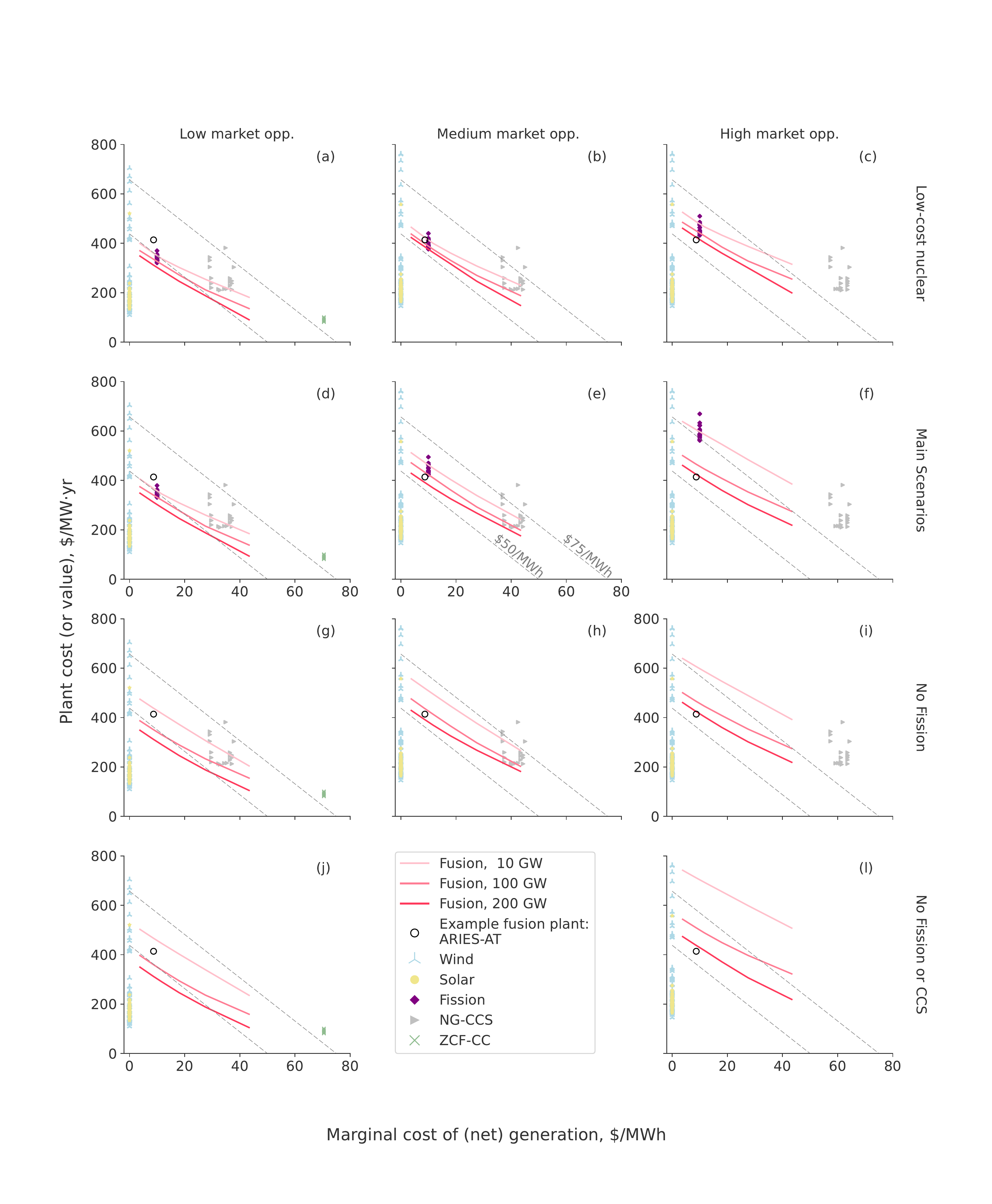}
\caption{The value of fusion depends on its marginal cost of net generation, and is anchored by the capacity cost of competitors with similar marginal generation costs. Here, capacity costs of wind and solar are normalized by their availability.
The fusion resources here are based on the mid-range reference plant, but with the core variable operations and maintenance (VOM) cost altered, to increase or decrease the marginal cost of net generation.
In (j) and (l), the scenarios without new fission or gas with CCS, fusion competes with a combination of variable renewables and storage---see Figs.\ \ref{fig:mix_highoppnonucorccs} and \ref{fig:displacementhighoppnonucorccs}.
The ZCF-CC and ZCF-CT generators are not shown (except for the former, and in the left column only) as they are off the right side of the plots; see Table~3 for their variable costs.
}\label{fig:morecompetition}
\end{figure}

Figure~\ref{fig:morecompetition} expands Figure~4 of the main paper to additional scenarios.
It illustrates how the value of fusion is set, in large part, by the costs of the other firm generators. 

\begin{figure}
\centering
\includegraphics[width=0.95\textwidth]{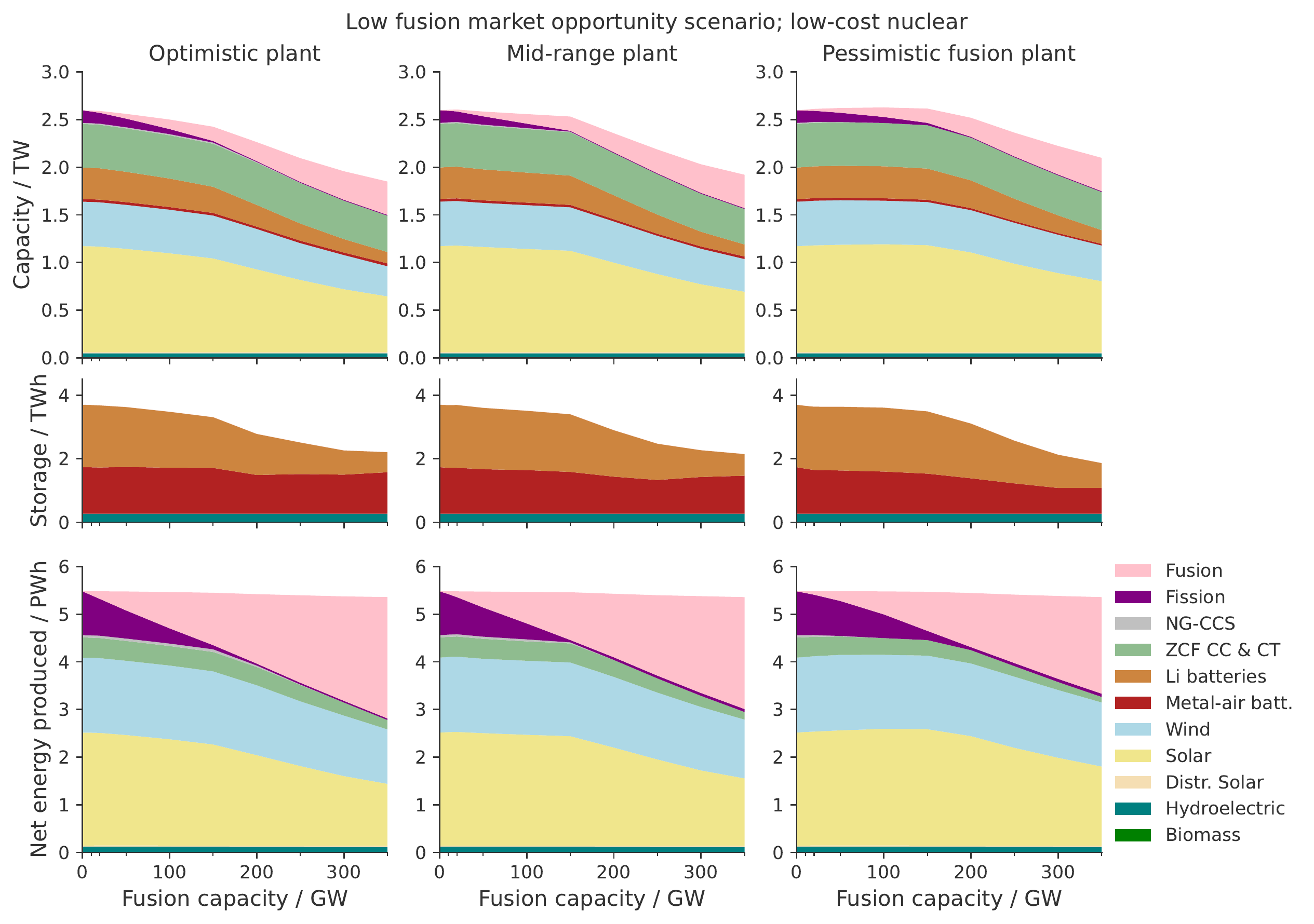}
\caption{Generation capacities, storage capacities, and energy production mixes in the Low fusion market opportunity scenario with low-cost fission, for the three reference plant designs without thermal storage.
}\label{fig:mix_LowOppNoNuclear}
\end{figure}

Figures~\ref{fig:mix_LowOppNoNuclear}--\ref{fig:mix_CRHighOpp} show the mixes of generation capacity, storage capacity, and net energy production of the resources in the system, for several scenarios, as a function of fusion's capacity.
These are provided to show the outcomes of the resource mix especially for the variant scenarios with low-cost fission, without fission or NG-CCS, and with constrained quantities of renewables.
In the cases with low-cost fission or constrained renewables, somewhat more fission is built than in the base cases. 
Again, fusion that is built is being built instead of new fission.
Metal-air storage is generally built only in the Low market opportunity scenario and its variants, where it is least expensive.
At increasing penetrations of fusion energy, it is built instead of lithium battery storage, while the amount of metal-air long-duration storage constructed remains relatively fixed.

Figures~\ref{fig:displacementlowopplownuclear}--\ref{fig:displacementcrhighopp} show the same data in a manner which highlights the quantities of each resources displaced by fusion.
Values shown are differences between quantities from the base case (that without fusion) and the case with the specified fusion capacity.
Positive values mean the fusion has displaced this resource; negative values mean that \textit{more} of the resource was built (or used) than in the case without fusion.
This is seen to be the case for solar in Fig.~\ref{fig:displacementmedopp}, for example.
Solid lines are for cases without storage.
Dashed lines, shown in a limited set of scenarios, are for cases with the option to build mid-priced storage.
Figures~\ref{fig:displacementlowopp}--\ref{fig:displacementhighopp} show the impact of the option to build storage on the value of fusion for the main scenarios.
Fusion with storage displaces more wind, NG-CCS, ZCF-burning plants, and lithium batteries than fusion without storage.
However, it often results in slightly \textit{more} solar being built than would be built at the same fusion capacity penetration without storage.

Figures~\ref{fig:maplowopp}--\ref{fig:maphighopp} show where fusion and other resources are built in the three main scenarios, in cases with \SI{100}{\giga\watt} of fusion with storage, \SI{100}{\giga\watt} of fusion without storage, and the base case without fusion.
While we have not studied this in detail, in all three scenarios, allowing thermal storage results in fusion plants being sited slightly further west.
Note that parts (c)--(f) of Fig.~\ref{fig:maphighopp} are identical to Figure~10; these are provided again in this format for better comparison with the other market scenarios and the case with thermal storage.
Note that in cases with storage, the fusion capacity shown is the long-run net generation capacity of the plant, not the peak generation capacity.

Each plot also shows the existing transmission capacity between zones at the start of the 2036--2050 run as a narrow black line, the final capacity as the width of the red line, and the maximum possible capacity as the width of the gray line.

Figure~\ref{fig:maptransvar} shows where fusion and other resources are built in additional sensitivity cases with no new transmission and with unlimited new transmission. 
Forbidding new transmission does not seem to substantially alter where fusion is built, but allowing unlimited transmission results in fusion built only along the eastern seaboard.

Figures~\ref{fig:hourly_nostor} and~\ref{fig:hourly_withstor} show the same type of data as Figure~8 of the main paper, but compares behavior between the three reference reactors.
It can be seen that the optimistic reactors operate slightly more often than the pessimistic reactors, and their net power output (when storage is used) is more steady.

\clearpage
\begin{figure}
\centering
\includegraphics[width=0.95\textwidth]{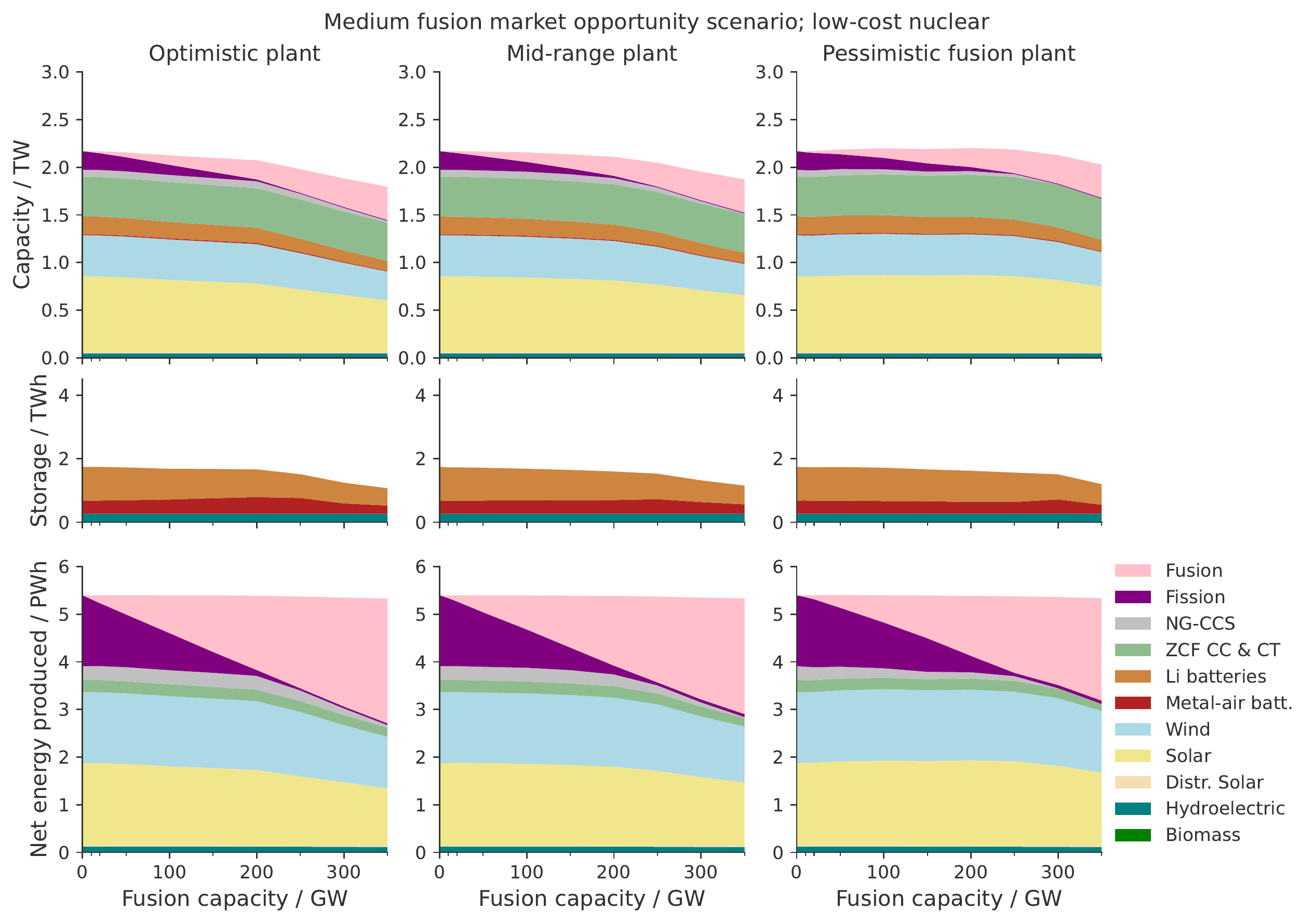}
\caption{Generation capacities, storage capacities, and energy production mixes in the Medium fusion market opportunity scenario with low-cost fission, for the three reference plant designs without thermal storage.
}
\end{figure}

\clearpage

\begin{figure}
\centering
\includegraphics[width=0.95\textwidth]{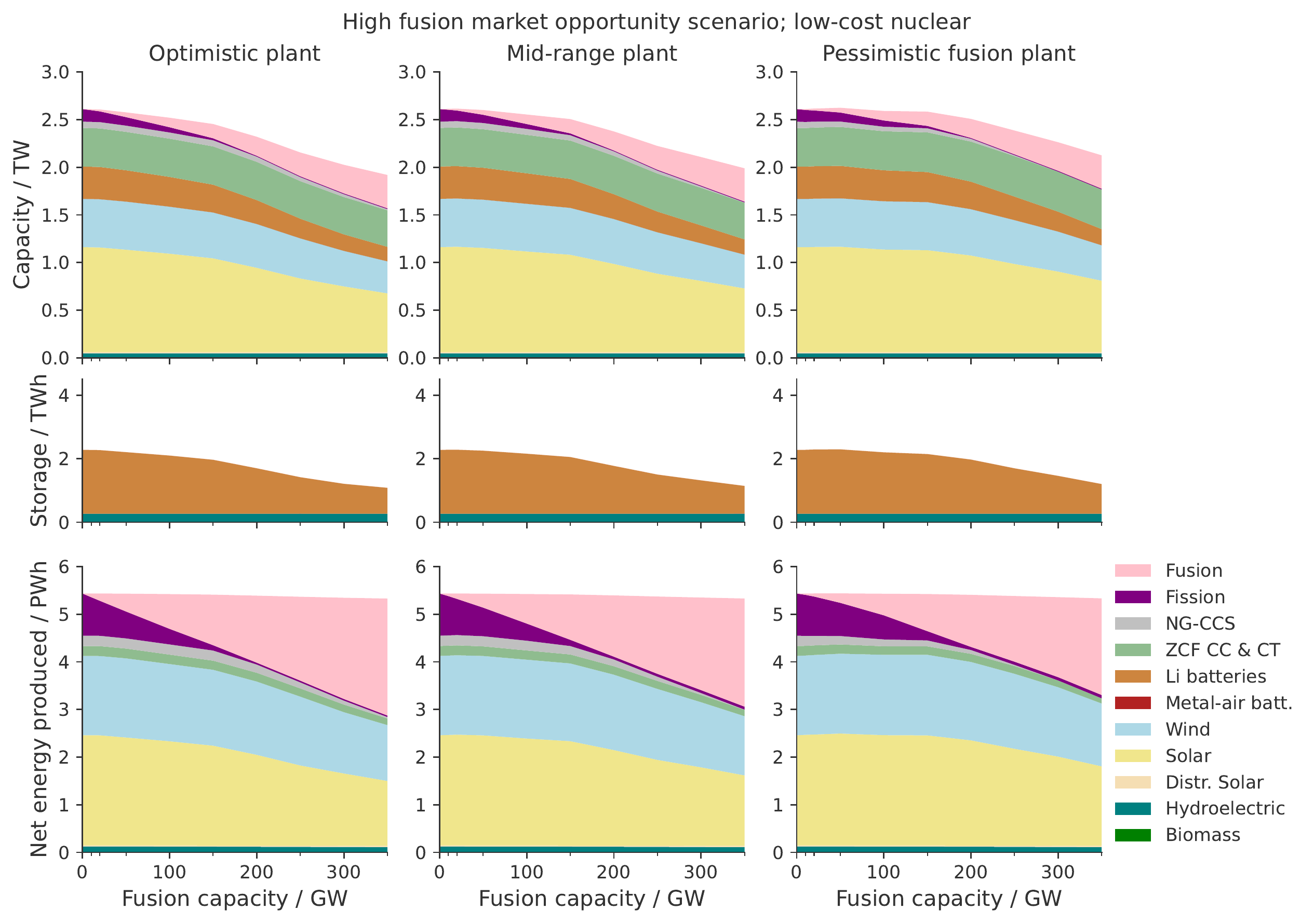}
\caption{Generation capacities, storage capacities, and energy production mixes in the High fusion market opportunity scenario with low-cost fission, for the three reference plant designs without thermal storage.
}
\end{figure}

\clearpage

\begin{figure}
\centering
\includegraphics[width=0.95\textwidth]{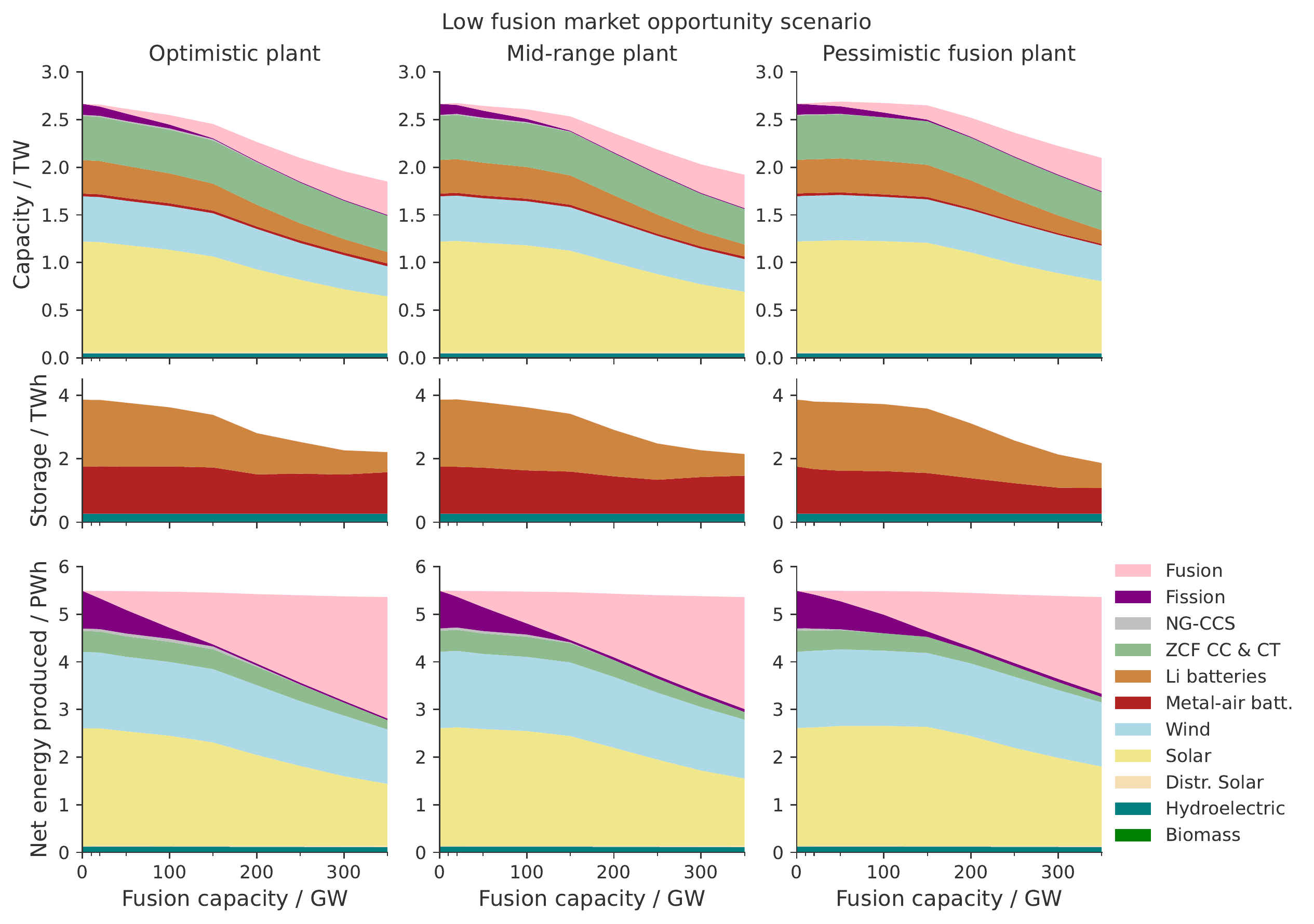}
\caption{Generation capacities, storage capacities, and energy production mixes in the Low fusion market opportunity scenario, for the three reference plant designs without thermal storage.
}
\end{figure}

\clearpage

\begin{figure}
\centering
\includegraphics[width=0.95\textwidth]{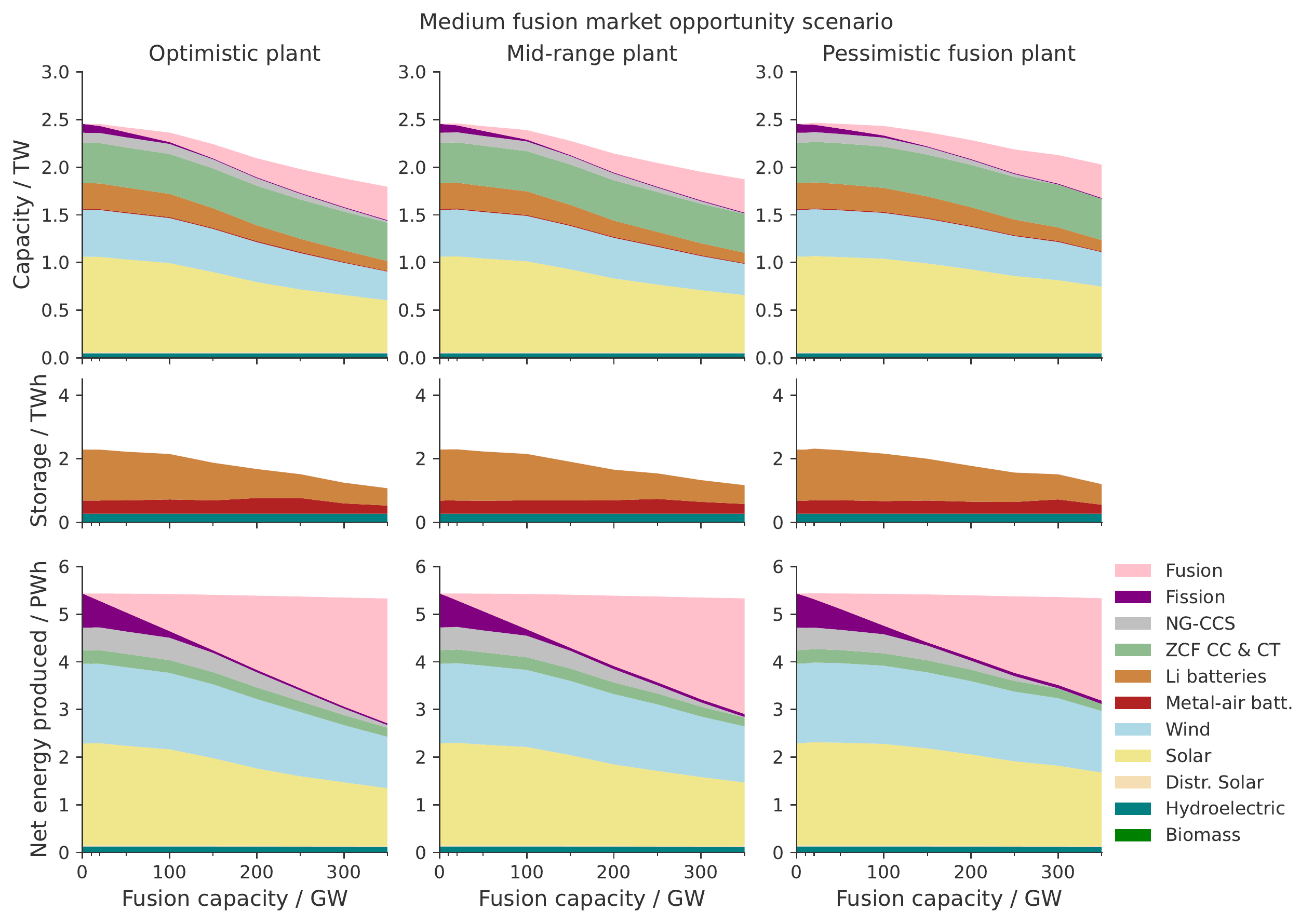}
\caption{Generation capacities, storage capacities, and energy production mixes in the Medium fusion market opportunity scenario, for the three reference plant designs without thermal storage. 
}
\end{figure}

\clearpage

\begin{figure}
\centering
\includegraphics[width=0.95\textwidth]{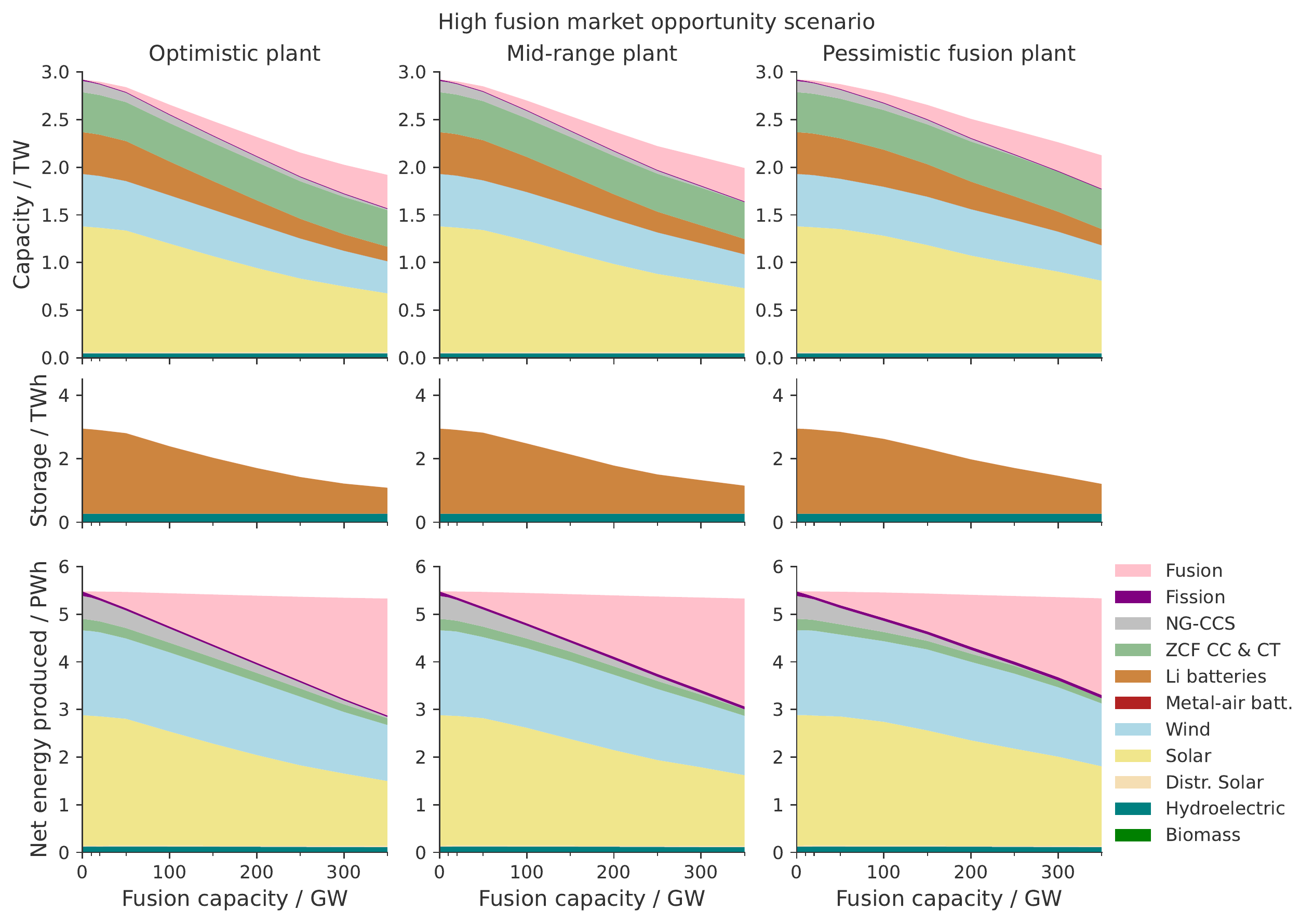}
\caption{Generation capacities, storage capacities, and energy production mixes in the High fusion market opportunity scenario, for the three reference plant designs without thermal storage.
}
\end{figure}

\clearpage

\begin{figure}
\centering
\includegraphics[width=0.95\textwidth]{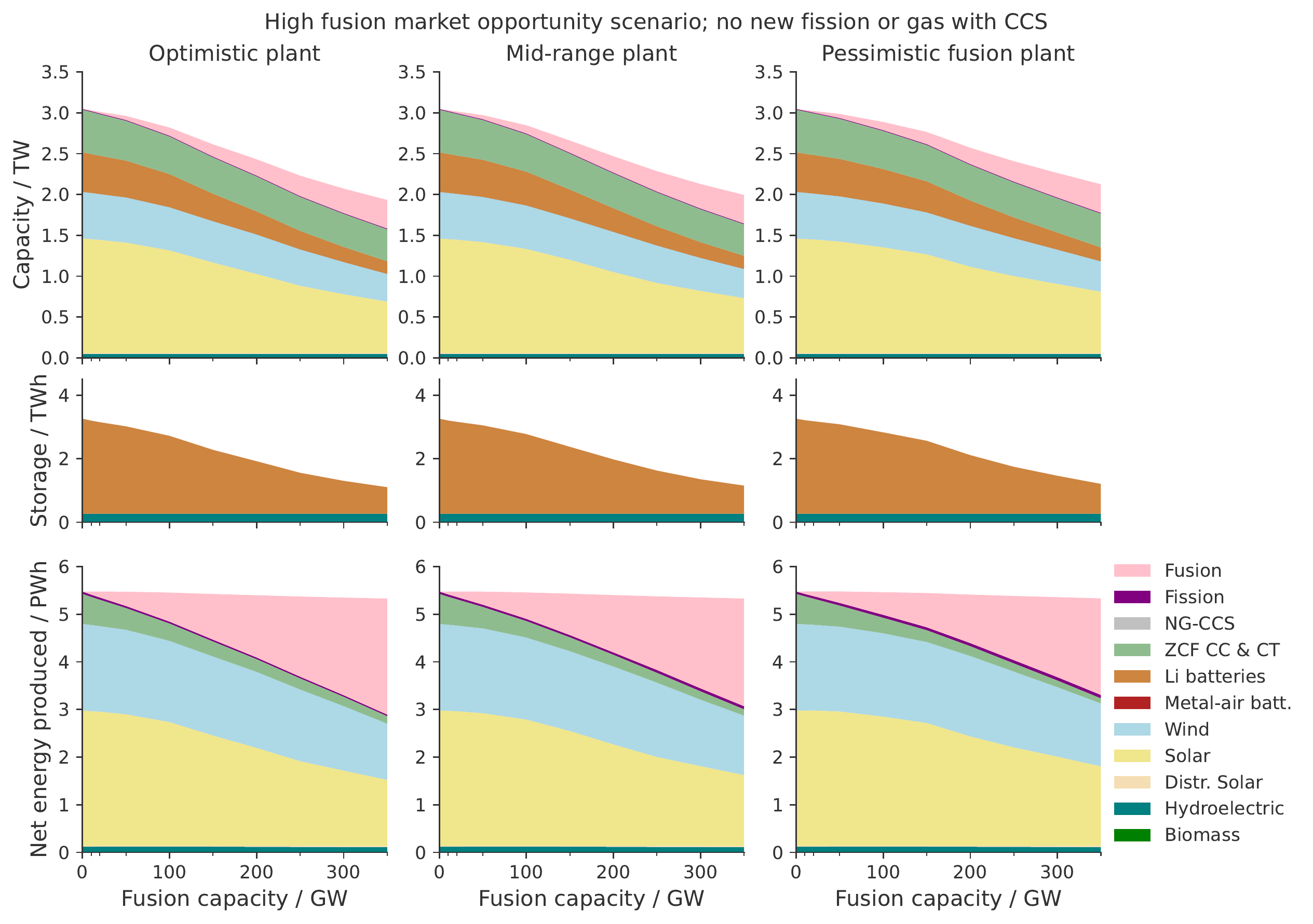}
\caption{Generation capacities, storage capacities, and energy production mixes in the High fusion market opportunity scenario, with no new fission or gas with CCS (NG-CCS),
for the three reference plant designs without thermal storage.
}\label{fig:mix_highoppnonucorccs}
\end{figure}
\clearpage
\begin{figure}
\centering
\includegraphics[width=0.95\textwidth]{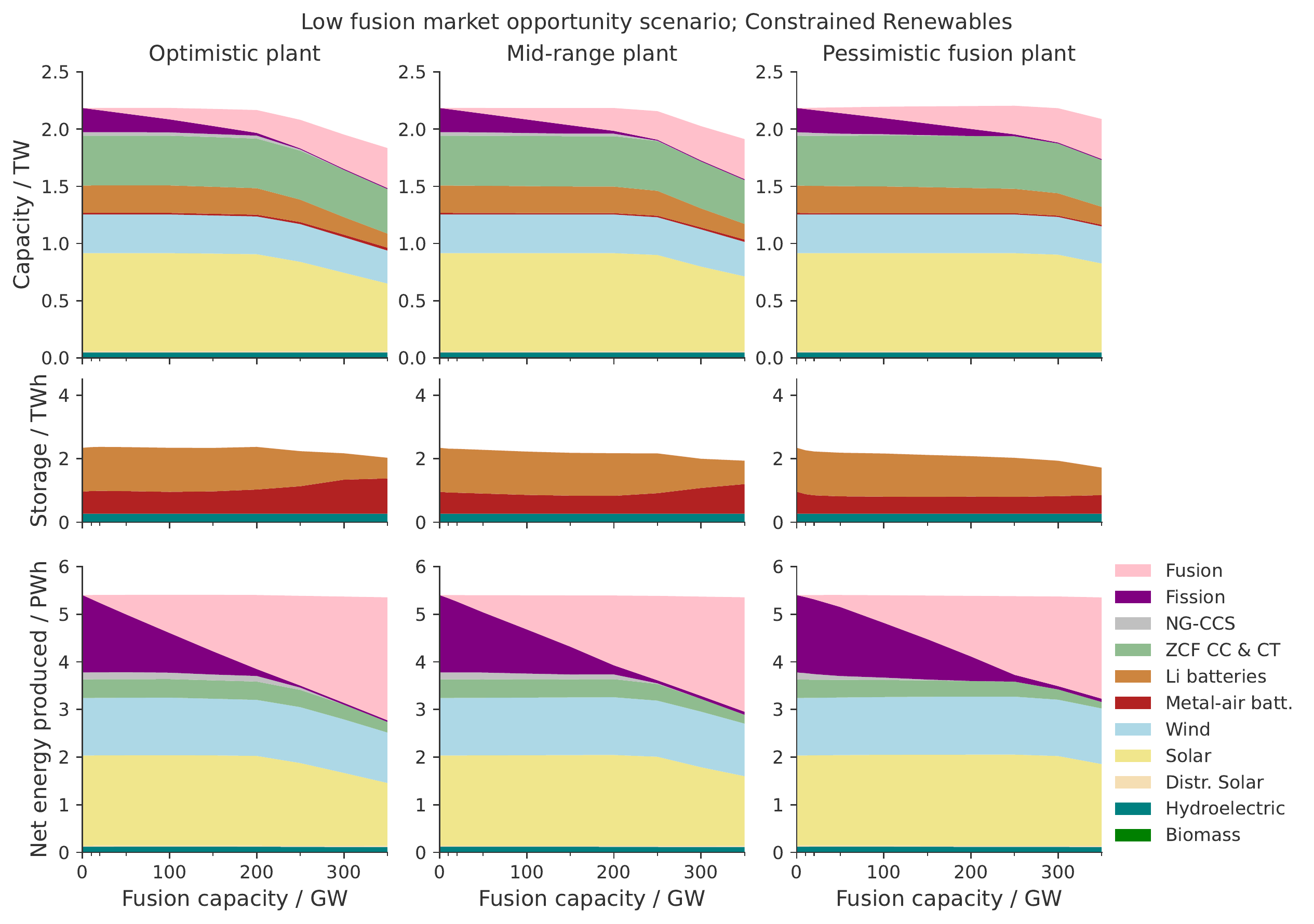}
\caption{Generation capacities, storage capacities, and energy production mixes in the Low fusion market opportunity scenario with constrained renewables, for the three reference plant designs without thermal storage.
}
\end{figure}

\clearpage

\begin{figure}
\centering
\includegraphics[width=0.95\textwidth]{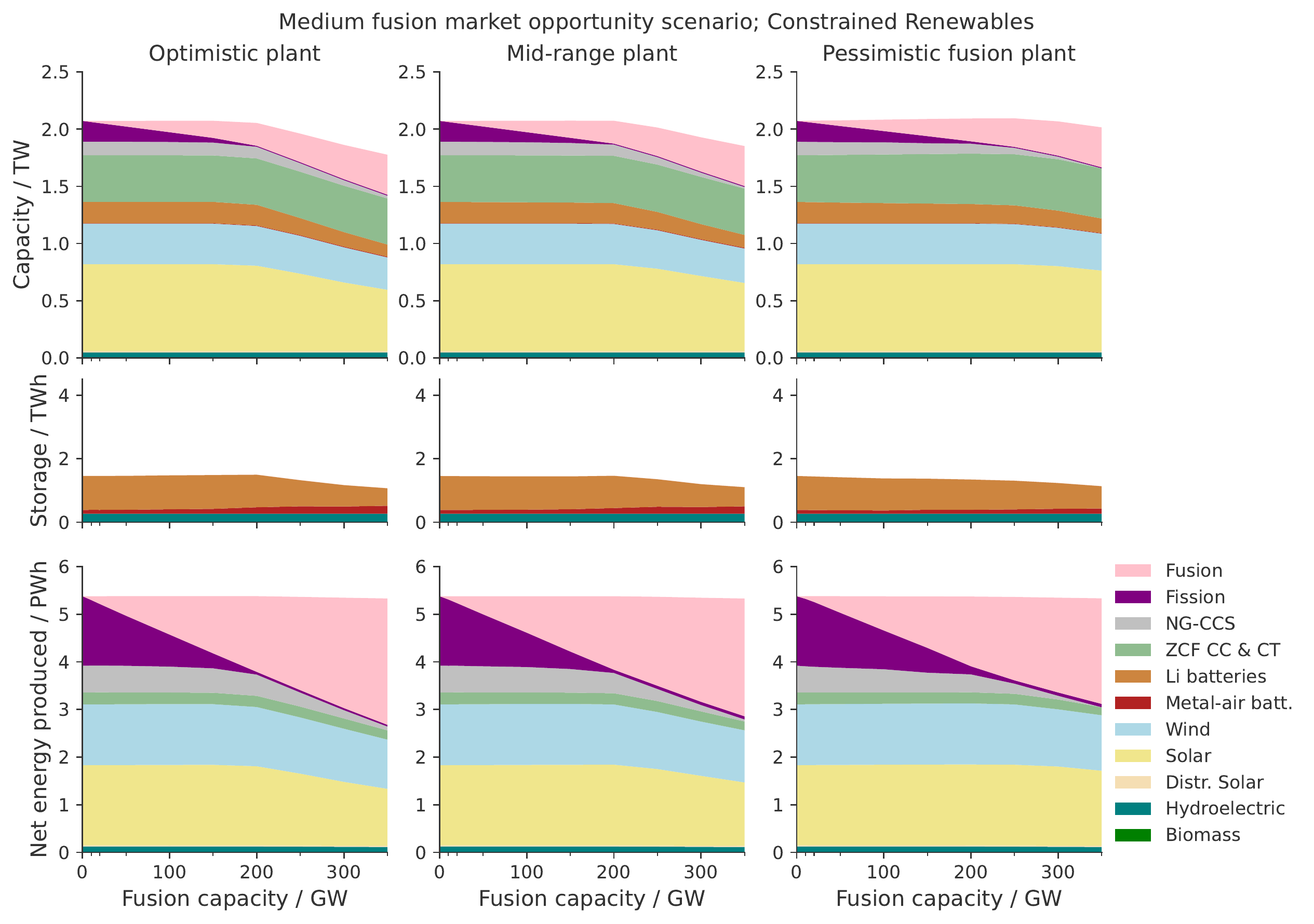}
\caption{Generation capacities, storage capacities, and energy production mixes in the Medium fusion market opportunity scenario with constrained renewables, for the three reference plant designs without thermal storage.
}
\end{figure}

\clearpage

\begin{figure}
\centering
\includegraphics[width=0.95\textwidth]{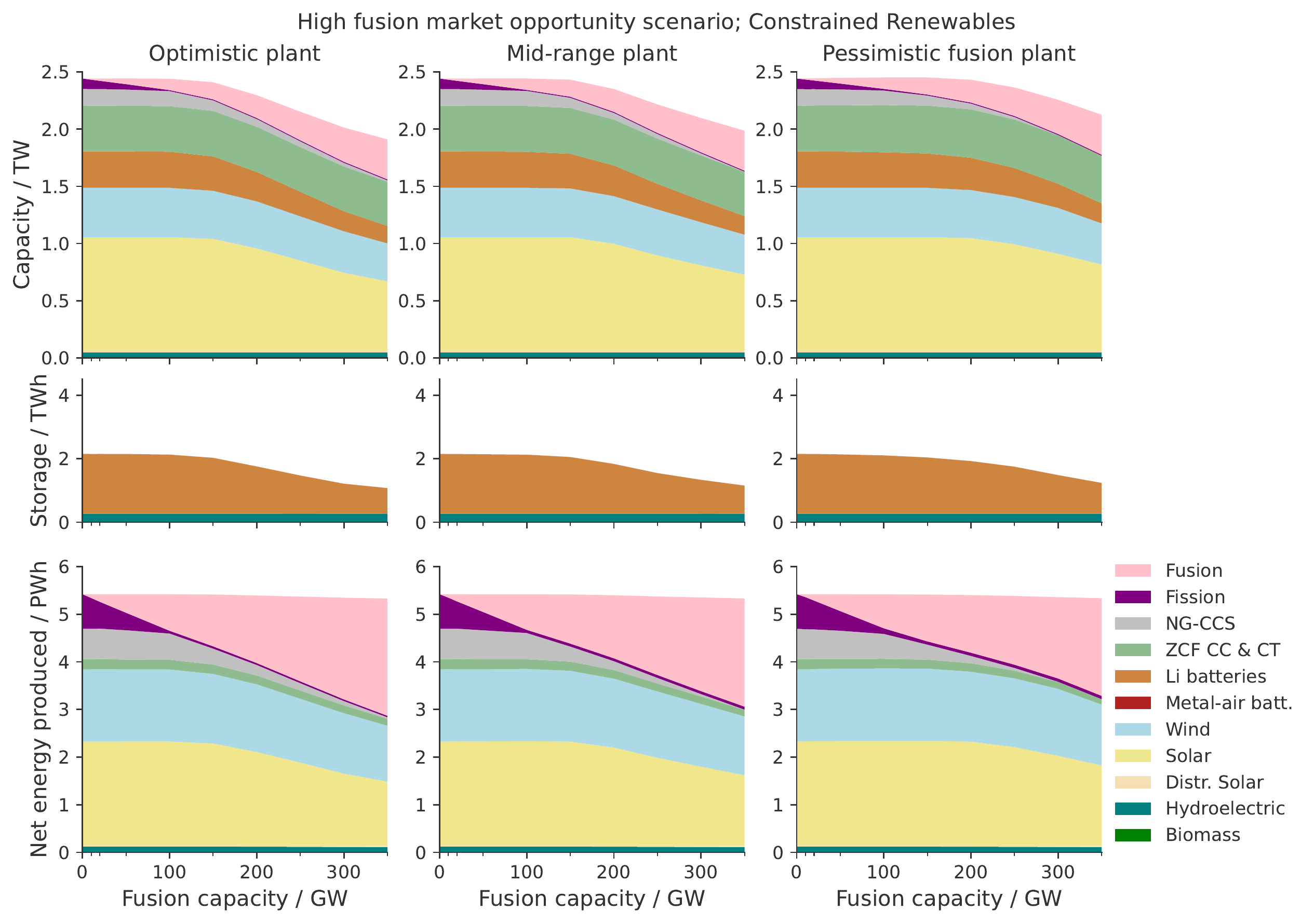}
\caption{Generation capacities, storage capacities, and energy production mixes in the High fusion market opportunity scenario with constrained renewables, for the three reference plant designs without thermal storage.
}\label{fig:mix_CRHighOpp}
\end{figure}

\clearpage

\begin{figure}
\centering
\includegraphics[width=0.95\textwidth]{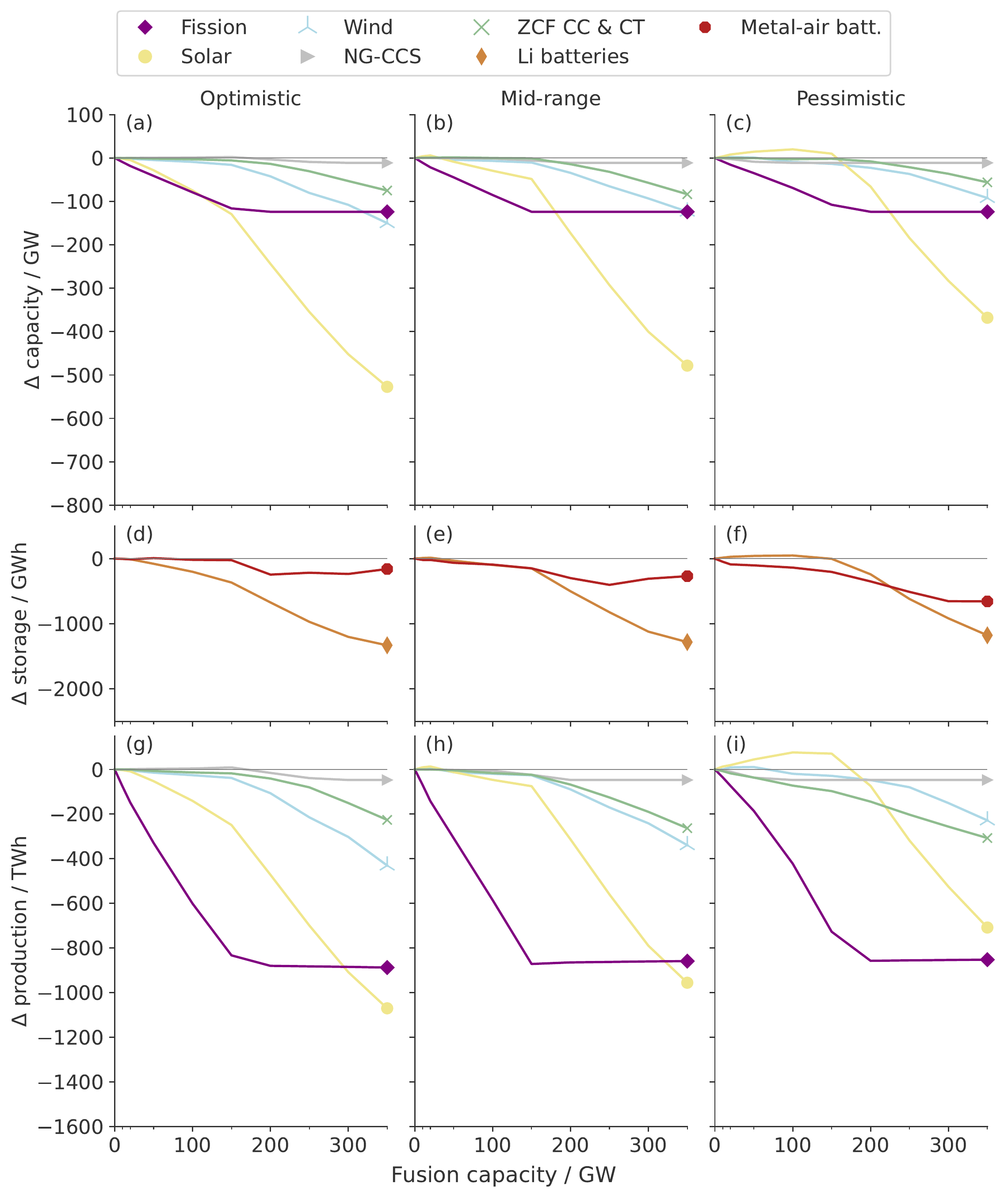}
\caption{Displaced generation capacity, storage, and energy production of other resources as fusion capacity increases in the Low market opportunity scenario with low-cost fission.
Dashed lines (if any) are for cases where fusion plants may include mid-priced storage.
}\label{fig:displacementlowopplownuclear}
\end{figure}

\clearpage

\begin{figure}
\centering
\includegraphics[width=0.95\textwidth]{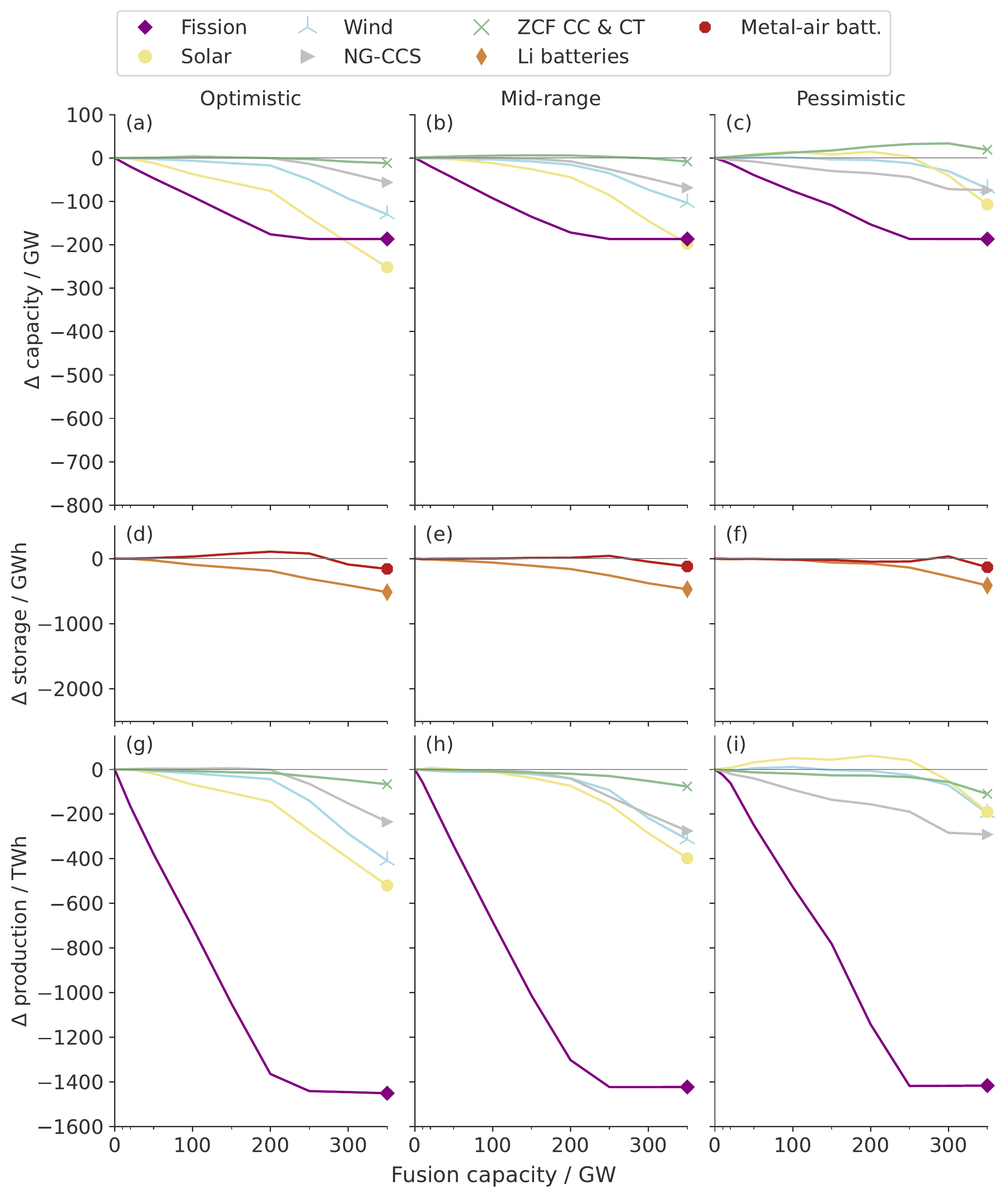}
\caption{Displaced generation capacity, storage, and energy production of other resources as fusion capacity increases in the Medium market opportunity scenario with low-cost fission.
Dashed lines (if any) are for cases where fusion plants may include mid-priced storage.
}\label{fig:displacementmedopplownuclear}
\end{figure}

\clearpage

\begin{figure}
\centering
\includegraphics[width=0.95\textwidth]{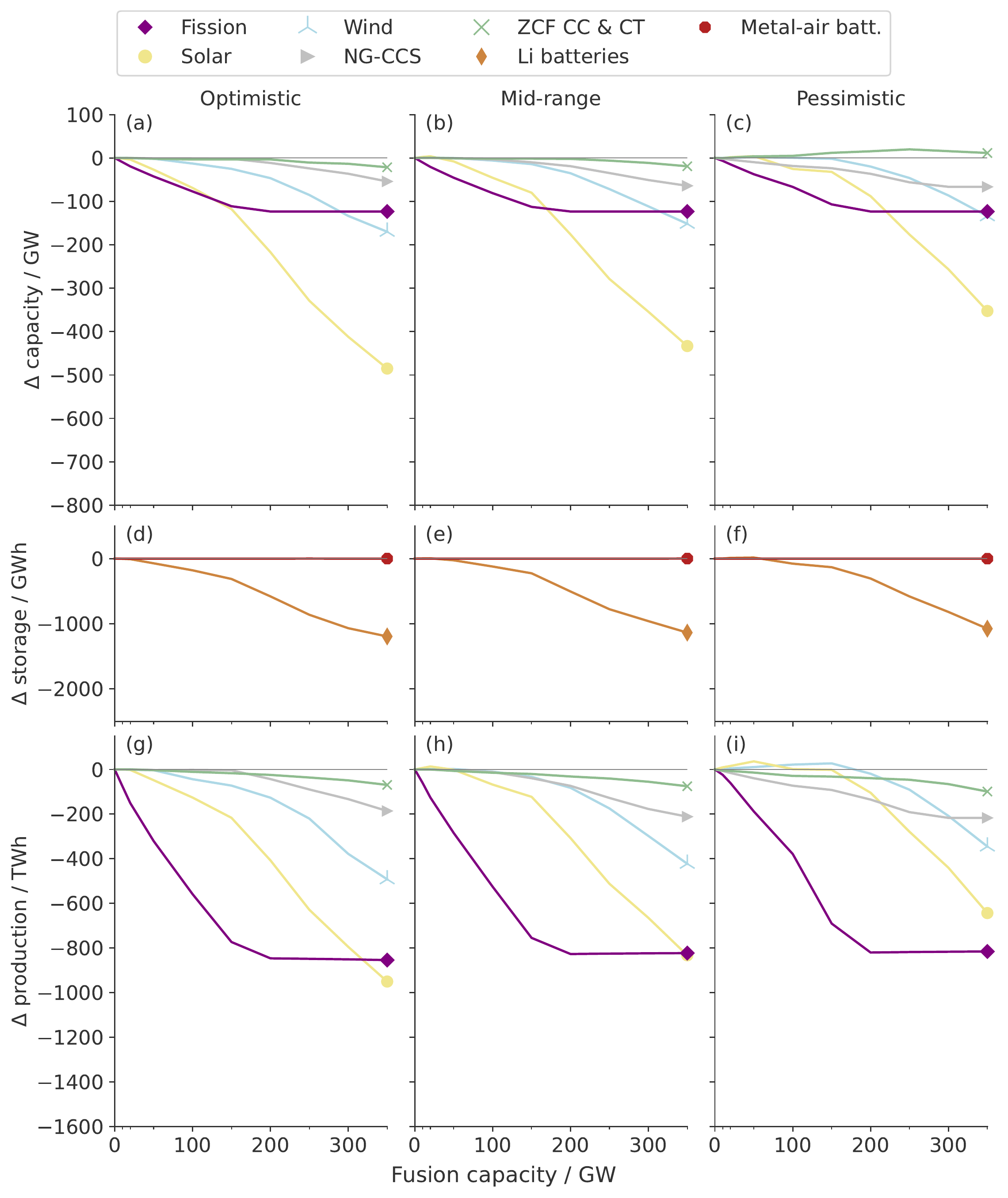}
\caption{Displaced generation capacity, storage, and energy production of other resources as fusion capacity increases in the High market opportunity scenario with low-cost fission.
Dashed lines (if any) are for cases where fusion plants may include mid-priced storage.
}\label{fig:displacementhighopplownuclear}
\end{figure}

\clearpage

\begin{figure}
\centering
\includegraphics[width=0.95\textwidth]{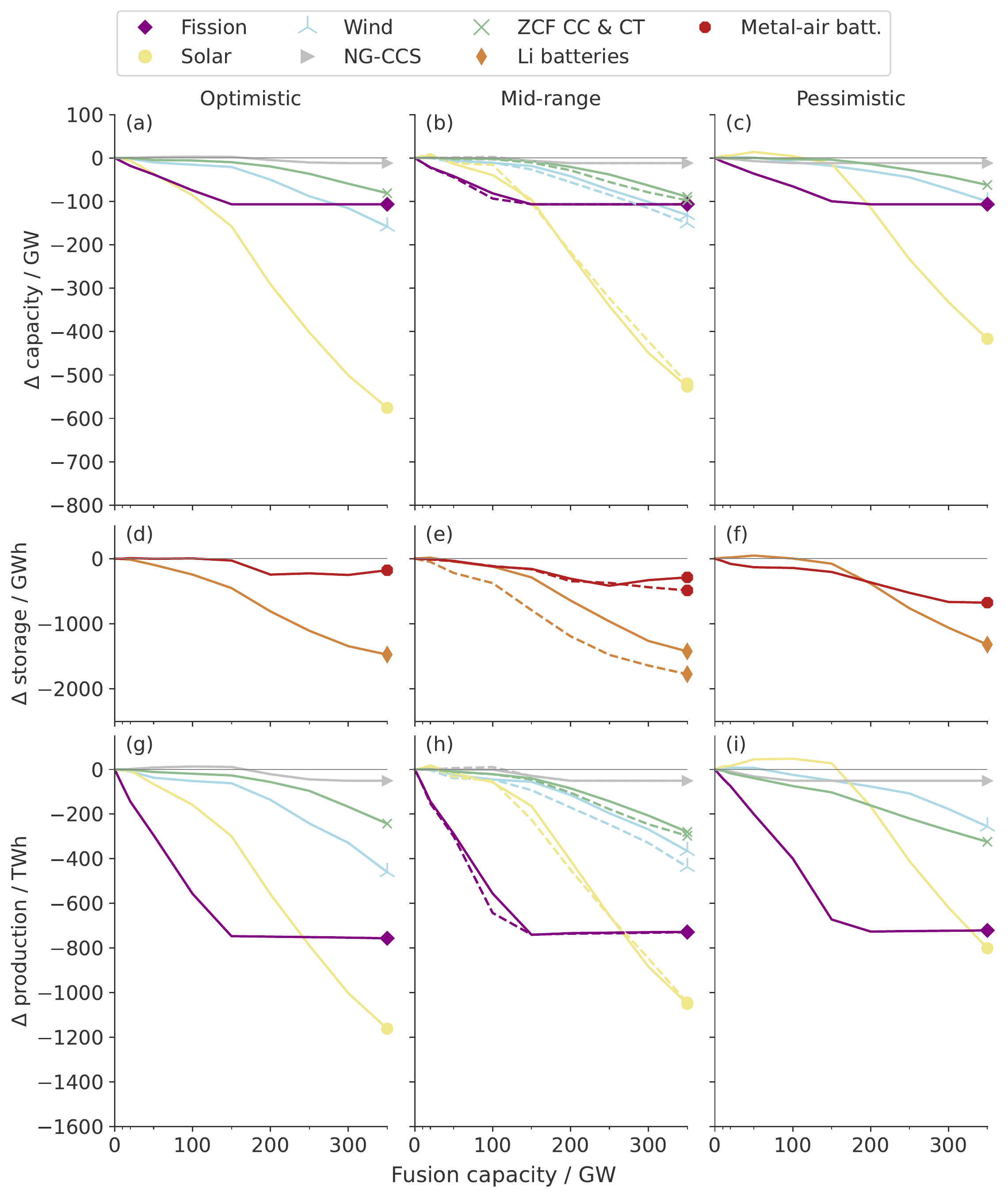}
\caption{Displaced generation capacity, storage, and energy production of other resources as fusion capacity increases in the main Low market opportunity scenario.
Dashed lines (if any) are for cases where fusion plants may include mid-priced storage.
}\label{fig:displacementlowopp}
\end{figure}

\clearpage

\begin{figure}
\centering
\includegraphics[width=0.95\textwidth]{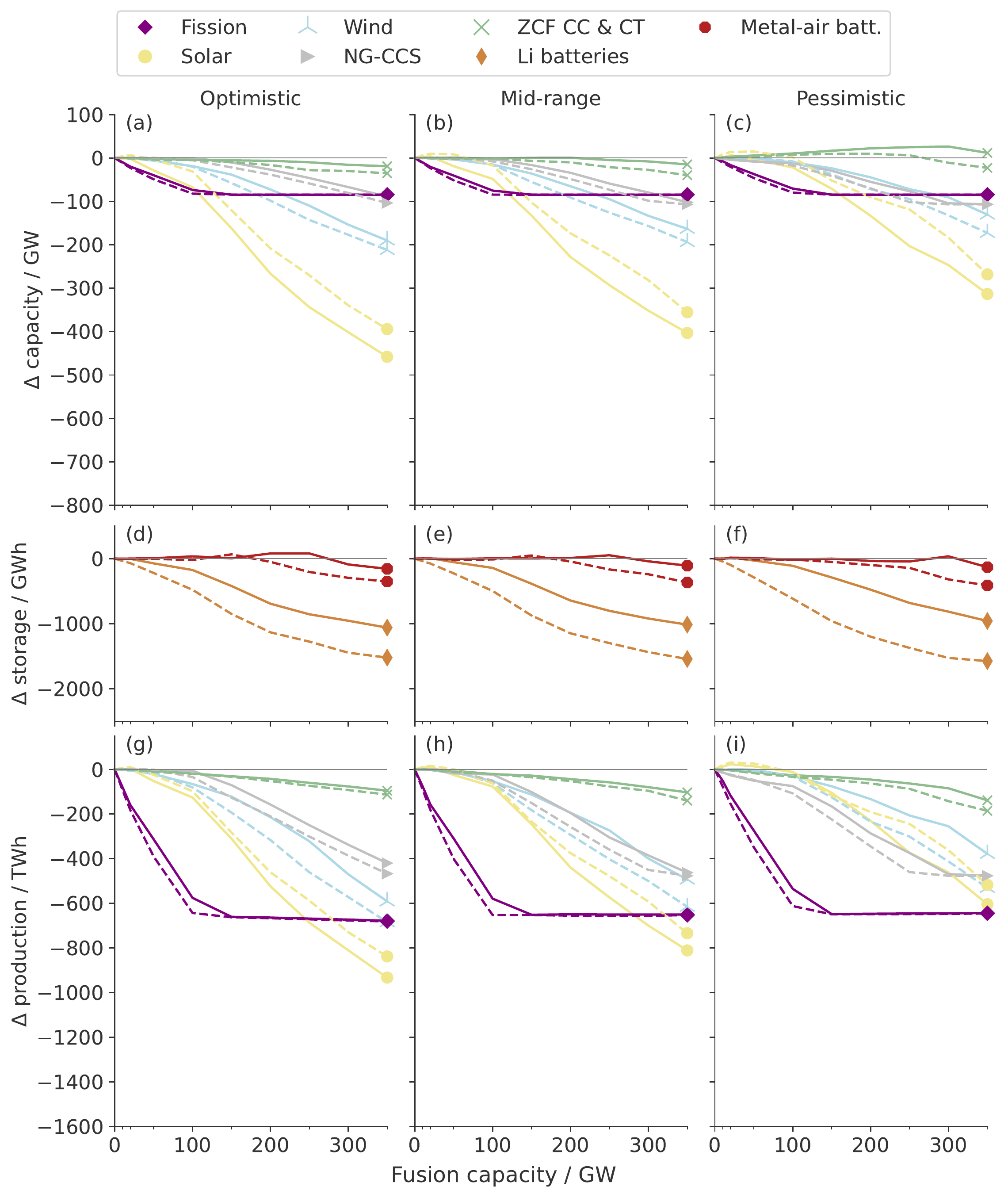}
\caption{Displaced generation capacity, storage, and energy production of other resources as fusion capacity increases in the main Medium market opportunity scenario.
Dashed lines (if any) are for cases where fusion plants may include mid-priced storage.
}\label{fig:displacementmedopp}
\end{figure}

\clearpage

\begin{figure}
\centering
\includegraphics[width=0.95\textwidth]{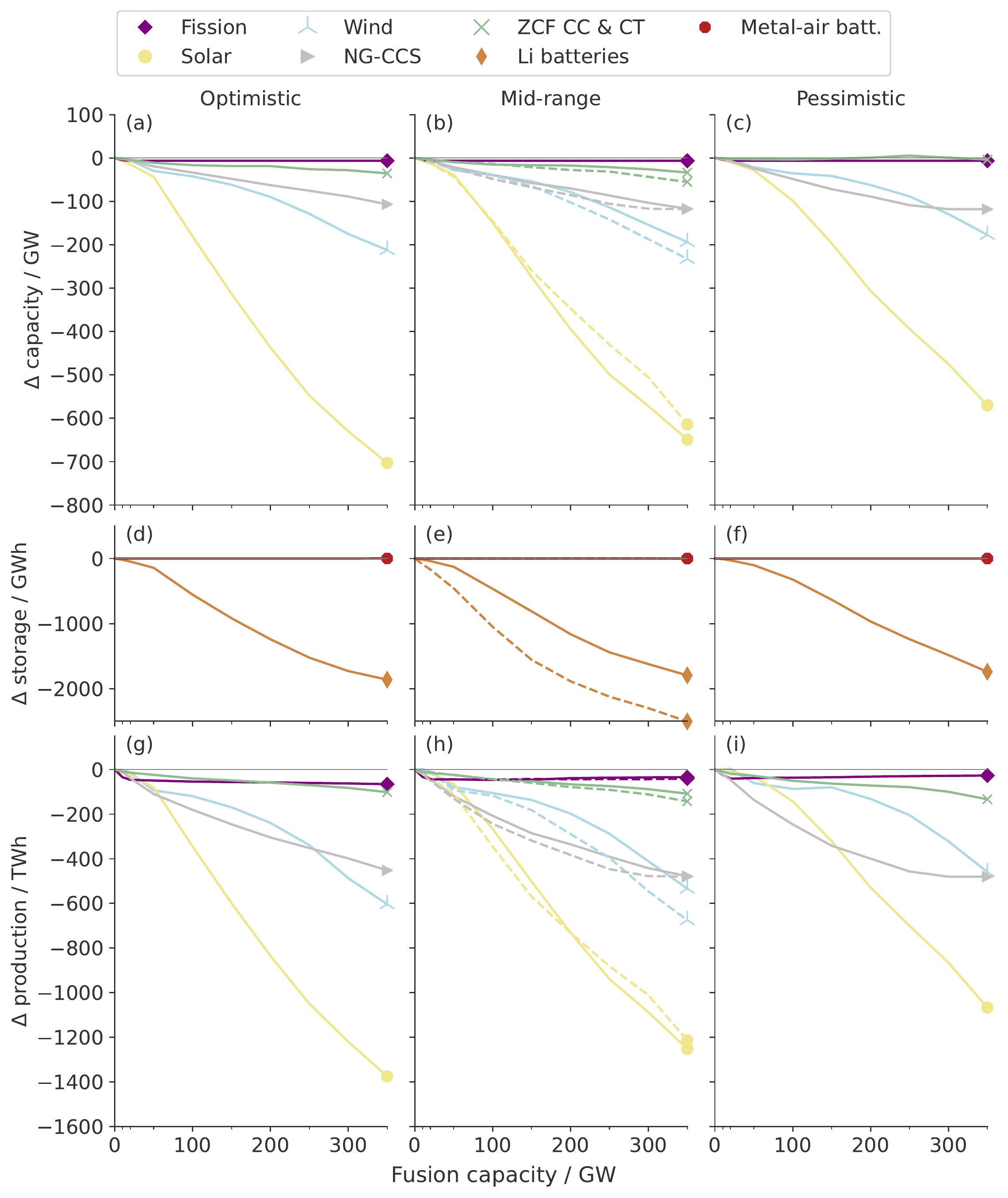}
\caption{Displaced generation capacity, storage, and energy production of other resources as fusion capacity increases in the main High market opportunity scenario.
Dashed lines (if any) are for cases where fusion plants may include mid-priced storage.
}\label{fig:displacementhighopp}
\end{figure}

\clearpage

\begin{figure}
\centering
\includegraphics[width=0.95\textwidth]{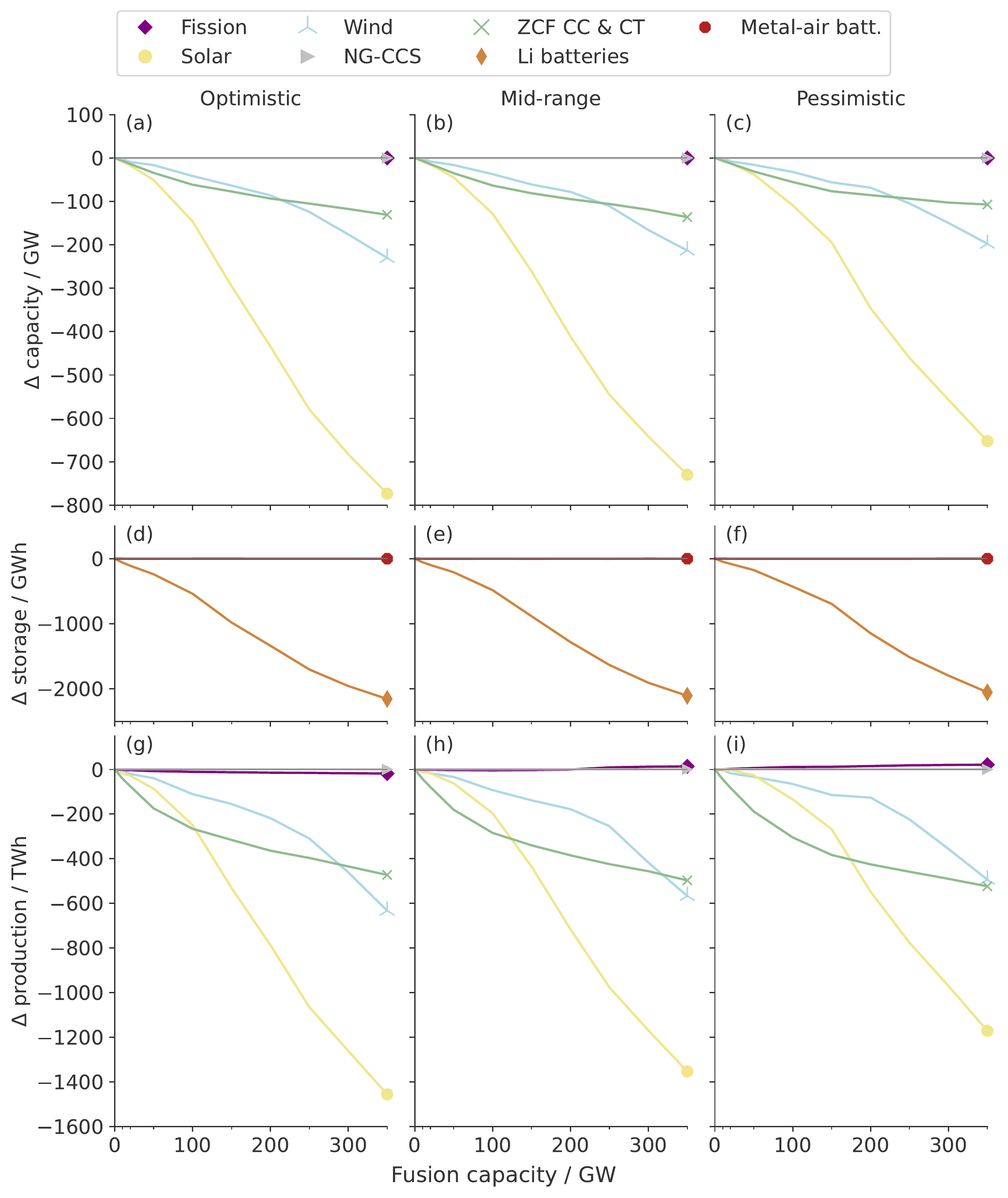}
\caption{Displaced generation capacity, storage, and energy production of other resources in the High market opportunity scenario without new fission or NG-CCS.
Dashed lines (if any) are for cases where fusion plants may include mid-priced storage.
}\label{fig:displacementhighoppnonucorccs}
\end{figure}

\clearpage

\begin{figure}
\centering
\includegraphics[width=0.95\textwidth]{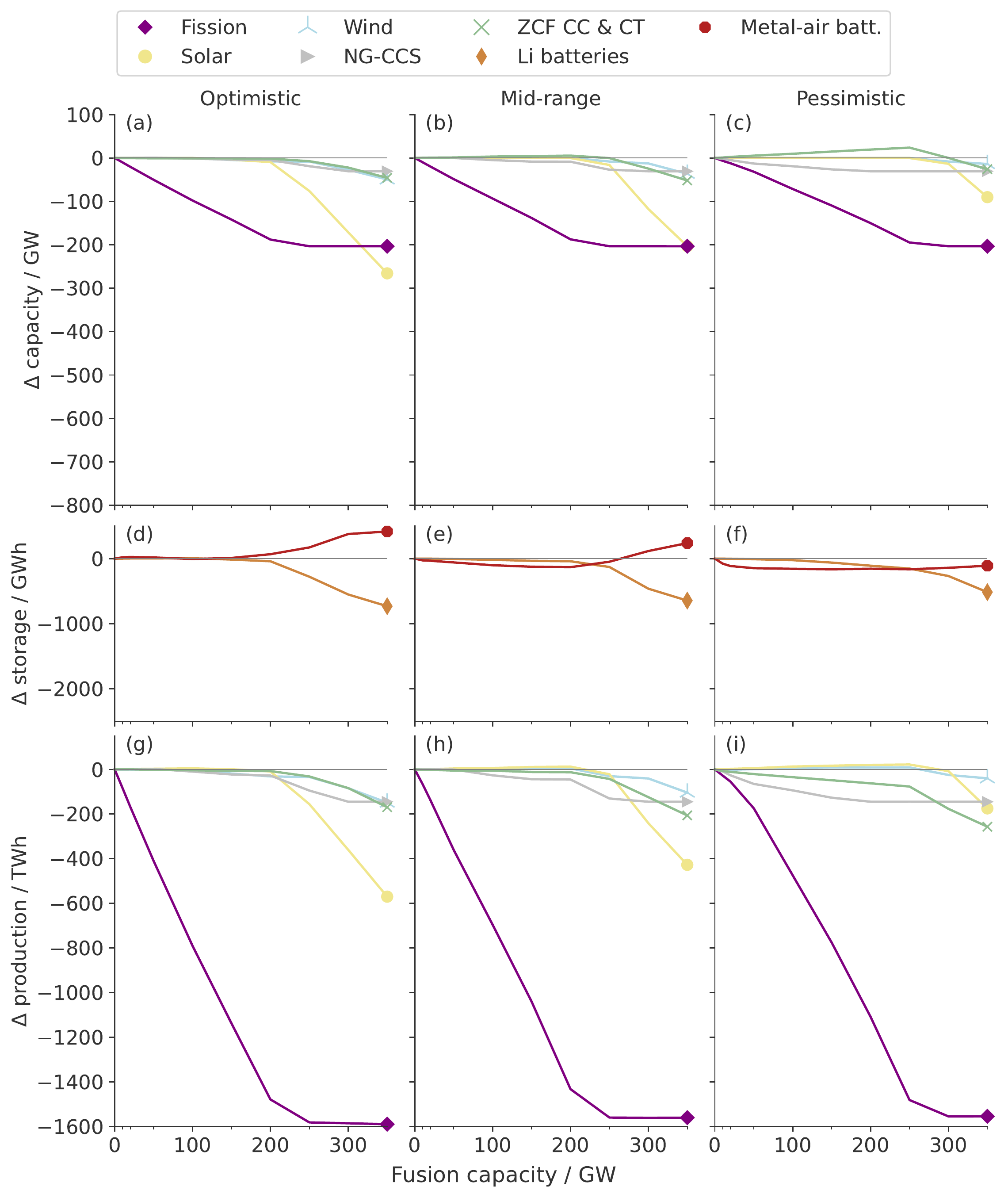}
\caption{Displaced generation capacity, storage, and energy production of other resources in the Low market opportunity scenario with constrained renewables.
Dashed lines (if any) are for cases where fusion plants may include mid-priced storage.
}\label{fig:displacementcrlowopp}
\end{figure}

\clearpage

\begin{figure}
\centering
\includegraphics[width=0.95\textwidth]{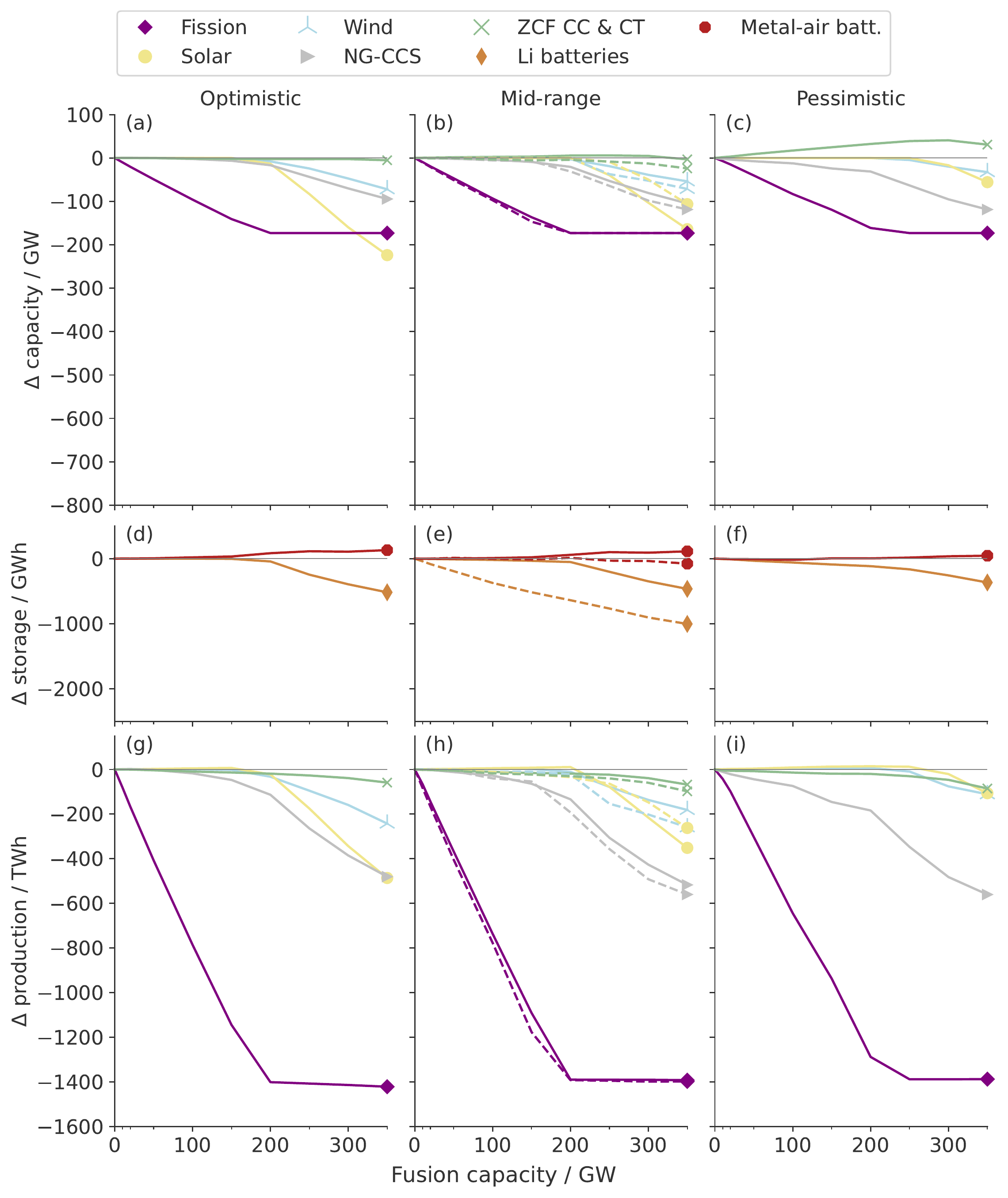}
\caption{Displaced generation capacity, storage, and energy production of other resources in the Medium market opportunity scenario with constrained renewables.
Dashed lines (if any) are for cases where fusion plants may include mid-priced storage.
}\label{fig:displacementcrmedopp}
\end{figure}

\clearpage

\begin{figure}
\centering
\includegraphics[width=0.95\textwidth]{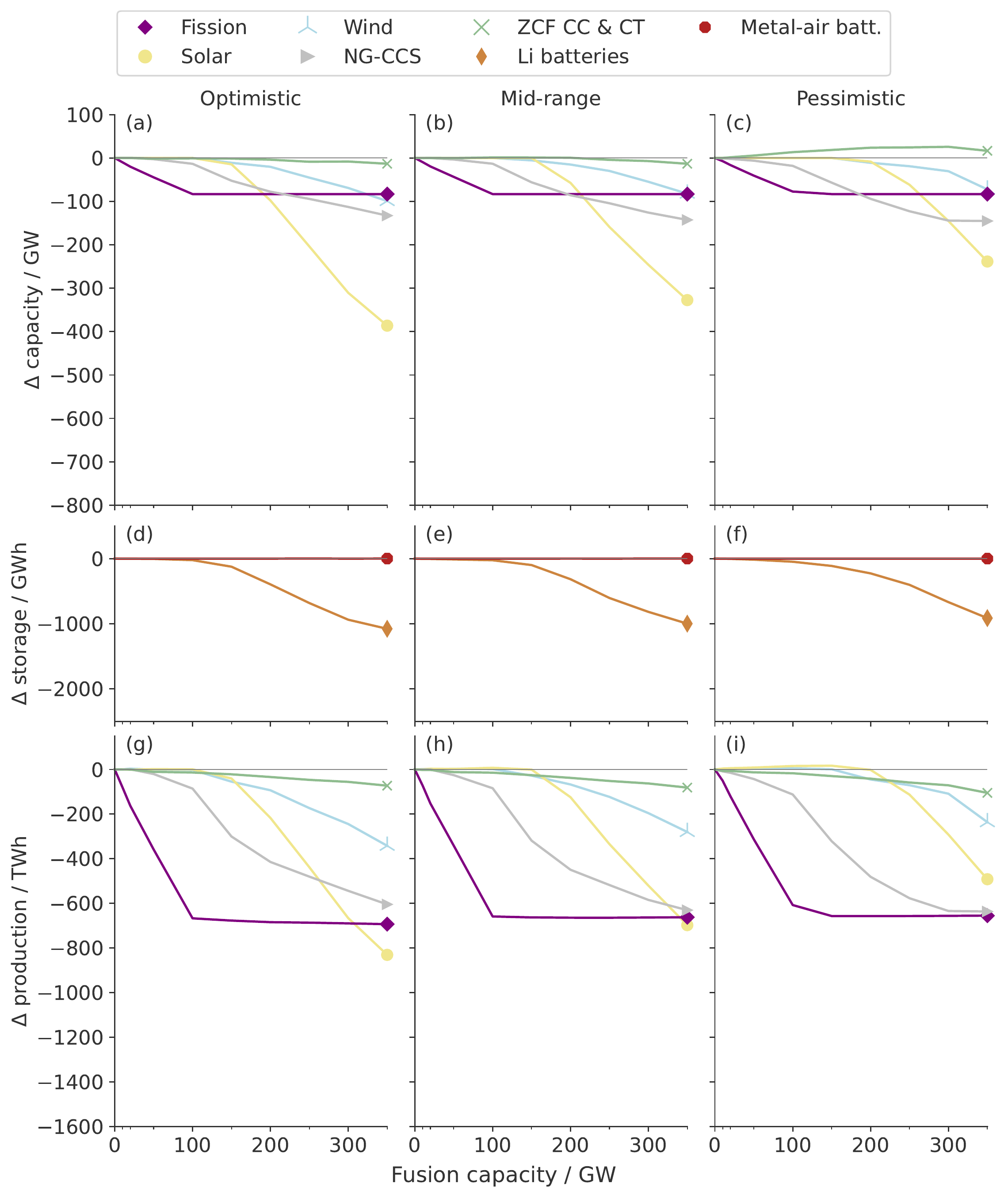}
\caption{Displaced generation capacity, storage, and energy production of other resources in the High market opportunity scenario with constrained renewables.
Dashed lines (if any) are for cases where fusion plants may include mid-priced storage.
}\label{fig:displacementcrhighopp}
\end{figure}

\newpage

\begin{figure}
\centering
\includegraphics[width=0.98\textwidth]{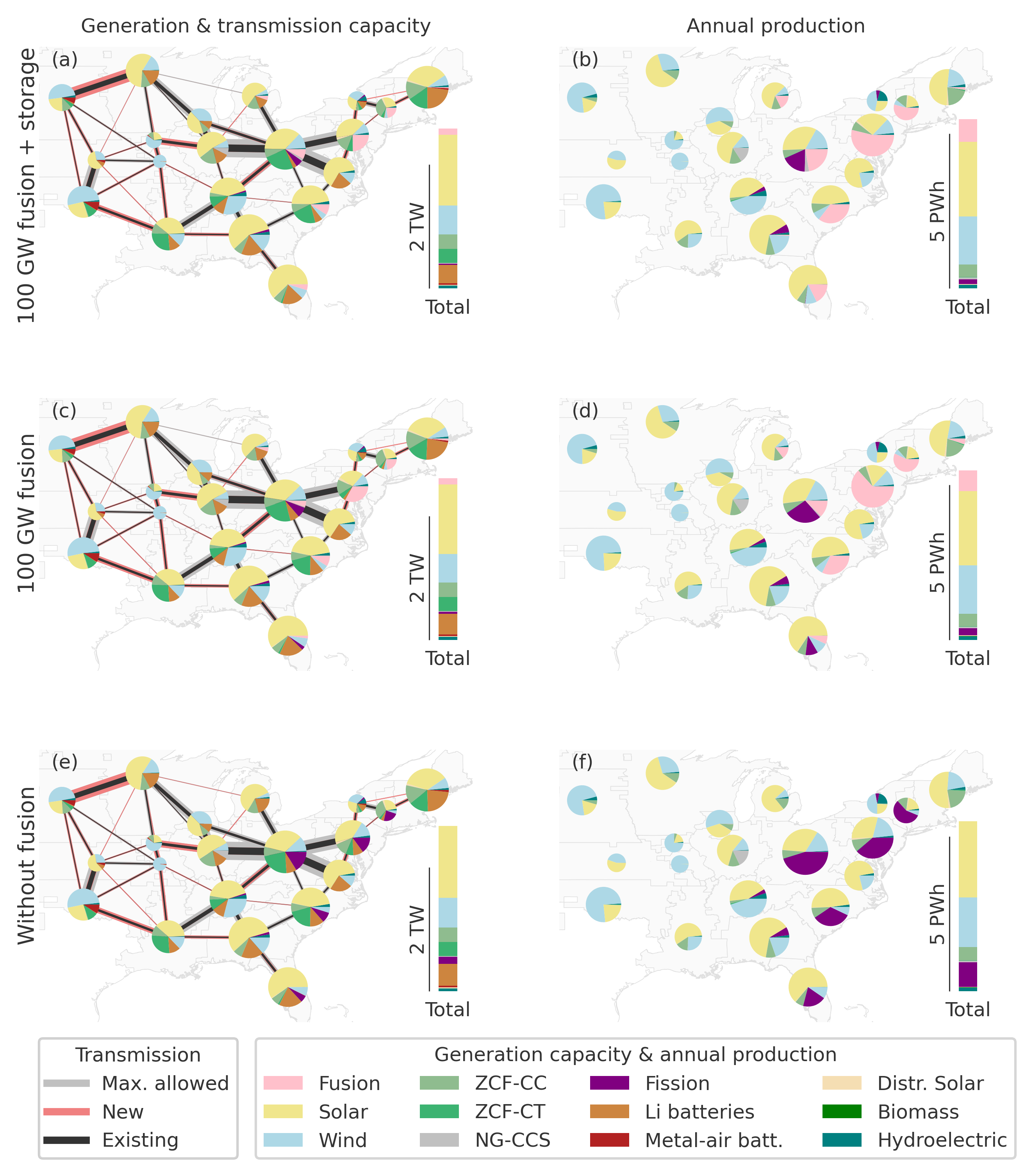}
\caption{Generation capacity and transmission capacity (left) and annual energy production (right) in each zone, sorted by resource type, for Low market opportunity cases with fusion that has a TSS, fusion without thermal storage, and without fusion.
This data are from LowOpp cases 10024, 10004, and 0.
}\label{fig:maplowopp}
\end{figure}

\clearpage

\begin{figure}
\centering
\includegraphics[width=0.98\textwidth]{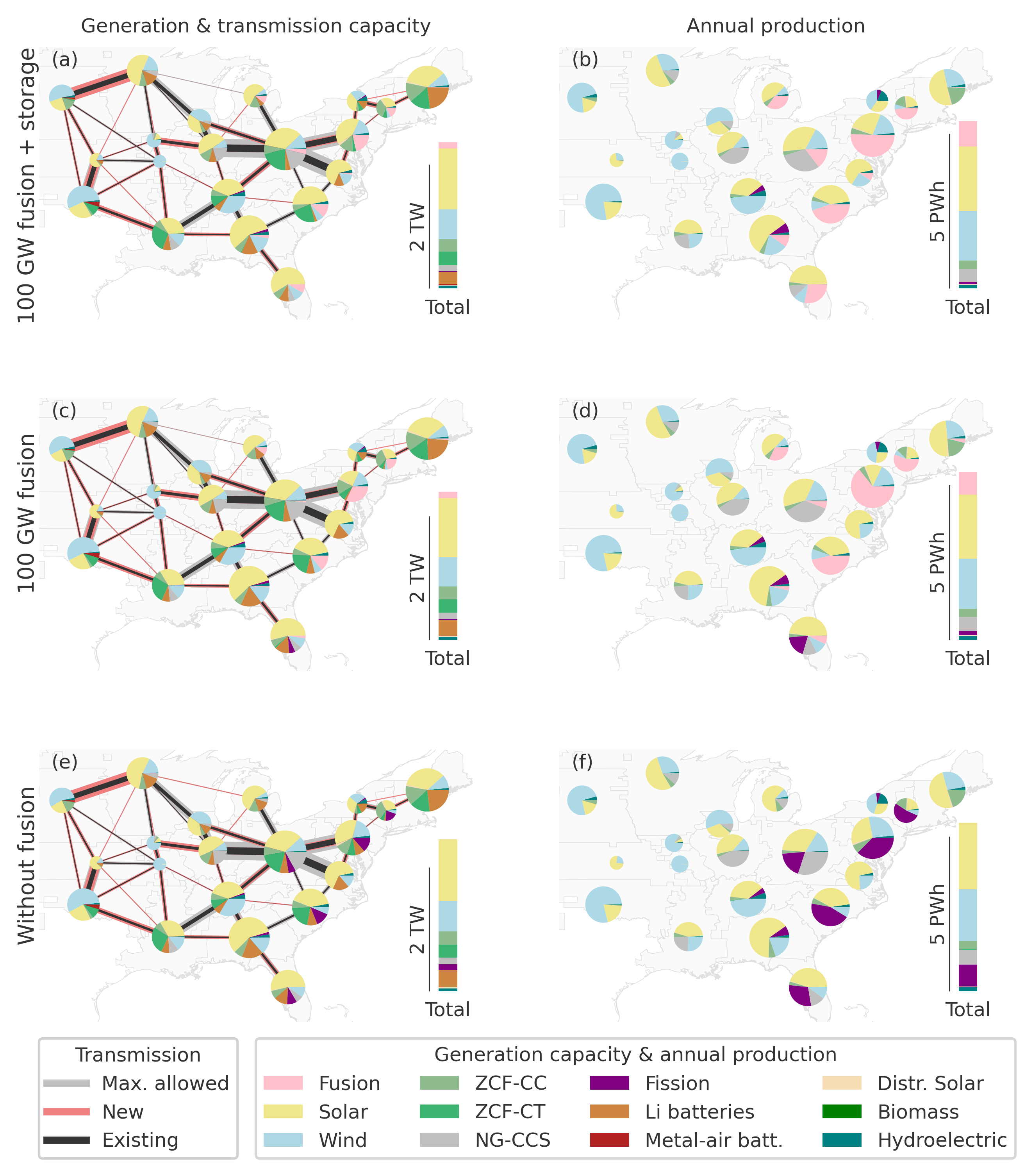}
\caption{Generation capacity and transmission capacity (left) and annual energy production (right) in each zone, sorted by resource type, for Medium market opportunity cases with fusion that has a TSS, fusion without thermal storage, and without fusion.
This data are from MedOpp cases 10024, 10004, and 0. 
}\label{fig:mapmedopp}
\end{figure}

\clearpage

\begin{figure}
\centering
\includegraphics[width=0.98\textwidth]{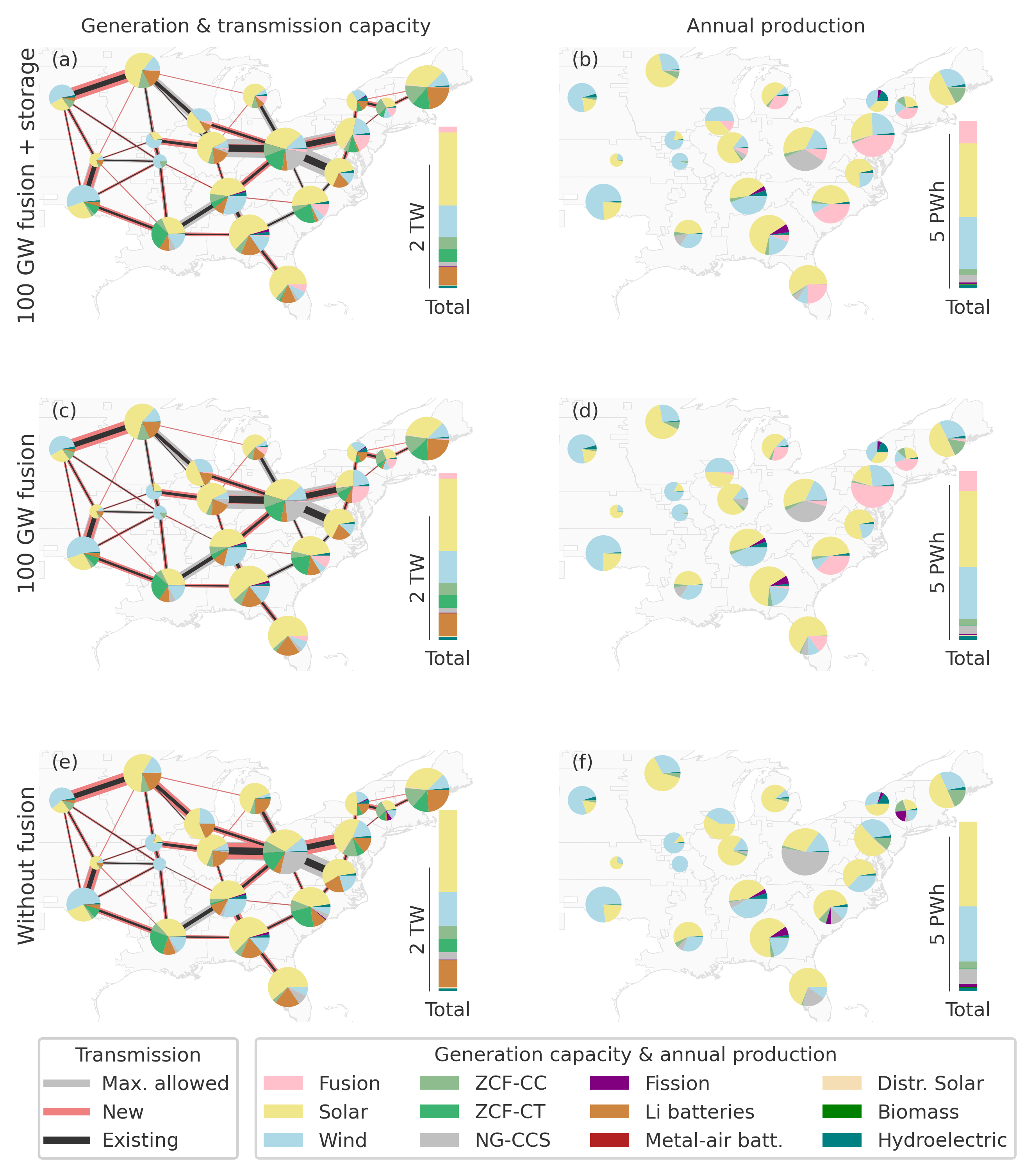}
\caption{Generation capacity and transmission capacity (left) and annual energy production (right) in each zone, sorted by resource type, for High market opportunity cases with fusion that has a TSS, fusion without thermal storage, and without fusion.
This data are from HighOpp cases 10024, 10004, and 0.
}\label{fig:maphighopp}
\end{figure}

\clearpage

\begin{figure}
\centering
\includegraphics[width=0.98\textwidth]{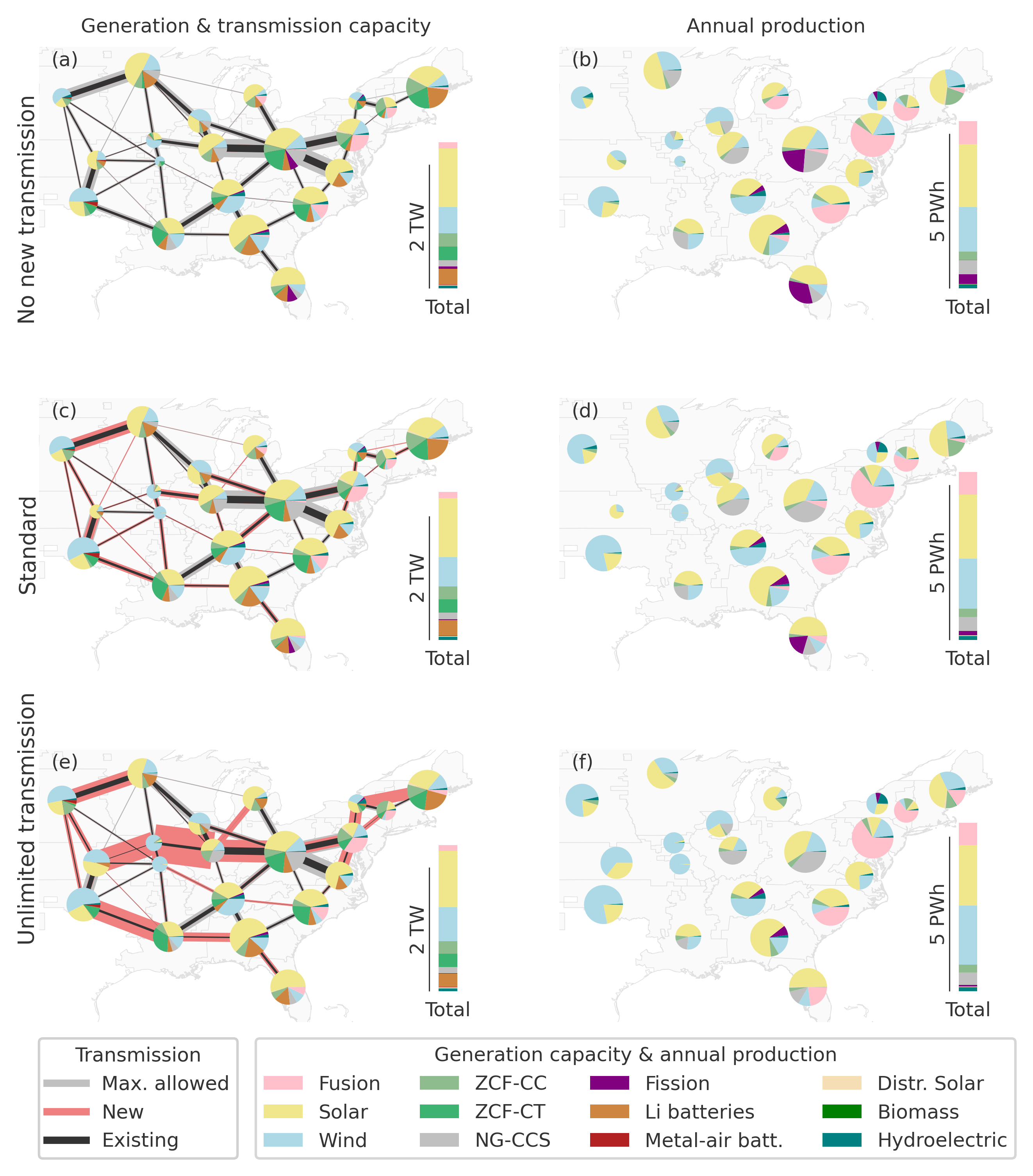}
\caption{Generation capacity and transmission capacity (left) and annual energy production (right) in each zone, sorted by resource type, for Medium market opportunity cases without new transmission, for the standard case (also seen in the middle row in Fig.~\ref{fig:mapmedopp}), and for a case with no constraints on transmission.
This data are from MedOpp cases 40004, 10004, and 30004.
}\label{fig:maptransvar}
\end{figure}

\newpage

\begin{figure}
\centering
\includegraphics[width=0.95\textwidth]{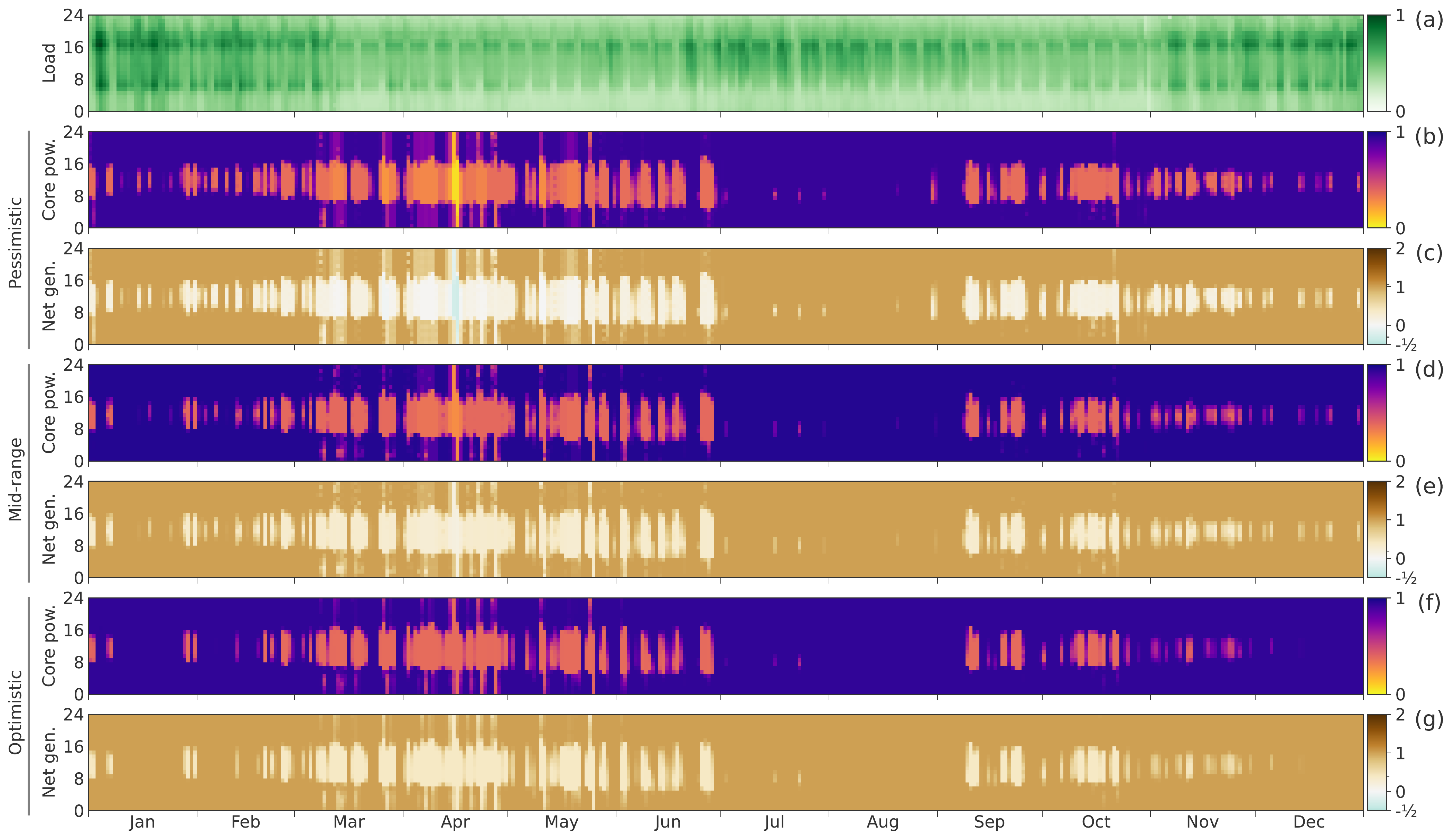}
\caption{Hourly operation of the three reference fusion plants without storage, in the Medium market opportunity scenario and in cases with \SI{100}{\giga\watt} of fusion capacity. Part (a) shows the load in the zone; (b), (d), and (f) show the normalized core power; and (c), (e), and (g) show the normalized net power output of the plant, where 1 is the maximum long-run output of the plant. See cases 304, 10004, and 104.
}\label{fig:hourly_nostor}
\end{figure}

\clearpage

\begin{figure}
\centering
\includegraphics[width=0.95\textwidth]{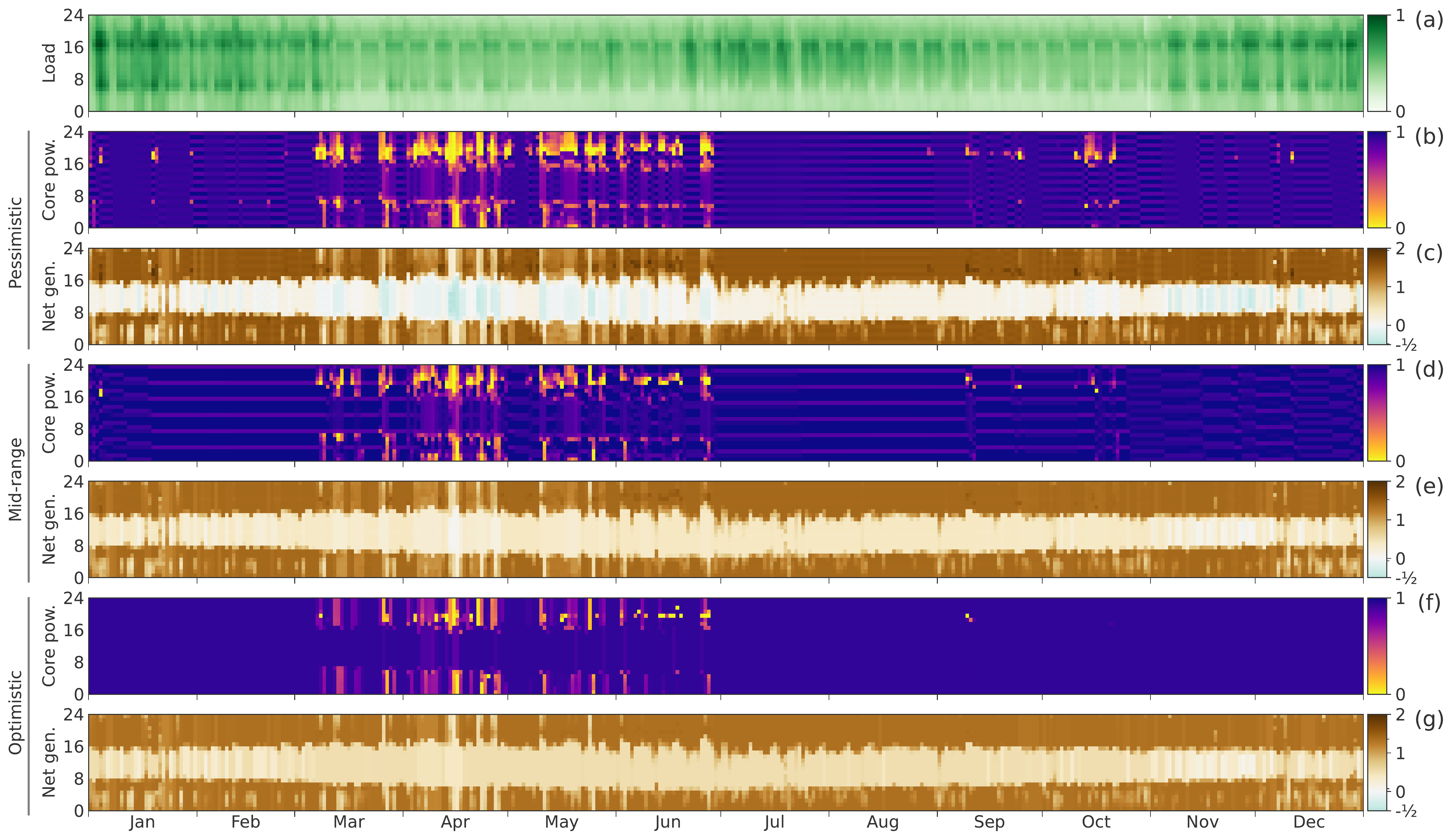}
\caption{Hourly operation of the three reference fusion plants with mid-price storage, in the Medium market opportunity scenario and in cases with \SI{100}{\giga\watt} of fusion capacity.  Part (a) shows the load in the zone; (b), (d), and (f) show the normalized core power; and (c), (e), and (g) show the normalized net power output of the plant, where 1 is the maximum long-run output of the plant.  See cases 324, 10024, and 224.
}\label{fig:hourly_withstor}
\end{figure}

\clearpage

\backmatter

%\bibliography{sn-bibliography}% common bib file
\bibliography{PlasmaControl}
%% if required, the content of .bbl file can be included here once bbl is generated
%%\input sn-article.bbl

%% Default %%
%%\input sn-sample-bib.tex%